\documentclass[
twocols
aps,
floats,
pre,
10pt,
groupedaddress,
twocolumn,
superscriptaddress,
nofootinbib]{revtex4-2}

\usepackage{amsmath,amssymb,amsthm,amsfonts,bm,bbm,braket,mathtools}
\usepackage[english]{babel}
\usepackage{comment}
\usepackage{natbib}
\usepackage{hyperref}
\usepackage[shortlabels]{enumitem}
\hypersetup{
    pdfauthor={Flavio Nicoletti, Matteo Negri, Francesco d'Amico},
    pdftitle={Stat_mech_vec_hop},
    colorlinks,
    linktocpage=true,
    pdfstartpage=1,
    pdfstartview=FitV,
    breaklinks=true,
    pageanchor=true,
    pdfpagemode=UseOutlines,
    plainpages=false,
    bookmarksnumbered,
    bookmarksopen=true,
    bookmarksopenlevel=1,
    hypertexnames=true,
    pdfhighlight=/O,
    urlcolor=blue,
    linkcolor=blue,
    citecolor=blue,
}

\allowdisplaybreaks

\usepackage[utf8x]{inputenc}	
\usepackage[T1]{fontenc} 	

\usepackage{graphicx}
\usepackage{array,booktabs}

\usepackage{soul}
\usepackage{textcomp}
\usepackage{nicematrix}
\usepackage{soul}
\usepackage{algorithm}
\usepackage{algpseudocode} 

\DeclareMathOperator{\arcos}{arcos}

\newcommand{\conf}[1]{\underline{\bm{#1}}}
\newcommand{\mm}[1]{\mathbb{#1}}
\newcommand{\sph}[1]{\mathcal{S}_{#1-1}}
\newcommand{\K}[1]{\mathcal{K}_{#1}}
\newcommand{\g}[1]{\mathrm{g}_{#1}}
\newcommand{\Oo}[1]{\mathcal{O}#1}
\newcommand{\Tr}[1]{\operatorname{Tr}(#1)}
\newcommand{\Green}{\mathcal{G}}

\def\Xint#1{\mathchoice
   {\XXint\displaystyle\textstyle{#1}}%
   {\XXint\textstyle\scriptstyle{#1}}%
   {\XXint\scriptstyle\scriptscriptstyle{#1}}%
   {\XXint\scriptscriptstyle\scriptscriptstyle{#1}}%
   \!\int}
\def\XXint#1#2#3{{\setbox0=\hbox{$#1{#2#3}{\int}$}
     \vcenter{\hbox{$#2#3$}}\kern-.5\wd0}}

\def\dashint{\Xint-}
\begin{document}

\title{Statistical mechanics of vector Hopfield network near and above saturation}

\author{Flavio Nicoletti}
\email[Corresponding author: ]{flavion@chalmers.se}
\affiliation{Department of Computer Science and Engineering Chalmers University of Technology and University of Gothenburg SE-41296 Gothenburg, Sweden}
\affiliation{Dipartimento di Fisica, Sapienza Università di Roma, 00185 Rome, Italy}

\author{Francesco D'Amico}
\email[Corresponding author: ]{francesco.damico@uniroma1.it}
\affiliation{Dipartimento di Fisica, Sapienza Università di Roma, 00185 Rome, Italy}
\affiliation{Institute of Nanotechnology, National Research Council of Italy, CNR\text{-}NANOTEC, Rome Unit}

\author{Matteo Negri}
\email[Corresponding author: ]{matteo.negri@uniroma1.it}
\affiliation{Dipartimento di Fisica, Sapienza Università di Roma, 00185 Rome, Italy}
\affiliation{Institute of Nanotechnology, National Research Council of Italy, CNR\text{-}NANOTEC, Rome Unit}

\date{\today}

\begin{abstract}
We study analytically and numerically a Hopfield fully-connected network with $d$-dimensional vector spins. These networks are models of associative memory that generalize the standard Hopfield with Ising spins, where $P$ examples are stored in a network of $N$ units as local minima in an energy landscape. We study the equilibrium and out-of-equilibrium properties of the system, considering the system in its retrieval phase $\alpha<\alpha_c$ and beyond, where $\alpha=P/N$ is the capacity of the system and $\alpha_c$ is its critical value, above which storage fails.

We derive the Replica Symmetric solution for the equilibrium thermodynamics of the system, together with its phase diagram: we find that the retrieval phase of the network shrinks with growing spin dimension, having ultimately a vanishing critical capacity $\alpha_c\propto 1/d$ in the large $d$ limit. As a trade-off, we observe that in the same limit vector Hopfield networks are able to denoise corrupted input patterns in the first step of retrieval dynamics, up to very large capacities $\alpha\propto d$.

We also study the static properties of the system at zero temperature, considering the statistical properties of soft modes of the energy Hessian spectrum. We find that local minima of the energy landscape related to memory states have ungapped spectra with rare soft eigenmodes: these excitations are localized, their measure condensating on the noisiest neurons of the memory state.

\end{abstract}

\maketitle

\section{Introduction}
\label{sec:introduction}
Associative memories are neural networks that can be thought in physical terms as models where the energy function is designed to have minima in correspondence of given "example" configurations, so that the dynamics of the model can reconstruct the examples if it is initialized close enough to one of those.
The presence of an energy function is a major conceptual advantage, as it allows to predict the  limits of functionality of these networks with the theoretical tools from the statistical mechanics of disordered systems.

Despite their importance as a theoretical model for biological networks, associative memories played a minor role during the first years of the deep learning revolution.  
However, many years after the classical studies on systems with higher order interactions \cite{PerettoNiez1986, E_Gardner_1987}, developments in machine learning have renewed interest in models with enhanced capabilities, the so-called modern Hopfield networks, that can store polynomial \cite{krotov2016dense} or even exponential \cite{Demircigil_2017,lucibello2024exponential} number of examples. Moreover, this class of associative memories appeared in multiple context of modern deep learning, in particular transformers architectures \cite{ramsauer2020hopfield} and generative diffusion \cite{ambrogioni_search_2024}, gathering the interest of the statistical physics community \cite{krotov2023new}.

The connection between associative memories and the attention mechanism at the base of transformers architectures \cite{vaswani2017attention} is particularly important: given that the success of transformers is still unexplained, the possibility of using statistical mechanics of associative memories to understand their performance is worth investigating. 
This connection relies on the interpretation of a single attention forward pass as one step of gradient descent of an energy function \cite{ramsauer2020hopfield}, which can be thought as the energy of an associative memory where the minima are learned from the data in a complex way rather that just assigned to single examples \cite{hoover2023energy}. This perspective is naturally useful for architectures that repeat the same layer multiple times \cite{devlin_bert_2019, jumper2021highly, damico2024self}, but was also found to be important for a single forward pass \cite{smart_-context_2025}.
The connection was further strengthened by the finding that associative memories can create additional minima in the energy that correspond to previously unseen examples \cite{kalaj2024randomfeatureshopfieldnetworks}, showing that even the mechanism for generalization may be theoretically accessible.

Still, to successfully apply the physics of associative memories to transformers architectures, we need to address a conceptual node: that transformers operate on sequences of tokens. A token is a vector in a learned embedding space that represents a concept, which can be decoded in different data spaces allowing multi-modal applications such as image generation from a text prompt. Such embedding vectors are often very high dimensional, reaching thousands of components in the largest, most successful architectures \cite{jumper2021highly, openai2022introducing}. 
Notably, such a large number of dimensions is an ideal setting for statistical mechanics, since it may be well described by a thermodynamic limit. A first attempt to understand transformers in terms of the statistical physics of vector spins has recently been made in \cite{bal2023spinmodeltransformers}.

High-dimensional vector spins also have neurobiological relevance, as they can represent groups of continuous-valued neurons (we will discuss this more at the end of Section~\ref{sec:discussion}).

\textbf{Contributions}. In this work, we introduce $d$-dimensional vector spins into a standard Hopfield network \cite{hopfield1982neural}. To the best of our knowledge, this is the first time that such degrees of freedom are introduced in an associative memory. Previously, 2-dimensional vector spins have been considered in oscillatory networks \cite{kymnoscillator}: these models, pioneered by so-called phasor networks \cite{noest1987phasor}, study the problem of features extraction in composite data, by extending results from resonator networks \cite{frady_resonator_2020, kent_resonator_2020} using complex number neurons. Moreover, artificial Kuramoto oscillators neurons were recently proposed as alternatives to threshold units in neural networks layers for machine learning \cite{miyato2025artificial}.  

We study the statistical mechanics of a vector Hopfield network in great detail, considering both its static and dynamical properties.
First, we characterize the equilibrium properties of the system, in terms of temperature $T$ and storage capacity $\alpha$: we compute the phase diagram for any $d$, by extending the classic calculations from \citet{amit1987statistical}, identifying spin glass and retrieval transitions;
we also identify in the high-dimensional limit $d\rightarrow\infty$ the correct scaling with $d$ of the capacity $\alpha$ for retrieval. 
We also study the stability of equilibrium states, in terms of $T=0$ linear low-energy excitations, represented by the lowest modes of the hessian of the energy function.
Hessian soft modes provide interesting insights on the physics of continuous-spins disordered systems with structured energy landscape, since they control their linear stability. In the present model, the study of hessian soft modes allows for the characterization of memory or Mattis states and spurious states in terms of their spectral behavior, potentially establishing a connection between the behavior of associative memory neural networks and glassy systems.
Finally, we study the dynamics of the model numerically and theoretically, focusing on the ability of the model to de-noise examples. We investigate the dynamics especially in the above saturation regime, where the network cannot store memories of patterns as local minima of the energy landscape.

\textbf{Summary of results.} 
While the main motivation for our work is to understand basic properties that can be extended to modern Hopfield networks in future research, among the results that we found there are some notable behaviors that are interesting per se.
The core findings of our work are the following:
\begin{itemize}
    \item \emph{Vector Hopfield networks can denoise patterns and mixtures above their storage capacity limits, through first-step retrieval or transient denoising} \cite{clark2025transientdynamicsassociativememory} (Section \ref{sec:thermodynamics} and Section \ref{sec:retrieval_dynamics}): in particular, the critical storage capacity for patterns vanishes with spin dimension $\alpha_c\propto 1/d$ (Section \ref{sec:thermodynamics}), whereas their first-step retrieval capacity increases $\widetilde{\alpha}\propto d$ (Section \ref{sec:retrieval_dynamics}).
    \item \emph{Memory states of Vector Hopfield networks are stable mean-field glasses} \cite{franz2022delocalization, franz2022linear} (Section \ref{sec:linear_stability}): the spectrum of the Hessian matrix of local minima of the energy landscape representing memory states does not present any gap ($\lambda_{min}\rightarrow 0$ for $N\rightarrow\infty$). Soft modes are rare and localized, spatially correlated with the noise generated by the storage process. Conversely, spurious states abound with delocalized and featureless soft modes.
\end{itemize}
Our results are discussed thoroughly in Section \ref{sec:discussion}.

\textbf{Outline of the paper.}
In section~\ref{sec:model_def} we give a precise definition of the model. In section~\ref{sec:thermodynamics} we compute the equilibrium properties of the model. In section~\ref{sec:retrieval_dynamics} we study the out of equilibrium properties. Finally, in section ~\ref{sec:discussion} we discuss our results and draw conclusions.

\section{Model definition}
\label{sec:model_def}

The Vector Hopfield Model (VHM) we study in this work is defined by the following energy function of $N$ spins and $P$ examples
\begin{eqnarray}
\label{eq:ModelHam}
    &\mathcal{H}(\conf{S};\{\conf{\xi}^{(\mu)}\}_{\mu=1}^P)\,=\,-\frac{1}{2}\sum_{i\neq j}^{1, N}\bm{S}_i\cdot \mm{J}_{ij}\bm{S}_j \\
    & \nonumber \\
    &\mm{J}_{ij}=\frac{1}{N}\sum_{\mu=1}^P\bm{\xi}_{i; \mu}\bm{\xi}_{j;\mu}^\intercal\nonumber.
\end{eqnarray}
Neuron spins $\{\bm{S}_i\}$ are vectors with $d$ components and fixed norm $\sigma$, while example spins $\{\bm{\xi}_{i}^{(\mu)}\}$ are independent and identically distributed (iid) random vectors with $d$ components and unit norm, drawn from the uniform distribution on the unit sphere in $d$ dimensions:
\begin{equation}
\label{eq:disorder_distr}
    P_{\xi}(\bm{\xi})=\frac{1}{\sph{d}(1)}\delta(|\bm{\xi}|-1)
\end{equation}
where $\sph{d}(1)=\frac{2\pi^{d/2}}{\Gamma(d/2)}$ is the surface of the unit hypersphere embedded in dimension $d$.
In the following, with the notation $\conf{S}=(\bm{S}_1,\dots, \bm{S}_N)^\intercal$ and $\conf{\xi}^{\mu}=(\bm{\xi}_1^\mu,\dots,\bm{\xi}_N^\mu)^\intercal$ we identify respectively configuration of spins and patterns.

Our definition in eqs. \eqref{eq:ModelHam},\eqref{eq:disorder_distr} is a natural extension of the standard Hopfield Model (SHM) to vector spins: the SHM Hamiltonian in fact is the particular case $d=1$ of \eqref{eq:ModelHam}, \eqref{eq:disorder_distr}. In section \ref{sec:thermodynamics} we study the thermodynamics of the model in \eqref{eq:ModelHam}.

Note that our choice of couplings in \eqref{eq:ModelHam} implies no global spin symmetries-the invariance under the $O(d)$ group in the case of vector spins. Conversely, the SHM in his original definition in \cite{hopfield1982neural, amit1985spin, amit1985storing} has a global spin-inversion symmetry.
In general, the VHM in \eqref{eq:ModelHam} is not the only possible generalization of the SHM to vector spins: an isotropic VHM with scalar couplings could also be defined. By preliminary studies of such VHM, we found that it bears no practical advantages over our model in \eqref{eq:ModelHam}, but rather unwanted features such as zero modes connected to the $O(d)$ symmetry. 

Eq. \eqref{eq:ModelHam} is a model of associative memory as the SHM studied in \cite{hopfield1982neural, amit1985spin, amit1985storing}: configurations of neurons correlated to examples can be stored as local minima of the energy in \eqref{eq:ModelHam}. Depending on how large is $P$, the energy landscape represented by \eqref{eq:ModelHam} may include so-called spurious minima, configurations of neurons unrelated to the original set of $P$ examples. 
The retrieval dynamics equations of the VHM can be obtained from the condition of stationarity of the energy function \eqref{eq:ModelHam}.
Stationary points can be identified by introducing Lagrange multipliers $\{\eta_i\}_{i=1}^N$ that enforce spins to have fixed norms:  
\begin{equation}
    \label{eq:stationarity_modelHam}
    \frac{\partial}{\partial \bm{S}_i}\left[\mathcal{H}(\conf{S})+\sum_{i=1}^N\eta_i\;(|\bm{S}_i|-\sigma)\right]\Bigl|_{\conf{S}=\conf{S}_*}=0
\end{equation}
so that
\begin{equation}
    \begin{cases}
        \eta_i=
        |\sum_{j:j\neq i}\mm{J}_{ij}\bm{S}_j^*|\\
        \; \\
        \bm{S}_i^* = \frac{\sigma\sum_{j:j\neq i}\mm{J}_{ij}\bm{S}_j^*}{|\sum_{j:j\neq i}\mm{J}_{ij}\bm{S}_j^*|}
    \end{cases}
\end{equation}
The Lagrange multipliers in \eqref{eq:stationarity_modelHam} are fixed to equate the norm of the local field vectors acting on the spins. The physical meaning of the $\{\eta_i\}$ is that they represent the strength of the input signal that each spin neuron $i$ receives from all the other spin neurons. In machine learning, local fields are often called pre-activations or receptive fields, depending on the context. In the following, when we use the term \textit{local field}, we will always refer to the Lagrange multipliers $\eta_i$, and not to the corresponding vectors in \eqref{eq:stationarity_modelHam}.
Eq. \eqref{eq:stationarity_modelHam} can be used to build a minimization algorithm for eq. \eqref{eq:ModelHam}. Choosing a syncronous update rule for the spins, we introduce the \emph{versor rule} for retrieval dynamics
\begin{equation}
\label{eq:retrieval_versor_rule}
    \bm{S}_i(t+1)=\frac{\sigma\sum_{j:j\neq i}\mm{J}_{ij}\bm{S}_j(t)}{|\sum_{j:j\neq i}\mm{J}_{ij}\bm{S}_j(t)|}=\frac{\sigma\bm{\eta_i}(t)}{|\bm{\eta_i}(t)|}
\end{equation}
According to \eqref{eq:retrieval_versor_rule}, at each time step each spin is aligned to its local field vector $\bm{\eta}_i(t)=\sum_{j:j\neq i}\mm{J}_{ij}\bm{S}_j(t)$. Eq. \eqref{eq:retrieval_versor_rule} is a greedy algorithm, since at each time step the spin configuration update follows the steepest descent direction, represented by the negative gradient of the energy.  
Note that \eqref{eq:retrieval_versor_rule} is the generalization to VHMs of the \emph{sign rule} introduced by Hopfield in \cite{hopfield1982neural}. In practice, the iteration of eq. \eqref{eq:retrieval_versor_rule} is stopped when the average overlap of the spins with the spins in the previous configuration is larger than a threshold $1-\epsilon$. In our simulations, we always use $\epsilon=10^{-12}$.
The retrieval dynamics of our VHM is discussed in greater detail in Appendix \ref{sec:details_retrieval_dynamics}. In addition,
in section \ref{sec:retrieval_dynamics} we study the retrieval dynamics just after initialization, unveiling the phenomenon of transient denoising.

At variance with the SHM, VHMs are continuous models
and thus allow for studying linear excitations around local minima, which are encoded in the $Nd\times Nd$ Hessian matrix of the energy function \eqref{eq:ModelHam}. The Hessian can be written as a block matrix in site indices
\begin{equation}
    \label{eq:Hessian_0}
    \mm{M}_{ij}(\conf{S})=\mm{P}_{\perp}(\bm{S}_i)\left(-\mm{J}_{ij}+\frac{\eta_i}{\sigma}\delta_{ij}\mm{I}_d\right)\mm{P}_{\perp}(\bm{S}_j)
\end{equation}
where $\mm{I}_d$ is the $d\times d$ identity matrix.
Orthogonal projectors $\mm{P}_{\perp}(\bm{S})=\mm{I}_d-\frac{\bm{S}\bm{S}^\intercal}{\sigma^2}$ suppress longitudinal fluctuations and are a consequence of fixed norm constraints on spins. In general, the Hessian \eqref{eq:Hessian} depends on the configuration on which it is evaluated, but in mean field systems it is well known that in the thermodynamic limit $N\rightarrow\infty$ physical hessians can be represented by random matrix ensembles: this simplification allows one to study analytically their properties.
Hessian matrices with a block structure similar to that in eq. \eqref{eq:Hessian_0} appear frequently in disordered systems; see, for instance, \cite{cicuta2018unifying, benetti2018mean, franz2022delocalization, franz2022linear, patil2024spectral}.
In section \ref{sec:linear_stability} we will explain how we study \eqref{eq:Hessian} through random matrix theory.
We are interested in studying the behavior of the spectrum of \eqref{eq:Hessian} close to the lower spectral edge: this region is the most relevant for the physics of the system, since Hessian soft modes rule both the long-time close-to-equilibrium relaxation and the linear stability of the system. The fixed points of retrieval,  corresponding to either Mattis states or spurious states, can thus be characterized according to the properties of their excitation spectra.

\section{The network below saturation}

In this section we study the properties of the VHM \eqref{eq:ModelHam} below saturation, $\alpha<\alpha_c$. In this regime, when initialized close to one of the patterns, the system can retrieve a memory of the pattern through retrieval dynamics \eqref{eq:retrieval_versor_rule}.Therefore, in this section we only consider the static properties of the system, deriving first the solution at thermodynamic equilibrium for generic temperatures $T$ and storage capacity $\alpha=P/N$ and then focusing on the linear stability of the system at zero temperature, by studying the statistics of soft modes of the Hessian \eqref{eq:Hessian}.

\subsection{Equilibrium thermodynamics}
\label{sec:thermodynamics}
Thermal averages with respect to the Boltzmann-Gibbs distribution are denoted by the symbol $\langle\cdots\rangle$, while averages over examples  \eqref{eq:disorder_distr} by $\overline{\cdots}$.

\subsubsection{Order parameters}

The order parameters describing the retrieval phase, as in the SHM, are the so-called Mattis magnetizations
\begin{equation}
\label{eq:Mattis_magn}
    m_{\mu}\,=\,\lim_{N\rightarrow\infty}\frac{1}{N\sigma}\sum_{i=1}^N \langle \bm{S}_i\rangle \cdot \bm{\xi}_i^{\mu}=\frac{1}{\sigma}\overline{\bm{S}_1\cdot\bm{\xi}_1^{\mu}}
\end{equation}
These quantities describe the correlation of the equilibrium state with the examples. Let us consider the vector $\bm{m}=(m_1,\dots,m_P)^{\intercal}$ to identify Mattis magnetizations.
When the equilibrium state is correlated to a single example, like $\bm{m}=(0,\dots,0,m_{\mu},0,\dots,0)^{\intercal}$, the equilibrium state is a \emph{Mattis} or \emph{memory} state; instead, if the equilibrium state is correlated with a finite number $s\ll P$ of examples, like $\bm{m}=(m_1,\dots,m_s,0,\dots,0)^{\intercal}$, we identify it as a mixture state; finally, states uncorrelated with all examples, $\bm{m}=\bm{0}$, are spin-glass states. Mixtures and spin glass solutions are termed \emph{spurious states} in the classical Hopfield literature \cite{amit1989modeling}. Recently, it was shown that mixture states can be a pathway to understanding learning and generalization in SHM with random features \cite{negri2023storage, kalaj2024randomfeatureshopfieldnetworks}.

When $P\propto N$ ($\alpha>0$), analogously to the SHM, we find that spurious states are spin glass states with Replica Symmetry Breaking (RSB): the order parameter in this case is the overlap distribution $P(q')=\overline{\sum_{\alpha\beta}w_{\alpha}w_{\beta}\delta(q'-q_{\alpha\beta})}$ between different pure states $\alpha, \beta$ \cite{mezard1987spin}. In this work, we consider the Replica Symmetry (RS) ansatz $P(q')\simeq \delta(q'-q)$: in this case, the order parameter is the Edward-Anderson overlap
\begin{equation}
\label{eq:EA_overlap}
    q\,=\,\lim_{N\rightarrow\infty}\frac{1}{N \sigma^2}\sum_{i=1}^N |\langle \bm{S}_i\rangle|^2=\frac{1}{\sigma^2}\overline{|\langle \bm{S}_1\rangle|^2}.
\end{equation}
This order parameter describes the persistence of spins in a disordered magnetic state. 
We found that for $\alpha>0$ the RS ansatz for the overlap is exact for Mattis states. For spurious states, the RS ansatz is exact only in the subextensive regime $P\ll N$ ($\alpha=0$). In this regime, one can show that $q=m^2$, where $m$ is the equilibrium Mattis magnetization.

As in the SHM, when $\alpha>0$ Mattis states have a magnetization $0<m<1$: the network is not able to perfectly retrieve examples, but rather a noisy memory of them. The noise on each Mattis state comes from the feedback of all the other Mattis states and has a typical value $\sqrt{\alpha r}$, where $r$ is the noise parameter
\begin{equation}
\label{eq:noise_order_parameter}
\begin{split}
    r&=\frac{1}{P}\sum_{\mu=2}^P\left(\frac{1}{\sqrt{N}}\sum_{i=1}^N\bm{\xi}_i^{\mu}\cdot\langle\bm{S}_i\rangle\right)^2 \\
    \, \\
    &=\overline{\left(\frac{1}{\sqrt{N}}\conf{\xi}^{(2)}\cdot\langle\bm{S}\rangle\right)^2}
\end{split}
\end{equation} 
As we show in Appendix \ref{sec:Replica_computation}, the noise parameter $r$ is not independent, but rather a function of $q$, therefore it cannot be strictly considered an order parameter.

\subsubsection{Equilibrium free energy}
 
We computed the free energy $f=-\lim_{N\rightarrow\infty}\frac{T}{N}\overline{\log \mathcal{Z}}$ using the Replica Method, a standard technique in the statistical physics of disordered systems \cite{mezard1987spin}: it is based on the renown Replica trick
\begin{equation}
\label{eq:replica_trick}
    \overline{\log\mathcal{Z}}=\lim_{n\rightarrow 0}\frac{\overline{\mathcal{Z}^n}-1}{n}.
\end{equation}
In practice, with \eqref{eq:replica_trick}, one introduces $n$ uncoupled copies of the original system and then evaluates the partition function of the replicated system.
This technique allows in mean field models to decouple the degrees of freedom (in our model, neuron spins) after the disorder average is performed, reducing the computation of the replicated partition function to a saddle-point integral
\begin{equation}
\label{eq:partition_function_evaluated}
\begin{split}
    \overline{\mathcal{Z}^n}\,&=\,\overline{\int D\conf{S}_1\cdots D\conf{S}_n e^{-\beta\sum_{a=1}^n \mathcal{H}(\conf{S}_a)}}\\
    &=\int d\mm{Q}
    \prod_{\mu=1}^s\prod_{a=1}^n (d\mathrm{m}_{\mu}^a
    )\;e^{N\,A_n(\{\mathrm{m}_{\mu}^a\}, 
    \,\mm{Q}
    )}.
\end{split}
\end{equation}
The replica action $A_n$ depends on the overlap matrix and the replicated Mattis magnetizations 
\begin{equation}
\begin{gathered}
    Q_{ab}=\frac{1}{N\sigma^2}\conf{S}_a\cdot\conf{S}_b\qquad a\neq b=1,\dots, n \\
    \; \\
    \mathrm{m}_{\mu}^a\,=\,\frac{1}{N\sigma}\conf{S}_a\cdot\conf{\xi}_{\mu}\qquad \mu=1,\dots, s\ll P
\end{gathered}
\end{equation}
In this work, we use the Replica Symmetric (RS) ansatz: $Q_{ab}=\mathrm{q}$ for any $a\neq b$ and $m_{\mu}^a=\mathrm{m}_{\mu}$ for any $a$. The free energy of the system is obtained after casting the saddle point evaluation of \eqref{eq:partition_function_evaluated} back into \eqref{eq:replica_trick}
\begin{equation}
\label{eq:free_energy_density_from_replica}
\begin{split}
     f(\alpha, \beta) =& -\frac{1}{\beta}\lim_{n\rightarrow 0}\frac{\max A_n(\{\mathrm{m}_{\mu}\}_{\mu=1}^s, \mathrm{q})}{n} \\
     \\
     =& \max \phi(\{\mathrm{m}_{\mu}\}_{\mu=1}^s, \mathrm{q}),  
\end{split}     
\end{equation}
where $\beta\equiv 1/T$ is the inverse temperature \footnote{Note that in last equation we used the known fact that when $n\rightarrow 0$, $A_n=O(n)$ and the saddle point extremizing the action switches from a maximum to a minimum}.
The replica order parameters extremizing the free energy correspond to the physical order parameters \eqref{eq:Mattis_magn}, \eqref{eq:EA_overlap}. The noise parameter \eqref{eq:noise_order_parameter} corresponds to a conjugated field of $\mathrm{q}$ in the replica computation.
The complete replica computation of the free energy is presented in Appendix \ref{sec:Replica_computation}, together with the expression of $A_n$.

Despite being non-rigorous, Replica method leads to the correct result in several spin glass models \cite{talagrand2006parisi}. Alternatively, the RS solution of the model can also be derived with the cavity method \cite{mezard1987spin, mezard2009information}, by generalizing the approach used in \cite{mezard1989space, mezard2017mean} to vector spins.

\subsubsection{Sub-extensive storage}

We found in the $\alpha=0$ case the free energy
\begin{equation}
\label{eq:eq_free_en_dens_a0}
\begin{split}
    f_{eq}(\beta)\,&=\,\frac{\sigma^2}{2}\sum_{\mu\leq s} m_\mu^2 \\
    &-
    \frac{1}{\beta}\overline{\ln \sigma^{d-1}\mathcal{K}_d\left(\beta \sigma^2 \left|\sum_{\mu\leq s}m_{\mu}\bm{\xi}_{\mu}\right|\right)}
\end{split}
\end{equation}
where Mattis magnetizations are given by the solution of the saddle point equations
\begin{equation}
\label{eq:sp_defm_a0}
    m_{\mu}=\overline{\g{d}\left(\beta \sigma^2 \left|\sum_{\mu=1}^s\;m_{\mu}\bm{\xi}_{\mu}\right|\right)\frac{\bm{\xi}_{\mu}\cdot\sum_{\nu=1}^s\;m_{\nu}\bm{\xi}_{\nu}}{\left|\sum_{\nu=1}^s\;m_{\nu}\bm{\xi}_{\nu}\right|}}.
\end{equation}
We introduced the functions
\begin{equation}
\label{eq:Kd_gd}
\begin{gathered}
    \K{d}(x)=\int d\bm{r}\delta(|\bm{r}|-1)e^{\bm{r}\cdot\bm{x}}=(2\pi)^{d/2}\frac{I_{d/2-1}(x)}{x^{d/2-1}} \\
    \; \\
    \g{d}(x)=\frac{1}{\K{d}(x)}\frac{d\K{d}(x)}{dx}=\frac{I_{d/2}(x)}{I_{d/2-1}(x)}
\end{gathered}
\end{equation}
where $I_{\nu}(x)$ is the modified Bessel function of order $\nu$. The function $\K{d}(x)$ is the partition function of a single spin under the action of a field of magnitude $x$, while the function $0\leq\g{d}(x)<1$ represents its related magnetization. For comparisons, in the SHM one has $\K{1}(x)=2\cosh(x)$ and $\g{1}(x)=\tanh(x)$. Note that for $d=1$, one recovers the equilibrium free energy and the saddle point equations in \cite{amit1985spin}.

For $\alpha=0$ the model has a second-order phase transition at $T_c=\sigma^2/d$, from a paramagnetic phase to a ferromagnetic retrieval phase. One can see this from eq. \eqref{eq:sp_defm_a0} in the case of Mattis states ($s=1$):
\begin{equation}
\label{eq:sp_defm_a0_s1}
    m\,=\,\g{d}(\beta\sigma^2 m).
\end{equation}
Since $\g{d}(x)\sim x/d$ for small $x$, a nonzero solution of \eqref{eq:sp_defm_a0_s1} exists if and only if $\beta>d/\sigma^2=\beta_c$. Note that in order to fix $T_c=1$ as in the SHM one needs spins with norm $\sigma=\sqrt{d}$.
The retrieval solution of \eqref{eq:sp_defm_a0_s1} defines Mattis states. The solution of eq. \eqref{eq:sp_defm_a0_s1} behaves as a Ferromagnetic spin magnetization close to the transition, $ m\simeq \frac{\beta_c}{\beta}\;\sqrt{1+\frac{2}{d}}\;\sqrt{1-\frac{\beta_c}{\beta}}$,
and converges as a power law to unit in the zero-temperature limit $m\simeq 1-\frac{d-1}{2\beta\sigma^2}$.
Note that this scaling contrasts with that of the SHM, where, because of the discrete nature of the spins, $m$ converges exponentially fast to unit.
At $T=0$, \eqref{eq:eq_free_en_dens_a0} converges to the energy density
\begin{equation}
    e=\frac{\sigma^2}{2}\sum_{\mu\leq s}m_{\mu}^2-\sigma^2\overline{\left|\sum_{\mu\leq s}m_{\mu}\bm{\xi}_{\mu}\right|}
\end{equation}
with the saddle-point equation
\begin{equation}
\label{eq:sp_defm_a0_T0}
    m_{\mu}=\overline{\left[\frac{\bm{\xi}_{\mu}\cdot\sum_{\nu=1}^s\;m_{\nu}\bm{\xi}_{\nu}}{\left|\sum_{\nu=1}^s\;m_{\nu}\bm{\xi}_{\nu}\right|}\right]}
\end{equation}
The $P$-degenerate ground state level of the system corresponds to Mattis states ($s=1$) and its value is $e_1=-\frac{\sigma^2}{2}$. 

We found that the solution of \eqref{eq:sp_defm_a0_s1} is stable at all temperatures $T<T_c=\sigma^2/d$. In contrast, we found that mixture solutions of \eqref{eq:sp_defm_a0} are unstable at all temperatures. We focused on the case of symmetric mixtures, $m_{\mu}=m$ if $\mu\le s$ and zero otherwise, and we determined that the instability of the solution is related to symmetry-breaking orthogonal fluctuations $m_{\mu}=m+\delta m_{\mu}$. We observed numerically that no differences occur in asymmetric mixtures, therefore we conclude that, at variance with the SHM, in VHMs mixtures of examples are not stored as spurious states. 
This different behavior is a consequence of the different nature of the degrees of freedom between VHM (continuous) and SHM (discrete). The instability of mixture states is related to the presence of transverse fluctuations of $\bm{z}=\sum_{\mu\leq s}\bm{\xi}_{\mu}$, which in the $d=1$ case are absent.
It was already observed by Amit et al. in \cite{amit1985spin} that when considering continuous disorder (e.g. examples sampled from the uniform distribution in a symmetric interval) mixture solutions destabilize also in the SHM. 
In the large $d$ limit we found for any $s>1$ that mixture solutions become marginally stable. We provide detailed stability computations for Mattis states and mixtures in Appendix \ref{sec:Instability_mixtures}.
\subsubsection{Extensive storage}

We consider the case $\alpha>0$:
since mixture solutions are unstable at $\alpha=0$, in the following, we will consider only non-retrieval solutions ($s=0$) and retrieval ones ($s=1$).

We used the Replica Symmetric ansatz for the replica overlap matrix and Mattis magnetizations
\begin{equation}
\label{eq:RS_ansatz}
    Q_{ab}^*=q\qquad a\neq b,\qquad m_{a}^*=m
\end{equation}
where $0<q<1$ is the Edward-Anderson order parameter introduced in \eqref{eq:EA_overlap} and $m$ is the Mattis magnetization; with this ansatz, we obtained the free energy density
\begin{equation}
\label{eq:free_energy_density_RS}
\begin{split}
    f_{eq}(\alpha,\beta)\,=&\,\frac{\sigma^2}{2}m^2 - \frac{1}{\beta}\log \sigma^{d-1}   \\
    -&\frac{1}{\beta}\int_0^{\infty}dh\; P_h\left(h\right)\log \K{d}\left(\beta \sigma h\right)\\
    +&\frac{\alpha}{2}\left\{
    \frac{\sigma^2}{d}
    +\frac{1}{\beta}\log\left[1-\frac{\beta \sigma^2}{d}\left(1-q\right)\right]\right.  \\
    +& \left. \frac{\beta (1-q)q}{\left[\frac{d}{\sigma^2}-\beta\left(1-q\right)\right]^2}
    -\frac{q}{\frac{d}{\sigma^2}-\beta\left(1-q\right)}
    \right\}
\end{split}
\end{equation}
where $r, m, q$ satisfy the saddle point equations
\begin{equation}
\label{eq:sp_eqs_T}
\begin{split}
    r &= \frac{q}{\sigma^2 \left[\frac{d}{\sigma^2}-\beta (1-q)\right]^2} \\
    \; \\
    q &= \overline{\int \frac{d \bm{h}\, e^{-\frac{h^2}{2\alpha r}}}{[2\pi\alpha r]^{d/2}}\g{d}\Bigl(\beta \sigma\Bigl|\bm{h}+\sigma m\bm{\xi}\Bigr|\Bigr)^2}\\
    &=\int_0^{\infty}dh\; P_h(h)\g{d}(\beta\sigma h)^2 \\
    \; \\
    m =& \overline{\int \frac{d \bm{h}\,e^{-\frac{h^2}{2\alpha r}}}{[2\pi\alpha r]^{d/2}}\frac{\left(\bm{h}+\sigma m \bm{\xi}\right)\cdot\bm{\xi}}{\Bigl|\bm{h}+\sigma m \bm{\xi}\Bigr|}} \\
    &\overline{\times\g{d}\left(\beta \sigma\Bigl|\bm{h}+\sigma m \bm{\xi}\Bigr|\right)} \\
    &=\int_0^{\infty}dh\; P_h(h)\g{d}\left(\frac{m \sigma h}{\alpha r}\right)\g{d}(\beta\sigma h).
\end{split}
\end{equation}

The equilibrium free energy and the saddle point equations reported above for $d=1$ are the same obtained in \cite{amit1985storing}.
The effective field $\bm{h}$ appearing in eqs \eqref{eq:free_energy_density_RS}, \eqref{eq:sp_eqs_T} can be related to the concept of cavity vector: for a given spin $\bm{S}_i$, its local fields vectors $\bm{\eta}_i$ can be decomposed into two contributions, one coming from the interaction with the other spins independently from the value of $\bm{S}_i$ (the cavity contribution) and one that takes into account the polarizing effect that $\bm{S}_i$ has on them (Onsager polarization \cite{mezard1987spin}). In mean-field disordered systems, the two contributions are independent.
We will come back to Onsager polarization later in section \ref{sec:linear_stability}.

The probability density function (PDF) of $h=|\bm{h}|$ can be computed after performing the average over $\bm{\xi}$ and reads
\begin{equation}
\label{eq:cavity_fields_pdf}
    P_h\left(h\right)\,=\,
    \begin{cases}
        \frac{h^{d-1}e^{-\frac{h^2}{2 \alpha r}}}{(2\pi \alpha r)^{d/2}}e^{-\frac{m^2\sigma^2}{2\alpha r}}\K{d}\left(\frac{m \sigma}{\alpha r}h\right)\qquad m>0 \\
        \; \\
        \frac{h^{d-1}\sph{d}e^{-\frac{h^2}{2 \alpha r}}}{(2\pi \alpha r)^{d/2}}\qquad m=0
    \end{cases}
\end{equation}
\begin{figure*}
    \centering
    \includegraphics[width=0.95\linewidth]{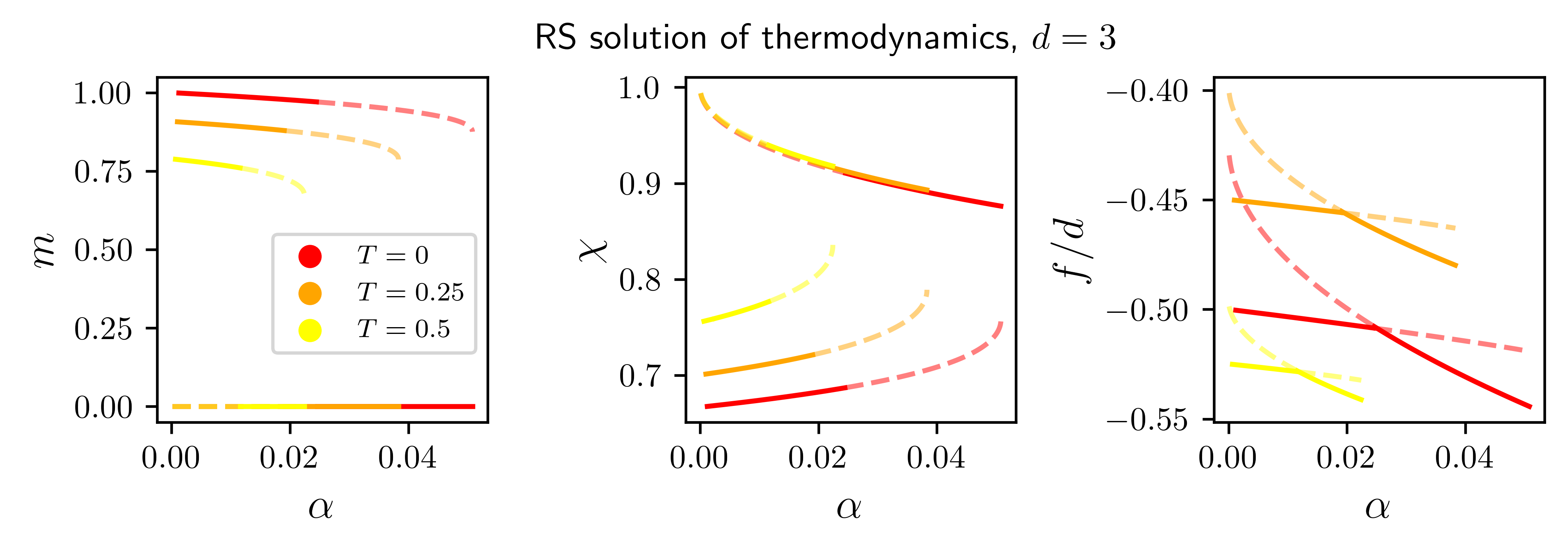}
    \caption{\textbf{The model is a genuine Hopfield associative memory}: Solution of the $d=3$ VHM \eqref{eq:ModelHam} in the RS approximation. We show the Mattis magnetization $m$ (left), the linear susceptibility $\chi$ (center) and the free energy density (right) as functions of the capacity $\alpha$, for temperatures $T=0., 0.25, 0.5$. The behavior is analogous to that of the SHM: for any $T$, the solution exists up to a critical capacity $\alpha_c$ and undergoes an equilibrium first order phase transition at $0<\alpha_m<\alpha_c$. We indicate equilibrium solution with continuous lines, while dashed opaque lines identify metastable solutions. Here we show data for spins with norm $\sigma=\sqrt{d}$, thus the free energy density is normalized by a factor $d$ and $0<\chi<1$.}
    \label{fig:RS_solution_thermodynamics_d=3}
\end{figure*}
Note that the cavity pdf of Mattis states is strongly suppressed close to the lower edge $h=0$: in fact, one finds for small fields $P_h(h;m)/P_h(h;m=0)\simeq e^{-\frac{m^2\sigma^2}{2\alpha r}}\ll 1$. We will show later in section \ref{sec:linear_stability} that this property is related to a better stability of Mattis states against perturbations, because the abundance of low fields $h$ is a proxy for that of soft eigenmodes in the Hessian spectrum.

At $T=0$, the energy density of the system reads
\begin{equation}
\label{eq:energy_density_RS}
\begin{split}
    e(\alpha)=&\frac{\sigma^2}{2}m_0^2-\sigma\int_0^{\infty}dh\; P_h(h;T=0)\,h\\
    &+\frac{\alpha}{2}\left[
    \frac{\sigma^2}{d}
    +\frac{\beta \chi}{\left(\frac{d}{\sigma^2}-\chi\right)^2}
    -\frac{1}{\frac{d}{\sigma^2}-\chi}
    \right]
\end{split}
\end{equation}
We introduced the linear susceptibility $\chi=\beta (1-q)$: this quantity represents the average static linear response of a neuron spin to the action of an external field. In VHMs, a physical solution for the susceptibility is strictly positive and upper bounded, $0<\chi< d/\sigma^2$. 
The $T=0$ saddle point equations read
\begin{equation}
\begin{gathered}
\label{eq:sp_eqs_T0}
    r_0=\frac{1}{\sigma^2}\frac{1}{\left(\frac{d}{\sigma^2}-\chi\right)^2} \\
    \; \\
    \chi=\frac{(d-1)}{\sigma}\int_0^{\infty}dh\frac{P_h(h)}{h} \\
    \; \\
    m=\int_0^{\infty}dhP_h(h)\g{d}\left(\frac{m \sigma h}{\alpha r}\right)
    \end{gathered}
\end{equation}
where it is implicit that in these last equations the parameters of $P_h(h)$, whose expression is given by \eqref{eq:cavity_fields_pdf}, are the $T=0$ ones, appearing on the lhs.
The susceptibility of the non-retrieval solution is simply $\chi(m=0)\,=\,\frac{c d}{(\sqrt{\alpha}+c)\sigma^2}$, with $c=(d-1)\frac{\mathcal{S}_{d-1}(1)}{\sqrt{2 \pi}\mathcal{S}_{d-2}(1)}$.
Note that $m_0<1$ (compare the third of \eqref{eq:sp_eqs_T0} with the second of eq. \eqref{eq:Kd_gd}) for any $\alpha>0$, as expected.

We numerically solved equations \eqref{eq:sp_eqs_T}, \eqref{eq:sp_eqs_T0} as a function of $\alpha$.
In figure \ref{fig:RS_solution_thermodynamics_d=3}, we show the solution we obtained for $m$, $\chi$ and the free energy $f$ as functions of $\alpha$, for the $d=3$ VH model and temperature values $T=0., 0.25, 0.50$. The qualitative physical behavior we find for VH models is the same as that in the SH model: the retrieval solution $m>0$ exists up to a critical storage $\alpha_c$, it is a meta-stable solution for $\alpha_m<\alpha<\alpha_c$ and a stable one for $\alpha<\alpha_m$; at $\alpha=\alpha_m$, the system undergoes a first order phase transition.

\emph{Stability of the RS solution}: The stability of the RS solution is determined by perturbing the Replica Action. The lowest Hessian eigenvalue is known in the spin glass literature as \emph{Replicon} eigenvalue $\Lambda$. In our model, we find that this quantity has the following expression
\begin{equation}
    \label{eq:Replicon_allT}
\begin{split}
    \Lambda=1-\frac{\alpha\beta^2 r}{q}\int_0^{\infty}\;dh P_h(h)\Bigr[(\g{d}'(\beta \sigma h))^2\\
    +(d-1)\frac{(\g{d}(\beta \sigma h))^2}{\beta^2 h^2}\Bigr]    
\end{split}
\end{equation}
The Replicon is related to the divergence of the spin glass susceptibility $\chi_{sg}\propto 1/\Lambda$: only when $\Lambda>0$, the RS solution is stable, the solution of $\Lambda=0$ determining the critical line $T_c(\alpha)$ for the RSB spin glass transition.

We found that the RS solution for spurious states is unstable whenever
the equation for $q$ in \eqref{eq:sp_eqs_T}
admits a non-zero solution: this happens at the critical temperature $T_c(\alpha)=\frac{\sigma^2(\sqrt{\alpha}+\sqrt{d})}{d^{3/2}}$, as can be seen by expanding the $m=0$ equation for $q$ in \eqref{eq:sp_eqs_T} for small $q$. As in the SHM, spurious states of VHMs are Full Replica Symmetry Breaking (fRSB) spin glass states. In particular, at $T=0$ and for $d>2$ the Replicon is equal to $-\frac{1}{d-2}$: in the large $d$ limit, the no-retrieval phase becomes a marginally stable RS phase, as in the spherical spin glass \cite{Cugliandolo_fulldyn_1995, niedda2024probingmarginalstabilityspherical}. 

At variance with the SH model $(d=1)$, we observe that the retrieval phase of VHMs with $d>2$ is always RS. Indeed, we found that eq. \eqref{eq:Replicon_allT} and its $T=0$ limit
\begin{equation}
\label{eq:Replicon_T0}
    \Lambda_0\,=\,1-(d-1)\alpha r_0\int_0^{\infty}dh\frac{P_h(h)}{h^2} 
\end{equation}
return a positive Replicon in the whole retrieval phase when $d>2$. This is not the case for $d=2$, since the integral in \eqref{eq:Replicon_T0} for this dimension diverges to $-\infty$, due to $P_h(h)\propto h^{d-1}$ close to $h=0$. As a consequence, the $d=2$ VH model must undergo a spin glass transition at finite temperature: using eq. \eqref{eq:Replicon_allT}, we find the transition line $T_c(\alpha)\sim e^{-\frac{1}{2}e^{\frac{m^2}{\alpha r}}}$. In the SH model, the retrieval solution undergoes an RSB transition at $T_c(\alpha)\propto e^{-\frac{m^2}{2\alpha r}}$ \cite{amit1985storing}.
Thus, RSB corrections to the RS solution for $d=2$ are expected to be even smaller than those observed for $d=1$, since the transition onsets at very low temperatures: we report $T_c(\alpha_c)\simeq 10^{-6}$ for $d=2$.

\begin{figure*}
    \centering
    \includegraphics[width=0.95\linewidth]{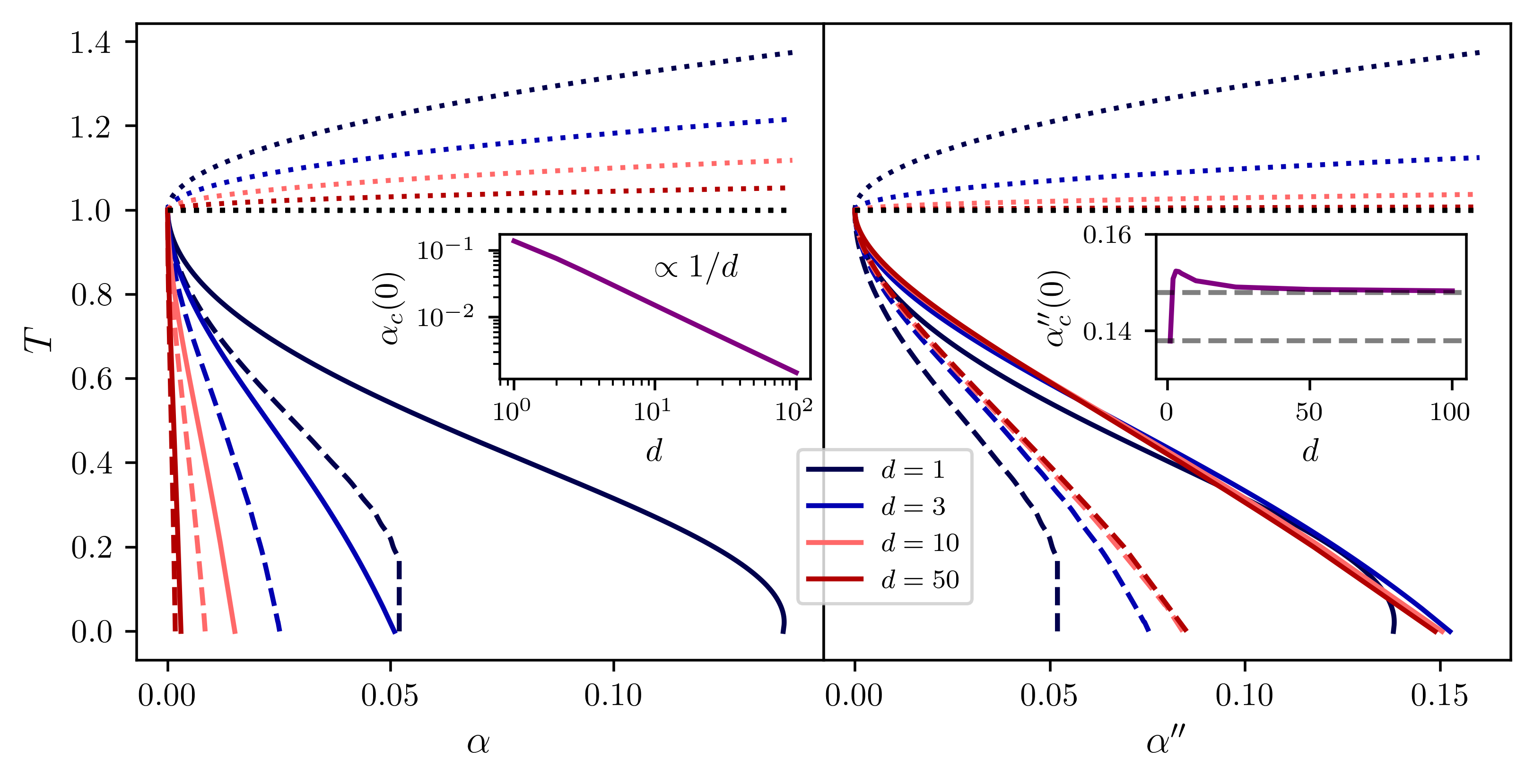}
    \caption{\textbf{The retrieval phase shrinks with growing spin dimension}: Phase diagrams of VHMs with $d=3, 5, 10, 50$, compared with that of the SHM ($d=1)$. Spins are chosen with norm $\sigma=\sqrt{d}$ in order to fix the no-retrieval spin glass temperature at $\alpha=0$ to unit.
    \textbf{Left}: the phase diagram is shown in the $T,\alpha$ plane. Retrieval line shrink for growing $d$, while the spin glass line for no-retrieval states converge to unit. In the inset, the zero-temperature critical capacity $\alpha_c(0)$ is shown to decrease as $1/d$. \textbf{Right}: the same phase diagram but with $\alpha''=\alpha d$ on the $x$-axis. Retrieval lines in this scaling regime seem to converge to non-trivial curves. In particular, we find that the critical capacity $\alpha_c(0)\rightarrow 4/27$ in the large $d$ limit, as shown in the inset.}
    \label{fig:phase_diag}
\end{figure*}

\emph{Phase diagram and scaling regimes}: In the preceding lines, we discussed the spin glass transition curves for no-retrieval and Mattis states.
The retrieval lines of the VHM \eqref{eq:ModelHam} are analogous to those of the SH: there is a spinodal line $T_c(\alpha)$ that marks the appearance of the retrieval solution as a metastable solution of thermodynamics, with spin glass states dominating the equilibrium measure, and a line $T_m(\alpha)<T_c(\alpha)$ that signals a first order equilibrium phase transition where the retrieval solution becomes the equilibrium one. At variance with the SHM \cite{amit1985spin, amit1985storing}, in the VHM there are no retrieval lines for solutions involving mixtures of examples, since we determined that solutions of mixtures of examples are always unstable (see Appendix \ref{sec:Instability_mixtures}). 

In the left panel of fig. \ref{fig:phase_diag} we show the phase diagram of VHMs with $d=3, 5, 10, 50$ in the $T,\alpha$ plane, portraying both the spin glass transition curves for no-retrieval states $T_{sg}^{(nr)}=\frac{\sigma^2(\sqrt{\alpha}+\sqrt{d})}{d^{3/2}}$ (dotted lines) and critical curves for retrieval, respectively the spinodal (full lines) and the first-order transition (dashed lines). We set $\sigma=\sqrt{d}$ for spins norms, in order to have $T_{sg}^{(nr)}(\alpha=0)=1$ independently from $d$. 
We derived the retrieval lines by numerically solving eqs \eqref{eq:sp_eqs_T}, \eqref{eq:sp_eqs_T0} for several $\alpha, T$. More in detail, the line $T_c(\alpha)$ is obtained by finding for a given capacity $\alpha$ the largest temperature such that the retrieval solution $m>0$ exists, while the line $T_m(\alpha)$ was computed by finding for any $\alpha$ the temperature at which the free energy of memory states is equal to that of spurious spin glass states. In VHM with $d\geq3$ the spinodal lines $T_c(\alpha)$ we derived are exact, having no reentrance phenomena as in the SHM \cite{amit1985storing, amit1987information}, while first-order transition lines $T_m(\alpha)$ computed by us through the RS ansatz are not exact, since spurious states are RSB spin glass states. 
We found that for increasing $d$, the storage capacity of our VHM is vanishing and always lower than that of the SHM, with both the $T=0$ critical capacities $\alpha_c,\alpha_m$ scaling as $1/d$ for large $d$: in the inset in the left panel of figure \ref{fig:phase_diag} we show the critical capacity $\alpha_c(0)$ as a function of $d$. 

In the right panel of fig. \ref{fig:phase_diag}, we show the phase diagram in the $T, \alpha''$ plane, with $\alpha''=\alpha d$; at this scale, we see that retrieval is maximum at $d=3$ and then apparently saturates to $\alpha''(d\rightarrow\infty)=4/27$ in the large $d$ limit, as shown in the inset: interestingly, this is the same critical capacity found in \cite{bolle2003spherical} for a spherical Hopfield model with both pairwise and four-body interactions.
The scaling $\alpha''$ corresponds to storing $P=\alpha'' \frac{N}{d}$ examples: so for a given system size $N$, as the number of dimensions of vector spins increases the number of memories that can be stored decreases linearly.  We also identified a scaling $\alpha'=\alpha/d$, corresponding to a regime in which the number of memories increases linearly with the dimension, $P=\alpha' N d$. With this scaling, the equilibrium free energy \eqref{eq:free_energy_density_RS} in the limit $d \rightarrow \infty$ converges to the free energy of a spherical Hopfield model with pairwise interactions. We will see in section \ref{sec:retrieval_dynamics} that the scaling $\alpha'=\alpha/d$ is also relevant for the very first steps of the retrieval dynamics in \eqref{eq:retrieval_versor_rule}, as it sets the limit capacity for transient retrieval.

We discuss the scaling regimes in spin dimension $d$ of the RS thermodynamic solution in Appendix \ref{sec:scaling_limits_computations}.

\subsection{Linear stability}
\label{sec:linear_stability}

In this section, we study the statistics of the Hessian spectrum of different local minima of the energy landscape, characterizing $T=0$ fluctuations of Mattis and spurious spin glass states. We obtain theoretical equations for the spectrum using random matrix techniques and we test our predictions with data obtained from exact numerical diagonalization. Numerical samples corresponding to local minima of the energy landscapes where obtained through eq. \eqref{eq:retrieval_versor_rule}, using respectively pattern and random initial conditions for memory and spurious states (see also Algorithm \ref{alg:retrievalAlgo} in Appendix \ref{sec:details_retrieval_dynamics}).

The Hessian matrix related to the energy function in \eqref{eq:ModelHam} reads
\begin{equation}
    \label{eq:Hessian}
    \mm{M}_{ij}(\conf{S})=\mm{P}_{\perp}(\bm{S}_i)\left(-\mm{J}_{ij}+\frac{\eta_i}{\sigma}\delta_{ij}\mm{I}_d\right)\mm{P}_{\perp}.(\bm{S}_j)
\end{equation}
The projectors $\mm{P}_{\perp}(\bm{S})=\mm{I}-\frac{\bm{S}}{|\bm{S}|}\frac{\bm{S}^\intercal}{|\bm{S}|}$ ensure that fluctuations keep the spins norms fixed.
Therefore, the nontrivial part of the Hessian in \eqref{eq:Hessian} is constituted by excitations that are orthogonal to the configuration $\conf{S}$ on which the Hessian is evaluated. The longitudinal sector of the spectrum is instead associated with the stationarity condition \eqref{eq:stationarity_modelHam}, defining the local fields $\eta_i$.
In the following, we will only consider the orthogonal sector of the spectrum.

We focus our analysis on the statistics of eigenvalues and eigenvectors in the $\lambda\rightarrow 0_+$ limit. This region of the spectrum is the most physically relevant for the static stability of the system against small linear perturbations. In particular, the softest modes rule the long-time relaxation dynamics of the network close to a fixed point.

\subsubsection{The spectral equation}

The statistical properties of the Hessian spectrum are encoded in the resolvent matrix $\mm{G}(z)=(\mm{M}-z\mm{I})^{-1}$, where $z$ is a complex number that lies outside the spectrum of the Hessian matrix $\mm{M}$ (eq. \eqref{eq:Hessian}). 

We derived an equation for the normalized trace of the resolvent $\Green(z)=\frac{d-1}{N d}\Tr{\mm{G}(z)}$, that is exact in the thermodynamic limit $N\rightarrow\infty$. Setting $z=\lambda-i0_+$, it reads\footnote{In its derivation, the Green function in eq. \eqref{eq:self_consistent_equation_resolvent} is implicitly regularized on the spectral line.}
\begin{equation}
\label{eq:self_consistent_equation_resolvent}
    \Green(\lambda)=\frac{(d-1)}{\sigma}\int_0^{\infty}\frac{d\eta \;P_\eta(\eta)}{\eta-\lambda\sigma-\frac{\alpha\sigma}{d}\frac{\Green(\lambda)}{\frac{d}{\sigma^2}-\Green(\lambda)}}
\end{equation}
where $P_\eta(\eta)$ is the distribution of the local fields $\eta_i$ on a stationary state. Equation \eqref{eq:self_consistent_equation_resolvent} was derived through a free convolution\,\cite{potters2020first} of the interactions and local field matrices that appear in the Hessian \eqref{eq:Hessian}, as explained in Appendix \ref{sec:derivation_spectral_equation}.
Equation \eqref{eq:self_consistent_equation_resolvent} is a self-consistent equation for $\Green(\lambda)$, in its real and imaginary parts. If the distribution $P_{\eta}(\eta)$ is known,  eq. \eqref{eq:self_consistent_equation_resolvent} can be solved numerically as explained in Appendix \ref{sec:spectral_equation_numerical}, allowing one to compute the distribution of eigenvalues and related eigenvectors, as explained later in the following paragraphs.

\subsubsection{The distribution of local fields}

To have verifiable analytical predictions for the spectrum through \eqref{eq:self_consistent_equation_resolvent}, it is therefore crucial to have a theoretical expression for the distribution $P_{\eta}(\eta)$. The information on the type of minimum on which the Hessian is evaluated (memory or spurious spin-glass states) is completely encoded in $P_{\eta}(\eta)$.
In disordered systems on dense networks, local fields $\eta_i$ are simply related to cavity fields $h_i$ in the thermodynamic limit, through $\eta_i=h_i+p$: the term $p$ is the Onsager reaction, which, as previously explained in \ref{sec:thermodynamics}, describes the response of the disordered system to the addition of a novel spin to the network \cite{mezard1987spin}. 
A nonzero Onsager reaction is a necessary condition for the stability of the system, so the PDF of local fields on local minima of \eqref{eq:ModelHam} must have a gap: $\eta_{min}=p>0$.
In mean field spin glasses, the Onsager reaction is exactly the linear susceptibility $p\equiv \chi$, whereas in our VHM \eqref{eq:ModelHam} we found
\begin{equation}
\label{eq:Onsager_reaction_term}
    p=\frac{\alpha\sigma}{d}\frac{\chi}{\frac{d}{\sigma^2}-\chi}
\end{equation}
where $\chi$ is the linear susceptibility defined in eq. \eqref{eq:sp_eqs_T0}.
The expression we obtained in this last equation is an extension to $d>1$ of that obtained in \cite{mezard2017mean} for the SHM ($d=1$). Based on these considerations, our theoretical prediction for the distribution of local fields is
\begin{equation}
\label{eq:local_fields_norms_pdf}
    P_\eta(\eta)=\vartheta\left(\eta-p\right)P_{h}\left(\eta-p\right)
\end{equation}
where $\vartheta(\cdot)$ is Heaviside Theta and $P_h(h)$ is the distribution of cavity fields introduced in \eqref{eq:cavity_fields_pdf}. 
We compared the PDFs predicted by eq. \eqref{eq:local_fields_norms_pdf} with the empirical pdfs of local fields of our numerical samples: our results can be found in Appendix \ref{sec:local_fields_pdf_study}. We found perfect agreement between theory and numerics for Mattis states minima, but poor agreement in the case of spin glass ones, as predictable from the fact that eq. \eqref{eq:local_fields_norms_pdf} uses the RS solution for $P_h(h)$ and $\chi$.

\subsubsection{The spectral density and its pseudogap}
We solved the spectral equation \eqref{eq:self_consistent_equation_resolvent} obtaining $\Green(\lambda)$ and computed the spectral density through Sokhotski–Plemelj formula \footnote{For a real quantity $x$, it holds $\frac{1}{x\pm i0_+}=\mathcal{P}\left(\frac{1}{x}\right)\pm i\pi\delta(x)$, where $\mathcal{P}(1/x)$ is the principal value integral of $1/x$. Equation \eqref{eq:spectral_density_theory} for the spectral density can be easily obtained from its definition as the Steltjes transform of the resolvent function $\Green(z)=\dashint \frac{\rho(\lambda)d\lambda}{\lambda-z}$.} as the resolvent imaginary part
\begin{equation}
\label{eq:spectral_density_theory}
\begin{gathered}
    \rho(\lambda)=\frac{1}{\pi}\Im \Green(\lambda-i0_+)
\end{gathered}
\end{equation}
In the following, we describe the properties of the spectral density obtained from \eqref{eq:spectral_density_theory}. We briefly discuss the bulk properties and then focus on the lower spectral edge.

\emph{The bulk}: as explained in Appendix \ref{sec:crossover_nontrivial_trivial_bulk}, our solution can be approximately decomposed as
\begin{equation}
\label{eq:spectral_density_effective}
    \rho(\lambda)\simeq \rho_{nt}(\lambda)\vartheta(\widehat{\lambda}-\lambda)+P_\eta\left(\lambda\sigma-\frac{\alpha\sigma}{d}\right)\vartheta(\lambda-\widehat{\lambda}).
\end{equation}
This decomposition results from the properties of the interaction matrix $\mm{J}$ in the Hessian in eq. \eqref{eq:Hessian}, which has rank $P<N(d-1)$ and thus its spectrum for $\alpha<(d-1)$ has a continuous and a singular part. The 'nontrivial' density $\rho_{nt}$ is given by the free convolution of the continuous part and the local fields distribution, while the trivial term in \eqref{eq:spectral_density_effective} comes from the free convolution of local fields with the singular part. Therefore, the eigenvalue $\widehat{\lambda}$ that marks the crossover between the two bulks corresponds to the rank-$P$ eigenvalue. Since the focus of this section is on low energy excitations, we continue the discussion on the bulk properties in Appendix \ref{sec:crossover_nontrivial_trivial_bulk}.
\begin{figure*}
    \centering
    \includegraphics[width=\linewidth]{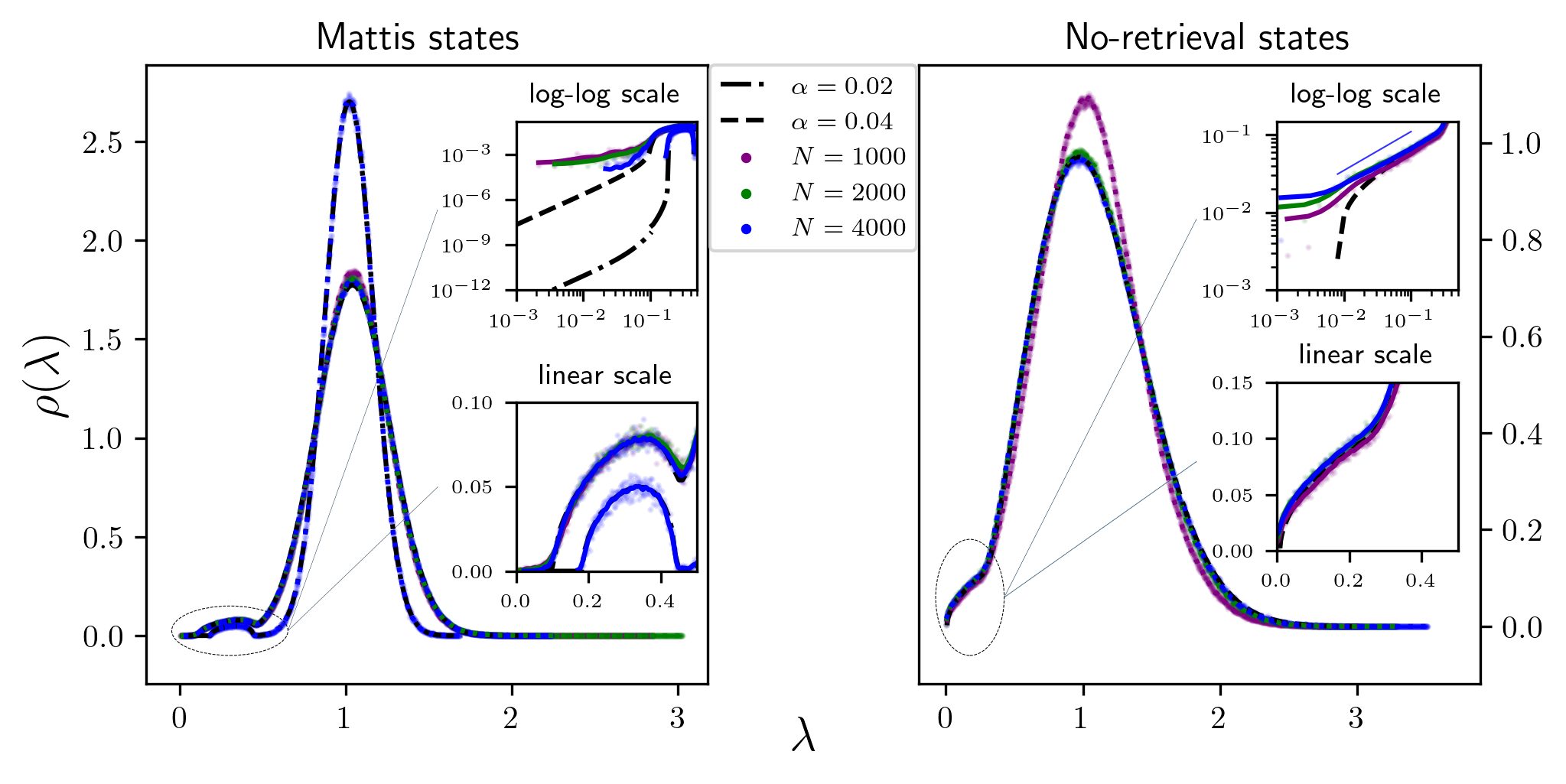}
    \caption{ \textbf{The pseudogap in the spectral density controls the stability of the system}. The spectral density obtained from the solution of eq. \eqref{eq:self_consistent_equation_resolvent} for the $d=3$ VHM, compared with empirical spectral densities computed from numerical measures. \textbf{Left}: the spectral density for Mattis states. With different dashed lines we show the theoretical predictons for capacities $\alpha=0.02, 0.04$, while continuous lines of different colors detailed in the legend are numerical measures for sizes $N=1000, 2000, 4000$. In the inset, we zoom on the non-trivial bulk close to the lower edge of the spectrum. The remainder of the spectral density has a shape that reminds of that of $P_\eta(\eta)$, as one can fully assess by comparing it with fig. \ref{fig:distribution_local_fields} in Appendix \ref{sec:local_fields_pdf_study}. There is a good agreement between theory and measures, with visible finite size effects only at low eigenvalues. In particular, while for capacity $\alpha=0.02$ the power tail close to $\lambda=0$ is not captured, for capacity $\alpha=0.04$ we see that the lower tail of the spectrum progressively approach the asymptotic curve for increasing sizes. \textbf{Right}: same but for spurious spin-glass states. Here we only compare theory and data for $\alpha=0.04$, to ease graph legibility. One can see systematic errors between the RS theory and numerical measures similar to those found in figure \ref{fig:distribution_local_fields} in Appnedix \ref{sec:local_fields_pdf_study}. In addition, the $\rho(\lambda)$ obtained from \eqref{eq:self_consistent_equation_resolvent} using the RS local fields distribution for no-retrieval states has an unphysical gap close to the lower edge: the numerical spectral density instead is clearly ungapped and seems to approach $\lambda=0$ as $\rho_{emp}(\lambda)\sim \sqrt{\lambda}$, consistently with one expects for the spectral density of spin glass states.}
    \label{fig:fig_spec_dens_Mattis_d=3_theor_vs_emp}
\end{figure*}

\emph{The lower edge}: we found that the lower spectral edge of the Hessian spectrum is vanishing in the thermodynamic limit.
Indeed, eq. \eqref{eq:self_consistent_equation_resolvent}, that describes the spectrum in the thermodynamic limit, is consistent if and only if the lower spectral edge is $\lambda_{min}=0$. We leave a proof of this claim to Appendix \ref{sec:asymptotic_expansions_spectrum_lower_edge} for interested readers.
We then expanded eq. \eqref{eq:self_consistent_equation_resolvent} for $\lambda\rightarrow 0$, as reported in Appendix \ref{sec:asymptotic_expansions_spectrum_lower_edge}, and derived the scaling of the lower tail of the spectral density of Mattis states of $d>2$ VH models:
\begin{equation}
\label{eq:spectral_density_close_to_lower_edge}
    \rho(\lambda)\approx a_d \lambda^{d-1},\qquad \lambda\ll\lambda_{pg}\propto \Lambda_0^2
\end{equation}
The spectral density follows a power law behavior, up to a scale $\lambda_{pg}$ controlled by the stability of the RS phase, through the Replicon $\Lambda_0$ defined in \eqref{eq:Replicon_T0}. Slightly above that scale, one has $\rho(\lambda)\simeq \sqrt{\lambda-\lambda_{pg}}$: the crossover between these two scaling behaviors is an example of a spectral \emph{pseudogap}, qualitatively identical to that observed in vector spin-glass models for glassy excitations in \cite{franz2022delocalization, franz2022linear}.
As a consistency check, note that the global spin glass susceptibility, related to the Hessian spectrum through $\chi_{SG}=\int d\lambda \rho(\lambda)/\lambda^2$, is finite for any $d>2$. This is in agreement with our previous observations in section \ref{sec:thermodynamics} that Mattis states are RS and thus should have $\chi_{SG}\propto 1/\Lambda_0<\infty$.
We did not explicitly derived the scaling for the $d=2$ VH networks, because they are fRSB and we do not know the correct distribution of local fields necessary to solve the spectral equation. However, by continuity, we claim that $\rho(\lambda)\approx \lambda$ in that case, up to logarithmic correcting factors. In fact, such scaling would return a divergent spin glass susceptibility.

The prefactor $a_d$ (whose expression is reported in Appendix \ref{sec:asymptotic_expansions_spectrum_lower_edge}) quantifies the abundance of low-energy excitations: indeed, using eq. \eqref{eq:spectral_density_close_to_lower_edge} one can see that the fraction of eigenvalues with at most energy $\lambda$ is $\int_0^{\lambda}d\lambda'\rho(\lambda')\simeq \frac{a_d}{d-1}\lambda^d$. We found that prefactors $a_d$ have, for the small values of $\alpha=\Oo(1/d)$ where the retrieval phase exists, exponentially low values, implying that for any reasonable size accessible by numerical simulations, low-energy excitations are quite rare to observe. It follows that soft eigenmodes observed in numerical simulations must be affected by significant finite size effects.

We tested our theoretical predictions by comparing the theoretical spectral density \eqref{eq:spectral_density_theory} with the empirical one:
\begin{equation}
\label{eq:spectral_density_empirical}
\begin{gathered}
    \rho_{emp}(\lambda)=\frac{1}{N (d-1)}\sum_{k=1}^{N (d-1)}\overline{\delta(\lambda-\lambda_k)}.
\end{gathered}
\end{equation}
In \eqref{eq:spectral_density_empirical} the symbol $\overline{(\cdot)}$ stands for average over numerical samples. We obtained local minima of the energy \eqref{eq:ModelHam} as fixed points of the retrieval dynamics \eqref{eq:retrieval_versor_rule}, and obtained their hessian spectra by exact numerical diagonalization.
In figure \ref{fig:fig_spec_dens_Mattis_d=3_theor_vs_emp} we report plots of the spectral density for the $d=3$ VH model: the left panel contains the theoretical and numerical spectral densities of Mattis states for some values of capacity $\alpha<\alpha_c(0)\simeq 0.0508$. We found excellent agreement between theory and numerical experiments far from the lower edge, but there are evident finite size effects close to the lower edge. Similar behaviors, which are related to the extreme value statistics of low eigenvalues, are completely analogous to those already observed for vector spin glasses in \cite{franz2022delocalization}: therefore, they will not be discussed in this work.
In the right panel, we show the spectral densities for the spectrum of spurious spin-glass local minima. We find, as expected, that our solution of the spectral equation \eqref{eq:self_consistent_equation_resolvent}, which is based on an RS approximation for the theoretical assessment of $P_{\eta}(\eta)$, produces a wrong result. In particular, close to the lower spectral edge, RS theory predicts a gap that is absent in numerical spectral densities. The correct lower edge behavior for spurious spin glass minima is expected to be $\rho(\lambda)\simeq \sqrt{\lambda}$ (see Appendix \ref{sec:asymptotic_expansions_spectrum_lower_edge}).

Note that the spectral density scales as the PDF of cavity fields (eq. \eqref{eq:cavity_fields_pdf}) close to the origin: $P_h(h)\approx h^{d-1}$ for small $h$. This behavior hints at the presence of a strong linear correlation between low-rank eigenvalues and low-rank cavity fields that, as shown in \cite{franz2022delocalization} for vector spin glasses and proven rigorously in \cite{lee2016extremal} in the context of the deformed Wigner ensemble of random matrices, is related to the localization of low-rank eigenvectors. In the following paragraph, therefore, we study the localization properties of eigenvectors of our Hessian \eqref{eq:Hessian}, again singling out their behavior for Mattis states and spin glass states.

\subsubsection{Localization of soft eigenvectors}

\begin{figure*}
    \centering
    \includegraphics[width=\linewidth]{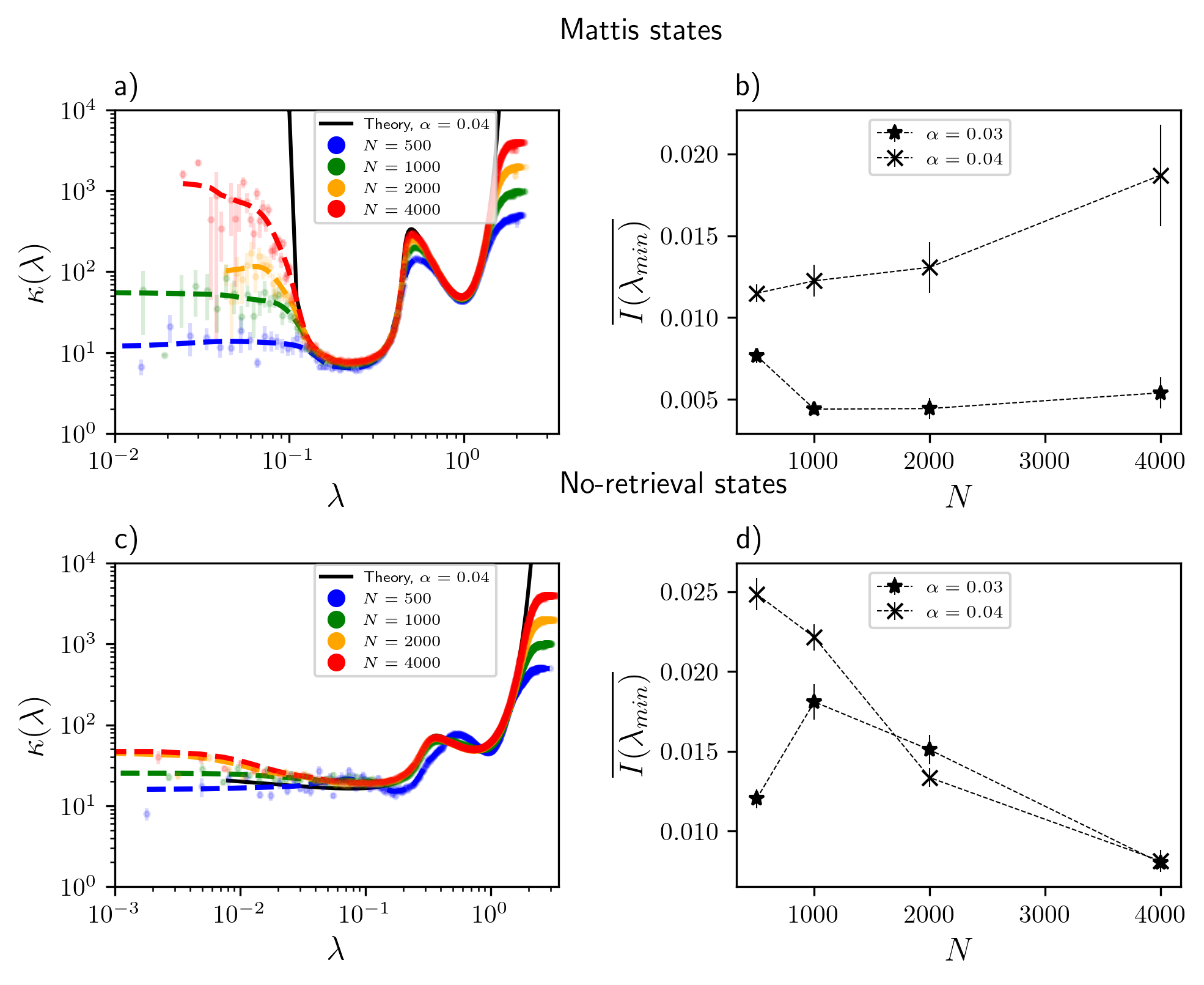}
    \caption{\textbf{Soft modes of memory states are localized.}. The rescaled IPR $\kappa=N I$ of the $d=3$ VHM, for Mattis states (top panel) and Spin glass states (bottom panel). \textbf{Top}: figure a) compares the theoretical prediction \eqref{eq:rescaled_IPR_theory} with empirical measures based on eqs \eqref{eq:IPR}. In the bulk there is an excellent agreement between theory and numerics, while at the spectral edges deviations consistent with eigenvectors localization appear. Figure b) shows the sample-averaged IPR of the softest mode for several values of $\alpha$ detailed in the legend. For increasing sizes, the IPR is not vanishing, validating the localization hypothesis.
    \textbf{Bottom}: same but for no-retrieval states. Figure c) shows the comparison between theoretical rescaled IPR and numerics: again, the RS theory is not correct since it predicts a non-physical spectral gap. The empirical rescaled IPR seems to be regular for $\lambda\rightarrow 0$. In figure d) we show the sample-averaged IPR of the smallest mode: this quantity is decreasing for increasing sizes in a range $N=\Oo(10^2):\Oo(10^3)$.}
    \label{fig:fig_rescaled_IPR}
\end{figure*}

The statistical properties of eigenvectors, in relation to their configuration on the underlying network, carry crucial physical information on how the system responds to small external perturbations. In mean field disordered systems on fully-connected graphs, the dominant local static responses are given by graph nodes where soft modes concentrate their measure. As an example, in terms of hessian eigenmodes, the linear and quadratic local susceptibilities read
\begin{equation}
    \label{eq:linear_susc}
    \chi_{ii}=\sum_{k=1}^{N}\frac{|\bm{\psi}_{i}(\lambda_k)|^2}{\lambda_k}\qquad \chi_{ii}^{sg}=\sum_{k=1}^{N}\frac{|\bm{\psi}_{i}(\lambda_k)|^2}{\lambda_k^2}
\end{equation}
Both quantities in this equation are dominated by the softest modes ($\lambda\rightarrow 0$) and the strongest weights ($|\bm{\psi}_i|^2=\Oo(1)$).
In equation \eqref{eq:linear_susc}, we identified eigenvectors related to eigenvalues $\lambda$ with the symbol $\conf{\psi}=(\bm{\psi}_1^j,\dots,\bm{\psi}_N^j)^\intercal$, where each $\bm{\psi}_i^j$ is a vector of dimension $d$, with the index $i=1,\dots, N$ identifying network sites and the index $j=1,\dots,N(d-1)$ identifying the rank of the mode. We consider by definition each eigenvector normalized to unit and adopt this notation for eigenvectors in the following.

In order to measure the localization properties of eigenvectors, we introduce the Inverse Participation Ratio (IPR) $I(\lambda)$ and fixed-rank Inverse Participation Ratios (frIPRs) $I_j$:
\begin{equation}
    \label{eq:IPR}
\begin{gathered}
    I(\lambda)=\frac{1}{N (d-1)}\sum_{j=1}^{N (d-1)}\overline{\delta(\lambda-\lambda_j)\mathrm{I}_j} \\
    \; \\
    \mathrm{I}_j=\sum_{k=1}^N|\bm{\psi}_k^j|^4
\end{gathered}
\end{equation}
The frIPRs in \eqref{eq:IPR} are descriptors of the degree of localization of eigenvectors. Because of the normalization constraints $\sum_{k=1}^N |\bm{\psi}_k^j|^2=1$, for any mode the typical value of its eigenvector weights is $|\bm{\psi}_k^j|^2=\Oo(\frac{1}{N})$: thus, when all the weights $|\bm{\psi}_k|^2$ bear such a contribution to normalization one has $I_j=\Oo(\frac{1}{N})$ and the eigenvector of rank $j$ is said \emph{delocalized} or \emph{extended}. Conversely, if one has $|\bm{\psi}_k^j|^2\gg 1/N$ for some sites, then one has either partially delocalized eigenvectors $I_j=\Oo(\frac{1}{N^a})$ for some $0<a<1$ or localized ones $0<I_j=\Oo(1)<1$. The IPR describes localization properties in the thermodynamic limit for fixed eigenvalues: therefore, it is an ideal tool to characterize localization in the bulk of the spectrum. It is convenient to define a rescaled IPR , $\kappa(\lambda)=N I(\lambda)$, where now a finite $\kappa(\lambda)$ identifies delocalized modes and a singular $\kappa$, diverging with system size, either partially delocalized ($\kappa=O(N^a)$, for $0<a<1$) or localized modes ($\kappa=O(N)$).

Eqs \eqref{eq:IPR} provide an empirical measure of delocalization. We obtained a theoretical prediction for the IPR in the bulk of the spectrum, in terms of the resolvent function computed from eq. \eqref{eq:self_consistent_equation_resolvent}: 

\begin{equation}
\label{eq:rescaled_IPR_theory}
\begin{split}
    \kappa(\lambda)&=3\left(1-\frac{1}{d}\right)\left[\frac{(d-1)\alpha}{\sigma^2|\frac{d}{\sigma^2}-\Green(\lambda)|_{\mathbb{C}}^2}\right]^2 \\
    &\times\int\frac{d\eta P_\eta(\eta)}{|\eta-\lambda\sigma-\frac{\alpha\sigma}{d}\frac{\Green(\lambda)}{\frac{d}{\sigma^2}-\Green(\lambda)}|_{\mm{C}}^{4}}
\end{split}
\end{equation}
Given a solution of eq. \eqref{eq:self_consistent_equation_resolvent}, the IPR can be computed from eq. \eqref{eq:rescaled_IPR_theory} as $I=\kappa/N$. We outline our derivation of eq. \eqref{eq:rescaled_IPR_theory} in Appendix \ref{sec:expansion_rescaled_IPR_close_to_lower_edge}.
We computed the rescaled IPR for several values of $\alpha$ of the $d=3$ VH model. In what follows, we outline our results for Mattis states and spin glass states.

\emph{Mattis states}: in the top panel of figure \ref{fig:fig_rescaled_IPR} we show on the left (fig. \ref{fig:fig_rescaled_IPR}.a) the rescaled IPR $\kappa(\lambda)$ as computed from eq. \eqref{eq:rescaled_IPR_theory} and its empirical counterpart $\kappa_{emp}(\lambda)=N I(\lambda)$, where $I(\lambda)$ is the IPR defined in \eqref{eq:IPR}, for capacities and sizes detailed in the legend. There is excellent agreement between our theoretical prediction and numerical data in the bulk of the spectrum, with finite size effects becoming increasingly smaller for increasing sizes $N$. However, we observe that, close to the edges of the spectrum, numerical data do not agree with the theoretical prediction: instead, $\kappa_{emp}(\lambda)$ seems to increase roughly linearly with $N$. This observation is a clear sign of localization for $\lambda\rightarrow 0$. To corroborate our claim, in figure \ref{fig:fig_rescaled_IPR}.b on the right we show the sample-averaged IPR of the first eigenvector versus $N$ for capacities $\alpha=0.03, 0.04$. In all cases, the IPR is not vanishing for increasing $N$, possibly saturating to a constant value for $N\rightarrow\infty$.

The type of localization occurring in the soft modes of Mattis states is a random matrix condensation phenomenon, similar to that observed in \cite{franz2022delocalization, franz2022linear, rainone2021meanfield} for mean field disordered systems or \cite{lee2016extremal, Ikeda2023Bose} for more general deformed Wigner matrices: for each low-rank (with rank $k$) eigenvector, a single component bears a finite contribution to normalization, while each of its remaining components are vanishing in the thermodynamic limit. In formulas, we obtain for the IPR at the lower edge, in the thermodynamic limit
\begin{equation}
    \label{eq:IPR_lower_edge}
    I(0)=\Lambda_0^2+O(N^{-1+\frac{2}{d}})
\end{equation}
so that the condensation phenomenon is tied to the stability of the RS phase ($\Lambda_0>0$), as already shown in \cite{franz2022delocalization}. The finite-size corrections to the asymptotic value in \eqref{eq:IPR_lower_edge} in the case $d=3$ analyzed by us are not negligible, being $O(N^{-\frac{1}{3}})$, thus hindering the condensation phenomenon, as previously described, up to very large sizes, inaccessible by numerical simulations.
We further discuss the condensation phenomenon in Appendix \ref{sec:asymptotic_expansions_spectrum_lower_edge}.

\emph{Spin glass no-retrieval states}: we show our results in the bottom panel of fig. \ref{fig:fig_rescaled_IPR}. Here, as expected, our RS theoretical prediction of the rescaled IPR leads to incorrect results, especially at low eigenvalues: numerical data clearly show that the spectrum is not gapped. The rescaled IPR appears to be regular in the limit $\lambda\rightarrow 0$, as is apparent from the progressive collapse of data curves as the size $N$ of the system increases. It follows that soft modes are delocalized, as it should be in a spin glass phase.

In this paragraph, we showed that Mattis states representing memories stored in the network feature localized soft modes. In the next paragraph, we unveil the physical meaning of such a phenomenon in this model, discussing the relation between eigenvectors, local fields, and noise in Mattis states.

\begin{figure*}
    \centering
    \includegraphics[width=\linewidth]{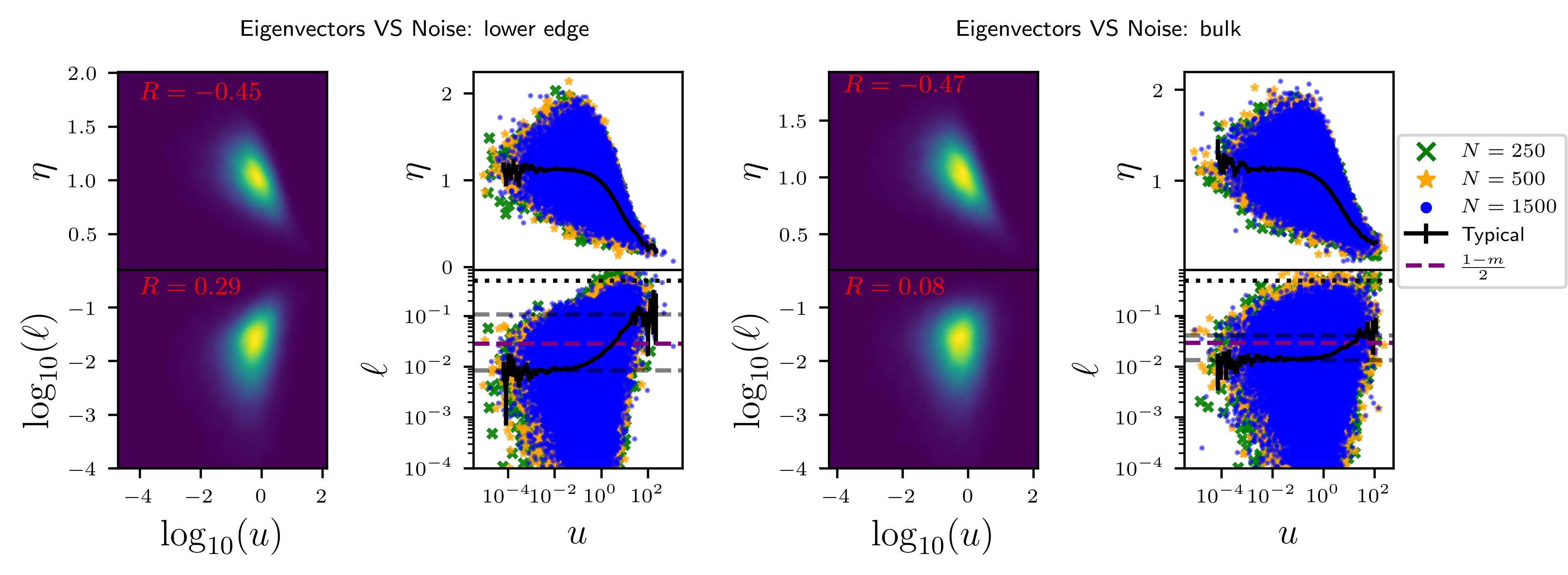}
    \caption{\textbf{Soft modes of Mattis states localize on noisy spins}. In this figure we show our measures of correlation for respectively local fields $\eta_i$ and rescaled weights $u_i=N|\bm{\psi}_i|^2$ and local noises $\ell_i$ and rescaled weights, for the $d=3$ VHM. We show heatmaps in panels a) and c) and scatter plots in figures b) and d). Superimposed to scatter plots we show with dark continuous lines typical curves of $\eta$ and $\ell$ at fixed $u$, computed respectively as $\eta_{typ}(u)=\exp\overline{\frac{1}{N}\sum_i\log (\eta_i)\delta(u-u_i)}$ and $\ell_{typ}(u)=\exp\overline{\frac{1}{N}\sum_i \log(\ell_i)\delta(u-u_i)}$. The dashed purple line marks the position of the sample-averaged local noise $\overline{\ell}=\frac{1-m}{2}$, while black dashed lines mark respectively the empirical lower and upper bounds of $\ell_{typ}$. Sizes are $N=250, 500, 1500$, capacity is $\alpha=0.04$. We considered a number of samples per size such that $N_s=1500\times\frac{100}{N}$.}
    \label{fig:relation_noise_spectrum_lowedge_and_bulk}
\end{figure*}

\subsubsection{Relation between noise and low energy modes}

The relation between the rescaled square components $u_{i,k}\equiv N|\bm{\psi}_{i,k}^2|$ of rank-$k$ eigenvectors and local fields $\eta_i$ is the following: the average $u_{i, k}$ for fixed $\eta_i$ reads
\begin{equation}
\begin{split}
    \label{eq:eigenvector_weights_local_fields}
    \langle u_{i,k}\rangle=\frac{\alpha/\sigma^2}{\left|\frac{d}{\sigma^2}-\Green(\lambda_k)\right|_{\mm{C}}^2}\frac{d-1}{\left|\eta_i-\lambda_k\sigma-\frac{\alpha\sigma}{d}\frac{\Green(\lambda_k)}{\frac{d}{\sigma^2}-\Green(\lambda_k)}\right|_{\mm{C}}^2}    
\end{split}
\end{equation}
Indeed, the components $\psi_{i,k}^a$ of any eigenvectors are in the thermodynamic limit gaussian variables with variance given by \eqref{eq:eigenvector_weights_local_fields}. Eq. \eqref{eq:eigenvector_weights_local_fields} suggests that the components of the eigenvectors and local fields are anticorrelated.
In order to characterize the distribution of the noise that the network experience in the retrieval of a pattern $\xi^1$ given its corresponding Mattis state $\bm{S}$, we introduce the local noises
\begin{equation}
\label{eq:local_noise_def}
    \ell_i = \frac{1}{2}-\frac{\bm{\xi}_i^{1}\cdot\bm{S}_i}{2\sigma}
\end{equation}
Local noises are bounded in the unit interval $0\leq \ell_i\leq 1$: spins highly correlated with pattern spins ($\frac{1}{\sigma}\bm{S}_i\cdot\bm{\xi}_i^1\simeq 1$) have very low local noise, while those weakly correlated ($\frac{1}{\sigma}\bm{S}_i\cdot\bm{\xi}_i^1\simeq 0$) have local noise close to $1/2$; finally, spins negatively correlated with pattern spins have $1/2<\ell_i\leq 1$. The average of local noises is simply connected to Mattis overlap in eq. \eqref{eq:sp_eqs_T0}: $\overline{\ell}=\frac{1-m}{2}$. The local fields $\eta_i$ can be simply related to local noises: by using the stationarity condition in eq. \eqref{eq:stationarity_modelHam}, one has
\begin{equation}
\label{eq:relation_local_fields_noise}
\begin{gathered}
    \eta_i = \frac{1}{\sigma}\sum_{j:j\neq i}\mm{J}_{ij}\bm{S}_j\cdot \bm{S}_i \simeq (1-2\ell_i)\sigma m+\bm{n}_i\cdot \bm{S}_i \\
    \; \\
    \bm{n}_i=\frac{1}{\sqrt{N}}\sum_{\mu=2}^P\left(\frac{1}{\sqrt{N}}\sum_{j:j\neq i}\bm{\xi}_j^{\mu}\cdot\bm{S}_j\right)\bm{\xi}_i^{\mu}
\end{gathered}
\end{equation}
The term $\bm{n}_i\cdot\bm{S}_i$ represents the local field in absence of pattern $\mu=1$: the vector $\bm{n}_i$ is in fact an isotropic gaussian vector with variance $\alpha r$ (compare with eq. \eqref{eq:noise_order_parameter}).
Eq. \eqref{eq:relation_local_fields_noise} suggests that local fields $\eta_i$ and local noises $\ell_i$ should be negatively correlated, and consequently thanks to eq. \eqref{eq:eigenvector_weights_local_fields} that local noises and rescaled weights $u_i$ are positively correlated. The vectors $\bm{n}_i$ are correlated to local noises in a non-trivial way, thus making the relation between local fields, weights and local noises hardly accessible through analytical techniques.

We claim that the correlation of local noises $\ell_i$ with rescaled weights $u_i$, or equivalently, the anticorrelation of local noises $\ell_i$ with local fields $\eta_i$, is stronger in lower edge modes than in bulk ones. We expect nodes of localization of soft modes to pinpoint spins of the Mattis state with high levels of noise ($\ell_i\approx 1/2$) with respect to the respective pattern spins. 
The reasoning behind our conjecture is the following: in mean-field disordered systems, localization is induced by extremal values of local fields, as it happens for Anderson localization and related models \cite{anderson1959absence, venturelli2023replica} or for soft modes of mean field models of glassy systems, such as the models studied in \cite{franz2022delocalization, franz2022linear, rainone2021meanfield} or the present model. In the latter case, spins with very low local fields are then on nodes where soft modes tend to localize in the thermodynamic limit. Since spins of the Mattis states well-aligned with their respective patterns spins will have on average local fields $\eta\approx 1$ (compare with eq. \eqref{eq:retrieval_versor_rule} with a spin $\bm{S}_i\simeq \sigma\bm{\xi}_i^1$), it follows that local fields $\eta_i\ll 1$ may be related to noisy spins in the Mattis state.

We tested this conjecture numerically, by measuring the Spearman correlation of local fields and local noises with the weights of the first eigenvector and the correlation of local fields and local noises with the weights of the eigenvector of rank $k=P/2\gg 1$. We considered several samples of the $d=3$ VHM at capacity $\alpha=0.04$, for sizes $N=250, 500, 1500$.
We report our findings in fig. \ref{fig:relation_noise_spectrum_lowedge_and_bulk}.
Figures in the top panel show that local fields are anti-correlated to eigenvector weights, with almost the same correlation for lower edge and bulk modes. 
The bottom panel discusses the correlation between local noises $\ell_i$ and rescaled weights $u_i$: it appears that the correlation between local noises and weights is stronger for lower edge modes ($R=0.29$) than for bulk ones ($R=0.08$).

Spins with high noise in the memory states can also have consequences on the computational costs of retrieval, as we discuss in Appendix \ref{sec:details_retrieval_dynamics}. Specifically, we study the correlation between the number of iterations needed for convergence of \eqref{eq:retrieval_versor_rule} and the Mattis magnetization of the final state. We study this dependence also locally, measuring the correlation between local noises and local convergence times, i.e. the different number of iterations necessary for the different spins to align to the desired accuracy. We found that, as expected, noisy spins are also slow spins. Computationally, the number of iterations needed to align the slowest spins seems to asymptotically have a very weak dependence on system size. 

\section{The network above saturation}
\label{sec:retrieval_dynamics}

In this section we study the network in the saturated regime $\alpha>\alpha_c$. No retrieval states exist as fixed point of eq. \eqref{eq:retrieval_versor_rule}. We thus concentrate on the out-of-equilibrium properties of the system, focusing on the transient behavior at the very first steps of the numerical retrieval dynamics. We study numerically the first-step, zero-temperature dynamics of the system and compare our results with theoretical predictions from a signal-to-noise analysis of eq. \eqref{eq:retrieval_versor_rule}: here, our analysis spans a wide range of capacities below and above saturation, considering both Mattis and mixture states as initial conditions. For numerical simulations of dynamics we use always systems with $N=1000$ because finite size effects are already negligible at this system size for the phenomenology studied, so that using values of $N = \{250,500,1000,2000\}$ results in overlapping curves.

\subsection{Transient denoising phenomenon}

\subsubsection{Mattis states above saturation}
\begin{figure*}
    \centering
    \includegraphics[width=0.49\linewidth]{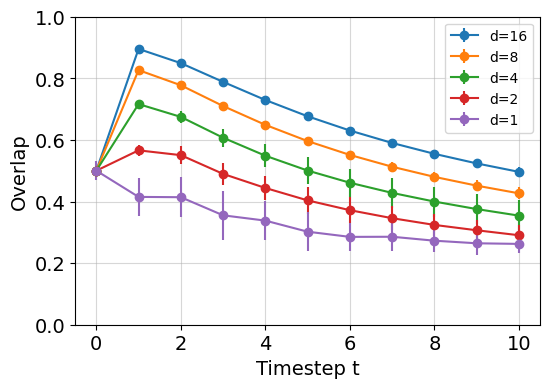}
    \includegraphics[width=0.49\linewidth]{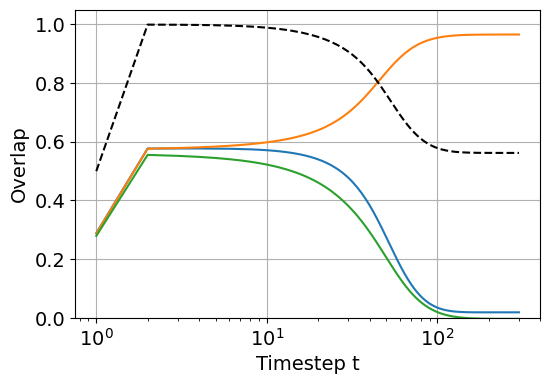}  

    \caption{\textbf{Transient denoising: the network can denoise examples and mixtures even above the storage capacity.} Panel a) shows the overlap with an examples as the function of (synchronous) time steps, for various dimensions of the spins $d$. Panel b) shows the overlap with a mixture of examples (black dashed line) as the function of (synchronous) time steps at $d=16$. The solid lines correspond to the overlaps with examples within the mixture.}
    \label{fig:m_t}
\end{figure*}

We studied Mattis states above saturation ($\alpha>\alpha_c$), out of the storage phase: here the dynamics initialized at a given pattern converges to a spin glass state with vanishing correlation with the same pattern.
\\We observed that, starting from a magnetization $0<m_0 < 1$ and small enough, $m_t$ has a peak at $t=1$, with $m_1 > m_0$. So, at the very first step the model has a denoising capability, even though the dynamics have not fixed points strongly correlated with examples. We define this phenomenon as \emph{transient} or \emph{first-step} denoising: it appears that this phenomenon is the stronger the larger $d$. 
In Fig.~\ref{fig:m_t}a we report an example of dynamics at capacity $\alpha=0.5>\alpha_c(d=1)\simeq 0.138$. In section \ref{sec:SNR_1step} we study this phenomenon more in depth.

\subsubsection{Mixtures states}
In section \ref{sec:thermodynamics} we stated that there is no stable RS solution involving mixture of examples and in Appendix \ref{sec:Instability_mixtures} we thoroughly discuss the origin of such instability. Our theoretical prediction is restricted to the case $\alpha=0$ and to symmetric mixtures: here we show numerically that mixture solutions are unstable also for $\alpha>0$.
\\In Fig. \ref{fig:m_t}b we show our results for dynamical runs starting from the symmetric mixture $\bm{S}_{ini}(0)=\frac{(\bm{\xi}_i^1+\bm{\xi}_i^2+\bm{\xi}_i^3)\sigma}{|\bm{\xi}_i^1+\bm{\xi}_i^2+\bm{\xi}_i^3|}$ and from an initial condition with overlap $m_{ini}=0.5$ with the same mixture. We chose $\alpha < \alpha_c(d=16)\approx 0.009$ in such a way there exist fixed points correlated to examples. In Fig. \ref{fig:m_t}b we show the magnetization $m_t=\frac{1}{N\sigma}\conf{S}(0)\cdot\conf{S}(t)$ during the dynamics. The dynamics always converges to a Mattis state related to one of the examples composing the mixture, therefore no mixtures stable solutions exist in VHMs. However a transient retrieval ability of the network happens also for mixtures states: at the very first step the dynamics is attracted by the mixture, but at later times it ends in one of the states composing the mixture.

\subsection{Signal-to-noise (STN) analysis at first step}
\label{sec:SNR_1step}
We study the first-step, zero-temperature dynamics of the system using the signal-to-noise analysis of eq. \eqref{eq:retrieval_versor_rule}. We consider the above saturation regime and noisy examples as initial condition. However the same computation holds also in the case of near saturation regime, and in Appendix \ref{sec:dyn_appendix} we discuss the case of mixtures as initial conditions.

\subsubsection{Mattis states}
We perform the STN analysis at the first step of the zero temperature dynamics of the model. The relevant parameters are spins dimension $d$, relative number of examples $\alpha$ and initial correlation $m_0$ with one of them. We choose as initial condition for the dynamics $\frac{1}{\sigma}\bm{S}_i(0)=m_0 \bm{\xi}_i^1+\sqrt{1-m_0^2}\;\bm{\pi}_i$ where $\bm{\pi}_i$ are unit vectors orthogonal to examples spins, $\bm{\pi}_i\cdot \bm{\xi}_i^1=0$. The initial overlap $m_0$ is a control parameter to quantify the initial amount of noise with respect to pattern $\conf{\xi}^1$. The local fields vectors at the first step of the dynamics read
\begin{equation}
\label{eq:local_fields_first_step}
\begin{split}
    &\frac{1}{\sigma}\sum_{j:j\neq i}\mm{J}_{ij}\bm{S}_j(0)=
    m_0\bm{\xi}_i^1+\bm{\mathrm{n}}_i \\
\end{split}
\end{equation}
The noise vectors $\bm{n}_i$ is an isotropic gaussian vector with zero mean and variance $\frac{\alpha}{d^2}$.

\begin{figure*}
    \centering
    \includegraphics[width=0.47\linewidth]{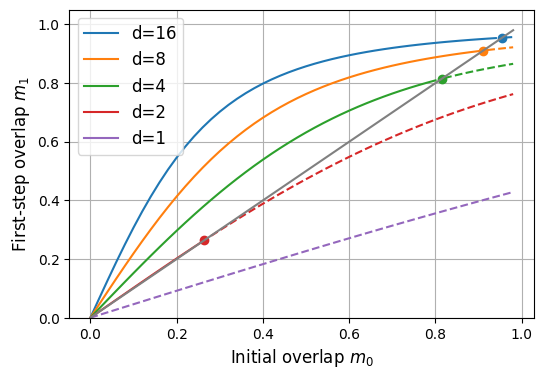}
    \includegraphics[width=0.49\linewidth]{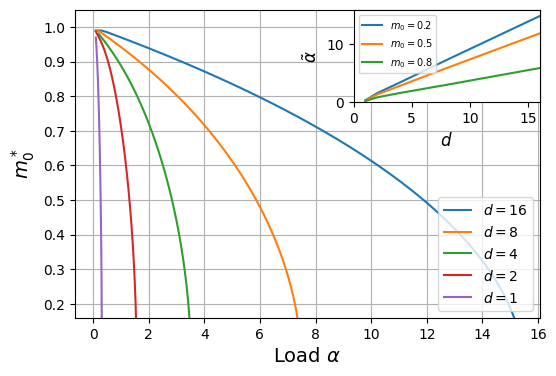}
    \caption{\textbf{The denoising performance improves with the dimension of the spins.} Panel a) We show the retrieval maps of examples for various $d$. We also show the line $m_1 = m_0$ (gray), to highlight where the model denoises effectively (solid lines, $m_1 > m_0$) and where it does not (dashed lines, $m_1 < m_0$). Panel b) the curves show the value of $m_0$ up to which $m_1 > m_0$ in the retrieval maps, as a function of the load $\alpha=P/N$. Higher values of $m_0$ mean better denoising.}
    \label{fig:m_star_alpha}
\end{figure*}
Thus, the first step Mattis magnetization is
\begin{equation}
\label{eq:m1}
\begin{split}
    m_1 &= \frac{1}{N\sigma}\sum_{i=1}^N \bm{\xi}_i^1\cdot \bm{S}_i(1) \\
    &= \int \frac{d\bm{\mathrm{n}}\; e^{-\frac{\mathrm{n}^2 d}{2(\alpha/d)}}}{[(2\frac{\pi}{d}) (\frac{\alpha}{d})]^{d/2}}\overline{\frac{m_0+\bm{\mathrm{n}}\cdot\bm{\xi}}{|m_0+\bm{\mathrm{n}}\cdot\bm{\xi}|}} \\
    \;\\
    &=\int_0^{\infty}\frac{d\mathrm{n}\;n^{d-1}e^{-\frac{\mathrm{n}^2}{2}-\frac{m^2 d}{2(\alpha/d)}}}{(2\pi)^{d/2}}\; \\
    &\,\,\,\,\,\times\K{d}\left(\frac{m_0\mathrm{n}\sqrt{d}}{\sqrt{\alpha/d}}\right)\g{d}\left(\frac{m_0\mathrm{n}\sqrt{d}}{\sqrt{\alpha/d}}\right)\\
    &=\Phi_d\left(\frac{m_0}{\sqrt{\alpha/d}}\right)
\end{split}
\end{equation}
In Fig. \ref{fig:transient_basins} (a) the analytical curves are obtained substituting in Eq. \ref{eq:m1} the values of interest, and they correctly predict the numerical simulations. In Appendix \ref{sec:dyn_appendix} we show more details on STN method and we explain how to deal with first step denoising of mixture states.
\\We introduced the function $\Phi_d\left(\cdot\right)$ to express in a more compact way the dependence of $m_1$ on $m_0, \alpha, d$. We note that $m_1$ depends only on $\frac{m_0}{\sqrt{\alpha/d}}$. 
In the special case $d=1$ of the SHM one can show that $\Phi_1(\cdot)=\operatorname{Erf}\left(\frac{\cdot}{\sqrt{2}}\right)$.
It is also worth to consider the large $d$ limit of \eqref{eq:m1}: we found that with the scaling $\alpha'=\alpha/d$ the first-step magnetization $m_1$ converges to a non-trivial function 
\begin{equation}
\label{eq:m1_limiting_behavior_large_d}
\begin{gathered}
   \lim_{d\rightarrow\infty}\Phi_d\left(\frac{m_0}{\sqrt{\alpha'}}\right)=\frac{m_0/\sqrt{\alpha'}}{\sqrt{1+(m_0/\sqrt{\alpha'})^2}}
\end{gathered}
\end{equation}
We define $x=m_0/\sqrt{\alpha'}$ and in Fig. \ref{fig:transient_basins} (b) we plot curves $y=\Phi_d(x)$ for different values of d, showing that the curves collapse rapidly to the $d \rightarrow \infty$ case. It means that the dependence on $d$ of Eq. \ref{eq:m1} is mainly in its argument rather than in $\Phi_d$ itself. Consequently, to compute first step value $m_1$ all parameters of the system $m_0, \alpha, d$ are relevant only through the combination $x$.

In Fig. \ref{fig:m_star_alpha} (a) we show retrieval maps at different $d$. The transient denoising, when present, occurs for values of $m_0 \in (0,m_0^*)$. This is the opposite of standard retrieval where the system is capable of magnetize a pattern if $m_0>\Tilde{m_0}$. In Fig. \ref{fig:m_star_alpha} (b) we show $m_0^*(\alpha)$ for various d. We obtain that transient denoising at a given $d$ can happen up to a certain $\alpha^*$, with such $\alpha^*$ increasing with d. The same is true not only for $m_0^*=0$, but for all values $k$ of $m_0^*=k$. In particular $\alpha^*$ increases linearly in $d$, with a coefficient depending on $k$.

\section{Discussion}
\label{sec:discussion}
Motivated by understanding the phenomenology of neural networks with vector variables, in this work we studied a classical Hopfield network with continuous vector spins. We first characterized the static properties of the system below saturation, solving the model in an RS ansatz and computing its phase diagram for varying storage capacity and temperature, identifying retrieval and spin glass lines. Secondly, we focused on the physics at $T=0$
and studied the low-energy excitations of the system, characterized by the soft modes of the Hessian matrix associated with the energy function. 
Finally, we studied the dynamical properties of the system, for patterns and mixture states, focusing on the behavior of the network in the early stages of the dynamics.

In the following, we discuss separately our two main results: i) Vector Hopfield networks trade off the classic retrieval of patterns, where memory states highly correlated with patterns are stored as local minima of the energy (static retrieval), for the ability of denoising patterns transiently \cite{clark2025transientdynamicsassociativememory}, in one dynamical step (first-step retrieval). The critical capacity for static retrieval shrinks with spin dimension, $\alpha_c\propto 1/d$, but the critical capacity associated with first-step retrieval grows, $\widetilde{\alpha}\propto d$. ii) Continuous associative memories with symmetric interactions, such as the Vector Hopfield model of this work, are equivalent to mean-field models of low-temperature glasses, in the behavior of the low-energy excitations of their local minima \cite{franz2022delocalization, franz2022linear}: minima related to memory states can be mapped to stable glasses, with soft modes localized on the noisiest spins of the memory state, while minima of spurious states are linked to marginal glasses, having delocalized and featureless soft modes.

Below we also discuss the possible future developments for each of these two results, which should be considered separately from the main development suggested in this work, that is to introduce vector spins in modern Hopfield networks.

\paragraph{First-step retrieval:} The phenomenon of transient denoising (in our model first-step retrieval) has recently been studied in great detail in \cite{clark2025transientdynamicsassociativememory} for very general continuous associative memory networks, through a Dynamical Mean Field Theory (DMFT) approach. The author interprets the phenomenon of transient retrieval as a geometric memory effect of the energy landscape in the vicinity of patterns, with attractive features for the dynamics that survive well beyond capacity saturation. We believe that such an explanation holds also for our model, even though at present we do not know why the first-step retrieval phase grows with spin dimension $d$: we reserve the test of the geometric memory hypothesis and the characterization of its dependence on spin dimension to future research. We point out that the landscape of our model in the vicinity of any memory state does not change sharply at saturation: indeed, since the Replicon eigenvalue in eq. \eqref{eq:Replicon_T0} remains finite at the critical load $\alpha_c=\mathcal{O}(1/d)$ for any $d$, the pseudogap width of the spectral density in eq. \eqref{eq:spectral_density_close_to_lower_edge} remains finite, and therefore there are no changes in the statistics of low energy modes at saturation. We expect that stable memory states are destabilized smoothly, with the appearance at saturation of a single negative outlier eigenvalue, together with a non-zero gradient direction\footnote{Since there is no stationary (but unstable) retrieval solution of the thermodynamics beyond saturation, we do not expect that memory states become saddle points of the landscape.}.

The author in \cite{clark2025transientdynamicsassociativememory} suggests that biological memory circuits may take advantage of dynamical transients to perform memory retrieval. In fact, transients may be fundamental for neural computations\,\cite{turner2021chartingnavigatingspacesolutions}.
Dynamical denoising and its trade-off with static storage is also relevant for current applications in Machine Learning (ML), considering the relation observed recently between the self-attention mechanism in Transformers \cite{vaswani2017attention} and modern Hopfield networks \cite{ramsauer2020hopfield, damico2024self, lucibello2024exponential}. In such ML architectures, the computation of neuron activations through Self-Attention can be seen as a single-step update for the retrieval dynamics of an exponential Hopfield network \cite{ramsauer2020hopfield}. In \cite{damico2024self} transient denoising is observed for Self-Attention layers of Visual Transformers, in a simplified setting where the image denoising task is performed by repeating a single layer of Attention.
Moreover we highlight the relation between denoising and most common machine learning tasks such as autoregression or classification in Transformer models: special mask vectors in the input have to be mapped to correct labels, in a denoising task where all of the noise is in that mask vectors while all other input vectors have not been corrupted. Interestingly, a simple scaling argument shows that state-of-the-art Large Language Models (LLMs) have typical parameters compatible with that phenomenon in our model. In a typical pre-trained Transformer for language modeling \cite{GPT3,LLAMA}, the order of magnitudes are $P\sim10^8$ training sequences $N\sim10^3$ tokens of $d\sim10^4$ dimensions. With those values, the order parameter for denoising assumes a reasonable value $\alpha'=\frac{P}{Nd}\sim 10$, while the memorization order parameter assumes an unreasonable scale $\alpha''=\frac{Pd}{N}\sim10^9$, which would rule out memorization phenomena. While LLMs have many differences with respect to a single update of a vector Hopfield network, it would be interesting to observe if transient denoising still plays a role in the performances of those models (as suggested in \cite{damico2024self}), especially because the task of masked token prediction is similar to aligning a chosen vector spin to the local field produced by the other spins. If this is the case, we note that the high dimensionality of variables itself could produce one-step retrieval of a large number of training data (proportional to the vector dimension as $P \sim Nd$), without the need for the higher-order interactions of modern Hopfield networks.

There are several interesting follow-ups that may be worth exploring to better understand first-step retrieval in Vector Hopfield networks.
 \begin{itemize}
     \item First, a necessary development of our research would be to go beyond random patterns and consider hidden-manifold models for their generation process, such as those studied in \cite{negri2023storage, kalaj2024randomfeatureshopfieldnetworks} for classical Hopfield networks with Ising spins.
     Since we showed that Vector Hopfield networks can perform first-step retrieval also for mixtures of patterns, transient denoising in Vector Hopfield models with structured data could be possible also for mixtures of hidden features, allowing for retrieval and generation of patterns up to very large capacities.
     \item It would be interesting to extend the pseudo-likelihood approach to associative memories of \cite{damico2025pseudolikelihood} to vector variables. Couplings tensor $\mathbb{J}$ would be obtained from minimization of pseudo-likelihood cost instead of being fixed with Hebb's rule. A interesting question is whether or not scalings with $d$ of equilibrium fixed points and transient denoising would be the same. 
    
     \item Another perspective could be to check if first-step retrieval of Vector Hopfield models is enhanced or weakened by algorithms that increase the capacity of Hopfield networks, such as daydreaming algorithms \cite{serricchio2025daydreaming} or Hebbian unlearning \cite{benedetti2022supervised}.
     \item Finally, there is also the possibility of studying first-step retrieval in the case of hetero-associative Hopfield networks \cite{agliari2025generalized, agliari2025networksneuralnetworksdifferent}, generalized to vector spins. These networks can store sequences of patterns, allowing for more complex computations than those performed by ordinary Hopfield networks.
 \end{itemize}
 
\paragraph{Vector Hopfield memories as stable glasses:} The nature of low-energy excitations of glasses is still debated: recent years have seen many numerical efforts to understand the density of states (DoS) of amorphous through computer glass models, being still unclear whether the DoS exhibits quasi-localized modes with a universal quartic scaling at vanishing frequencies, beyond the lowest phonon excitation \cite{lerner2021low, moriel2024experimental}, or rather if the universal quartic DoS scaling law is an artifact of numerical simulations \cite{schirmacher2024nature}. Mean-field analytical approaches on phenomenological \cite{rainone2021meanfield} and first-principle \cite{franz2022delocalization, franz2022linear} glassy models have shown that key features of glassy soft modes can be reproduced by means of random matrix theory: these models have Hessians of their respective energy functions that, in the thermodynamic limit, are instances of a deformed Wigner ensemble, where a gaussian matrix with vanishing $\Oo(1/\sqrt{N})$ entries (interaction matrix or second derivatives of the energy) is perturbed by a diagonal matrix with heterogeneous $\Oo(1)$ random entries (local fields or stiffness parameters). From the knowledge of the distribution of diagonal entries, by using free convolution of random matrices \cite{potters2020first} it is possible to evaluate the spectrum of the deformed Wigner matrix. Depending on the relative strength of the diagonal matrix with respect to the off-diagonal one, rigorous results \cite{lee2016extremal} predict the existence of localization for eigenmodes of deformed Wigner ensembles close to the spectral edges, provided that the distribution of the diagonal elements satisfies specific scaling conditions close to the edges of their support. The type of localization occurring in these models, which are defined on densely connected networks, is a condensation phenomenon, where edge eigenvectors concentrate their measure on single nodes, following a mathematical mechanism reminiscent of Bose-Einstein condensation \cite{Ikeda2023Bose}. Note that this type of localization is different from Anderson localization \cite{anderson1959absence} and related models \cite{truong2016eigenvectors, venturelli2023replica}, where bulk modes at the center of the spectrum undergo a localization transition driven by strong spatial disorder, and also from Lifshitz tails in the spectrum of the adjacency matrices of sparse locally tree-like heterogeneous networks \cite{bapst2011lifshitz, slanina2012localization}, which are induced by anomalous fluctuations in the degrees of the nodes.

The Hessian matrix \eqref{eq:Hessian} studied in this work in the thermodynamic limit is a random matrix of a deformed Wishart ensemble, where a Wishart-like matrix (the interaction matrix in \eqref{eq:Hessian}) is added to a diagonal matrix (the matrix of local fields in \eqref{eq:Hessian}). To our knowledge, this is the first work where such a random matrix ensemble is studied in relation to a physical model. The observation that memory states behave in their low-energy excitations as stable mean-field glasses corroborates the link between continuous associative memories and glassy models. It is important to stress that localization and its physical meaning in the present model are an effect of the heterogeneity of the Lagrange multipliers in \eqref{eq:stationarity_modelHam}, i.e. of the local fields: in the present model, this property is a consequence of the local constraints on the norm of neuron spins, but for models with unconstrained soft neuron spins, such as that studied in \cite{kuhn1991statistical}, heterogeneity can be achieved through local site potentials, thus our results should be valid for several continuous models of associative memories. Note that in spherical models, such as that studied in \cite{bolle2003spherical}, the Hessian diagonal is homogeneous, therefore the rich phenomenology unveiled in this work is absent. Moreover, our results in Appendix \ref{sec:spectrum_infinite_d_limit} show that the nontrivial properties observed for the lower edge spectrum of the Hessian are lost in the high-dimensional $d\rightarrow\infty$ limit, therefore they are not relevant for applications in Transformers, where the dimension of token vectors is very large, but may be relevant for activations in neural network layers, as shown in \cite{miyato2025artificial}, where the authors use Kuramoto Oscillators (two-dimensional vector spins) to perform computations, instead of customary threshold units.
There are several possible developments for our research, relatively to the properties of low-energy excitations of finite-$d$ vector Hopfield models.
\begin{itemize}
    \item It would be interesting to consider how diluting the interaction matrix would affect the stability properties of the system. Neurons with few connections may be more sensitive to noise, possibly affecting the features of soft modes. There could be phases with concentrated but delocalized soft modes, as reported in \cite{franz2025soft}.
    \item Vector Hopfield models could be studied for dynamical models with non-symmetric interactions, as attractor associative memories in recurrent neural networks. It would be insightful to see how our findings on linear stability extend to attractor models.
    \item Higher-order interactions of modern Hopfield networks \cite{krotov2016dense} could enrich the phenomenology of low energy modes.
\end{itemize}

Finally, vector spins may be relevant for neuroscience applications. Potts spins, i.e. discrete degrees of freedom with multiple states available \cite{kanter_potts-glass_1988}, have been used to represent the state of groups of binary neurons \cite{naim_reducing_2018, boboeva_capacity_2018}. Vector spins can be seen as a continuous version of Potts spins (notice the tensor coupling common between the two), and therefore they may be useful to represent the state of groups of continuous-valued neurons -- for instance, graded response neurons \cite{kuhn1991statistical}.

\section{Acknowledgments}

The authors thank Federico Ricci-Tersenghi for insightful discussions; Elena Agliari, Dario Bocchi and Alberto Fachechi for feedbacks. 
MN acknowledges the financial support of PNRR MUR project PE0000013-FAIR.

\bibliography{biblio.bib}
\appendix

\onecolumngrid

\newpage

\section{Replica computation of the free energy}
\label{sec:Replica_computation}

The equilibrium free energy is evaluated using the replica trick
\begin{equation}
    f(\alpha, \beta)=-\lim_{N\rightarrow\infty}\frac{1}{N\beta}\overline{\log \mathcal{Z}(\alpha,\beta)}=-\lim_{N\rightarrow\infty}\frac{1}{N\beta}\lim_{n\rightarrow 0}\frac{\overline{\mathcal{Z}^n}-1}{n}
\end{equation}
which allows one to exchange the disorder average of the log with that of the $n$-th moments of $\mathcal{Z}$. If one assumes $n$ to be an integer, we can introduce $n$ fictitious replica of the original system and perform the average over the disorder on this system. We have  

\begin{equation}
\label{eq:Zn_1}
\begin{gathered}
   \overline{\mathcal{Z}^n} = \int D\conf{S}_1\cdots\int d\conf{S}_n\;\overline{e^{\frac{N\beta}{2 }\sum_{\mu=1}^s\left(\frac{1}{N}\sum_{j=1}^N \bm{S}_j^a\cdot \bm{\xi}_j^\mu\right)^2
   +\frac{\beta}{2}\sum_{\mu=s+1}^P\left(\frac{1}{\sqrt{N}}\sum_{j=1}^N\bm{S}_j^a\cdot \bm{\xi}_j^\mu\right)^2
   -\frac{\beta}{2 N}\sum_{\mu=1}^P\sum_{j=1}^N(\bm{S}_j^a\cdot\bm{\xi}_j^\mu)^2
   }} \\
   \;\\
   \propto\,e^{-\frac{\beta P n \sigma^2}{2\;d}}\int D\conf{m}D\conf{\hat{m}}
   \exp\Bigl(-iN\sum_{\mu\le s}\sum_{a}m_{\mu}^a \hat{m}_{\mu}^a
   +\frac{\beta N \sigma^2}{2}\sum_{\mu\le s}\sum_a\;(m_{\mu}^a)^2-i\sum_{\mu>s}\sum_{a}m_{\mu}^a \hat{m}_{\mu}^a
   +\frac{\beta}{2}\sum_{\mu>s}\sum_a\;(m_{\mu}^a)^2\Bigr) \\
     \\
   \times\int\prod_a D\conf{S}_a\;\int \prod_{\mu}\prod_{k=1}^N\left[d\bm{\xi}_k^\mu P(\bm{\xi}_k^\mu)\right] 
   \exp\left(\frac{i}{\sigma }\sum_{\mu\le s}\sum_a\;\;\hat{m}_{\mu}^a\sum_k\bm{\xi}_{k}^a\cdot \bm{S}_{k}^a
   +\frac{i}{\sqrt{N}}\sum_{\mu>s}\sum_a\;\;\hat{m}_{\mu}^a\sum_k\bm{\xi}_{k}^a \cdot\bm{S}_{k}^a\right)
\end{gathered}
\end{equation}
In the last equation we do not show prefactors that are not relevant for the final saddle point evaluation.
We introduced magnetizations order parameters through delta functions:
\begin{equation*}
    m_{\mu, a}=\begin{cases}
        \frac{1}{N\sigma}\sum_{k=1}^N \xi_{k}^{\mu}S_{k}^a\qquad \mu\leq s \\
        \; \\
        \frac{1}{\sqrt{N}}\sum_{k=1}^N \xi_{k}^{\mu}S_{k}^a\qquad \mu>s.
    \end{cases}
\end{equation*}
The disorder average over noncondensated magnetizations reads

\begin{equation}
\label{eq:disorder_avg_noncond}
    \int \prod_{\mu>s}\prod_{k=1}^N\left[d\bm{\xi}_k^\mu P(\bm{\xi}_k^\mu)\right]\exp\left(\frac{i}{\sqrt{N}}\sum_{\mu>s}\sum_a\;\;\hat{m}_{\mu}^a\sum_k\bm{\xi}_{k}^a \cdot \bm{S}_{k}^a\right)\,\simeq\,\exp\left(-\frac{1}{2 d N}\sum_{\mu>s}\sum_{ab}\hat{m}_{\mu, a}\hat{m}_{\mu, b}\sum_{k}\bm{S}_{k}^a\cdot \bm{S}_{k}^{b}\right)
\end{equation}
Next, we introduce the spin glass overlap matrix
\begin{equation}
    Q_{ab}=\frac{1}{N\sigma^2}\sum_{k=1}^N \bm{S}_k^a \cdot \bm{S}_k^b
\end{equation}
through delta functions. After the last passages, the replicated partition function reads
\begin{equation}
\label{eq:Zn_2}
\begin{gathered}
    \overline{\mathcal{Z}^n}\propto\,e^{-\frac{\beta P n \sigma^2}{2\;d}}
    \int D\conf{m}_{\mu\leq s}D\conf{\hat{m}}_{\mu\leq s}D\mathbb{Q}D\widehat{\mathbb{Q}}\exp\left(-iN\sum_{\mu\le s}\sum_{a}m_{\mu}^a \hat{m}_{\mu}^a
   +\frac{\beta N \sigma^2}{2}\sum_{\mu\le s}\sum_a\;(m_{\mu}^a)^2+iN\sum_{ab}\widehat{Q}_{ab}Q_{ab}\right) \\
   \; \\
    \int\prod_a D\conf{S}_a\;\int \prod_{\mu\leq s}\prod_{k=1}^N\left[d\bm{\xi}_k^\mu P(\bm{\xi}_k^\mu)\right] 
   \exp\left(\frac{i}{\sigma }\sum_{\mu\le s}\sum_a\;\;\hat{m}_{\mu}^a\sum_k\bm{\xi}_{k}^a\cdot \bm{S}_{k}^a-\frac{i}{\sigma^2}\sum_{ab}\widehat{Q}_{ab}\sum_{k=1}^N \bm{S}_k^a\cdot \bm{S}_k^b\right)   \\ 
   \; \\
   \times \int D\conf{m}_{\mu > s}\frac{D\conf{\hat{m}}_{\mu > s}}{(2\pi)^{P-s}}\exp\left(-\frac{\sigma^2}{2d}\sum_{\mu>s}\sum_{ab}Q_{ab} \hat{m}_{\mu}^a \hat{m}_{\mu}^b-i\sum_{\mu>l}\sum_{a} m_{\mu}^a \hat{m}_{\mu}^a+\frac{\beta}{2}\sum_{\mu>s}\sum_a\;(m_{\mu}^a)^2\right)
\end{gathered}
\end{equation}
The integrals over noncondensated magnetizations $\{m_{\mu, a}\}_{\mu>s}$ and their conjugated fields $\{\widehat{m}_{\mu, a}\}_{\mu>s}$, after the integration over disorder in \eqref{eq:disorder_avg_noncond}, are gaussian integrals and thus can be performed straightforwardly
\begin{equation}
\begin{gathered}
        \int D\conf{m}_{\mu > s} \exp\left(\frac{\beta}{2}\sum_{\mu>s}\sum_a\;(m_{\mu}^a)^2
        \right)\int \frac{D\conf{\hat{m}}_{\mu > s}}{(2\pi)^{P-s}} \exp\left(
        -\frac{\sigma^2}{2d}\sum_{\mu>s}\sum_{ab}Q_{ab} \hat{m}_{\mu}^a \hat{m}_{\mu}^b
        -i\sum_{\mu>s}\sum_{a} m_{\mu}^a \hat{m}_{\mu}^a\right) \\
        \; \\
        \,=\, \left(\det \frac{\sigma^2}{d}\mathbb{Q}\right)^{-\frac{(P-s)}{2}}
        \int \frac{D\conf{m}_{\mu > s}}{(2\pi)^{\frac{P-s}{2}}}
        \exp\left(-\frac{1}{2}\sum_{\mu>s}\sum_{ab}\left(\frac{d(Q^{-1})_{ab}}{\sigma^2}-\beta\delta_{ab}\right)m_{\mu}^a m_{\mu}^b\right)=\left[\det\left(\mathbb{I}_n-\frac{\beta \sigma^2}{d}\mathbb{Q}\right)\right]^{-\frac{(P-s)}{2}}
\end{gathered}
\end{equation}
We can easily eliminate $\{\widehat{m}_{\mu, a}\}_{\mu\le s}$, $Q_{aa}$ and $\widehat{\mathbb{Q}}$ through saddle point evaluations
\begin{equation*}
    Q_{aa} = 1\qquad i\widehat{m}_{\mu}^a=\beta\sigma^2 m_{\mu}^a\qquad i\widehat{Q}_{ab}=\frac{\alpha}{2}\frac{\partial}{\partial Q_{ab}}\log\det\left(\mathbb{I}_n-\frac{\beta \sigma^2}{d}\mathbb{Q}\right)= -\frac{\alpha \beta \sigma^2}{2 d}\left(\mm{I}_{n}-\frac{\beta \sigma^2}{d}\mm{Q}\right)^{-1}_{ab}\equiv -\frac{\alpha\beta^2\sigma^2}{2} R_{ab}
\end{equation*}
The replicated partition function can be finally written as a saddle point integral
\begin{equation}
    \label{eq:Zn_3}
\begin{gathered}
    \overline{\mathcal{Z}^n}\propto\int D\conf{m}_{\mu\le s}D\mathbb{Q}\exp(N A_n(\conf{m}_{\mu\le s}, \mathbb{Q})) \\
    \; \\
    A_n(\conf{m}_{\mu\le s}, \mathbb{Q})) = -\frac{\alpha\beta\sigma^2 n}{2d} -\frac{\beta \sigma^2}{2}\sum_{\mu\le s}\sum_a (m_{\mu}^a)^2-\frac{\alpha\beta^2\sigma^2}{2}\sum_{a b} R_{ab}Q_{ab}-\frac{\alpha}{2}\log\det\left(\mathbb{I}_n-\frac{\beta \sigma^2}{d}\mathbb{Q}\right)+\log W(\conf{m}_{\mu\le s}, \mathbb{R}) \\
    \; \\
    W(\conf{m}_{\mu\le s, a}, \mathbb{R})=\int \prod_{a=1}^n\left(dS_a\delta(S_a-\sigma)\right)\int \prod_{\mu\le s}\left(\frac{d\xi^\mu\delta(\xi^\mu-1)}{\mathcal{S}_{d-1}(1)}\right)e^{\beta\sigma\sum_a \sum_{\mu\le s} m_\mu^a\;\bm{\xi}_\mu\cdot \bm{S}_a+\frac{\alpha\beta^2}{2}\sum_{ab}R_{ab}\;\bm{S}_a\cdot \bm{S}_b}
\end{gathered}
\end{equation}
The original problem has been reduced to a problem of $n$ interacting spins $\{\bm{S}_a\}_{a=1}^n$: once the saddle point integral is evaluated, the free energy can be finally evaluated after continuing analytically to continuous $n$ and performing the $n\rightarrow 0$ limit. The saddle point equations related to \eqref{eq:Zn_3} read
\begin{equation}
    \label{eq:sp_eq_m_Zn3}
    m_{\mu}^a = \frac{1}{\sigma}\langle \bm{\xi}_{\mu}\cdot\bm{S}_a \rangle_{\text{eff}}
\end{equation}
\begin{equation}
    \label{eq:sp_eq_Q_Zn3}
    Q_{ab} = \frac{1}{\sigma^2}\langle \bm{S}_a\cdot\bm{S}_b \rangle_{\text{eff}}
\end{equation}
\begin{equation}
    \langle\cdot\rangle_{\text{eff}}=\frac{1}{W}\int \prod_{a=1}^n\left(dS_a\delta(S_a-\sigma)\right)\int \prod_{\mu\le s}\left(\frac{d\xi^\mu\delta(\xi^\mu-1)}{\mathcal{S}_{d-1}(1)}\right)e^{\beta\sigma\sum_a \sum_{\mu\le s} m_\mu^a\;\bm{\xi}_\mu\cdot \bm{S}_a+\frac{\alpha\beta^2}{2}\sum_{ab}R_{ab}\;\bm{S}_a\cdot \bm{S}_b}(\cdot)
\end{equation}
Unfortunately, saddle point equations \eqref{eq:sp_eq_m_Zn3}, \eqref{eq:sp_eq_Q_Zn3} cannot be solved for generic $\conf{m}_{\mu\le s}, \mm{Q}$, but an ansatz is needed. The natural choice is a Replica Symmetric ansatz, as the original problem is invariant under permutation of replica indices:
\begin{equation}
\label{eq:RS_ansatz}
    m_{\mu}^a=m_{\mu}\qquad Q_{ab}=\delta_{ab}+(1-\delta_{ab})q
\end{equation}
With this ansatz the action in \eqref{eq:Zn_3} can be further simplified, allowing us to finally evaluate the free energy of the system. First, the noise parameter $\mm{R}$ becomes
\begin{equation}
    R_{ab}\equiv \begin{cases}
        \frac{1}{\beta\sigma^2}\frac{1}{\frac{d}{\sigma^2}-\beta (1-q)}+\frac{q}{\sigma^2\left[\frac{d}{\sigma^2}-\beta (1-q)\right]^2}+O(n)\equiv r_d\qquad a=b \\
        \; \\
        \frac{q}{\sigma^2\left[\frac{d}{\sigma^2}-\beta (1-q)\right]^2}+O(n)\equiv r\qquad a\neq b
    \end{cases}
\end{equation}
The action becomes
\begin{equation}
\label{eq:RS_Action}
\begin{gathered}
A_{n}=-\frac{\alpha\beta\sigma^2 n}{2d} -\frac{\beta \sigma^2 n}{2}\sum_{\mu\le s}m_{\mu}^2-\frac{\alpha\beta^2\sigma^2n(n-1)}{2} rq-\frac{\alpha\beta^2\sigma^2 n}{2}r-\frac{\alpha}{2}\Bigl[\log\left(1-\frac{\beta\sigma^2}{d}(1+(n-1)q)\right) \\
\; \\
+(n-1)\log\left(1-\frac{\beta\sigma^2}{d}(1-q)\right)\Bigr]+\log \overline{\int_{(\mm{S}_{d-1}(1))^n}\prod_a (ds_a)\sigma^{n-1} e^{\beta\sigma^2\sum_{\mu\le s} m_\mu\;\bm{\xi}_\mu\cdot \sum_a\bm{s}_a+\frac{\alpha\beta^2\sigma^2 r}{2}\left(\sum_{a}\;\bm{s}_a\right)^2}}^{(\{\bm{\xi}_{\mu}\}_{\mu\le s})}    
\end{gathered}
\end{equation}
The first term with the log can be easily simplified
\begin{equation*}
\begin{gathered}
    \frac{\alpha}{2}\Bigl[\log\left(1-\frac{\beta\sigma^2}{d}(1+(n-1)q)\right)+(n-1)\log\left(1-\frac{\beta\sigma^2}{d}(1-q)\right)\Bigr] \\
    \; \\
    =\frac{\alpha n}{2}\log\left(1-\frac{\beta\sigma^2}{d}(1-q)\right)+\frac{\alpha\beta\sigma^2}{2 d}\frac{nq}{1-\frac{\beta\sigma^2}{d}(1-q)}+O(n^2)
\end{gathered}
\end{equation*}
The integral in the last entropic term can be further simplified through an Hubbard-Stratonovich transform
\begin{equation*}
    e^{\frac{\alpha\beta^2\sigma^2 r}{2}\left(\sum_a \bm{s}_a\right)^2}=\int\frac{d\bm{h}}{(2\pi \alpha r)^{d/2}}e^{-\frac{h^2}{2\alpha r}+\beta\sigma \bm{h}\cdot\sum_a \bm{s}_a}
\end{equation*}
With this last passage, we have 
\begin{equation*}
\begin{gathered}
\log \overline{\int_{(\mm{S}_{d-1}(1))^n}\prod_a (ds_a)\sigma^{n-1} e^{\beta\sigma^2 \sum_{\mu\le s} m_\mu\;\bm{\xi}_\mu\cdot \sum_a\bm{s}_a+\frac{\alpha\beta^2\sigma^2 r}{2}\left(\sum_{a}\;\bm{s}_a\right)^2}}^{(\{\bm{\xi}_{\mu}\}_{\mu\le s})} \\
\; \\
= n\int\frac{d\bm{h}}{(2\pi \alpha r)^{d/2}}e^{-\frac{h^2}{2\alpha r}}\,\overline{\log \K{d}\left(\beta\sigma\left|\bm{h}+\sigma \sum_{\mu\le s}m_{\mu}\bm{\xi}_{\mu}\right|\right)}^{(\{\bm{\xi}_{\mu}\}_{\mu\le s})}
\end{gathered}
\end{equation*}
where we used $\K{d}(x)=\int_{\mm{S}_{d-1}(1)}d\bm{s}e^{\bm{s}\cdot \bm{x}}$ defined in \eqref{eq:Kd_gd}.
Combining all these passages, using \eqref{eq:free_energy_density_from_replica} one can straightforwardly obtain \eqref{eq:free_energy_density_RS} and related saddle point equations \eqref{eq:sp_eqs_T} by extremizing \eqref{eq:free_energy_density_RS}.

\section{The instability of mixtures}
\label{sec:Instability_mixtures}

In this appendix, we study the stability of mixture solutions ($s>1$) of eq. \eqref{eq:sp_defm_a0}. 
We have numerical evidence that mixtures are unstable for any $\alpha$ and generic types of mixtures (see for instance figure \ref{fig:m_t}b). Here we focus on the sublinear case $\alpha=0$ and on symmetric mixtures and show analytically that these states are unstable at all temperatures.\\
First, for readers convenience, we write again the expression of the free energy density in the $\alpha=0$ case (eq. \eqref{eq:eq_free_en_dens_a0}), but for symmetric mixtures
\begin{equation}
\label{eq:eq_free_en_dens_a0_again}
    f_{eq}(\beta)\,=\,\frac{\sigma^2 s}{2}m^2 
    -
    \frac{1}{\beta}\overline{\ln \sigma^{d-1}\mathcal{K}_d\left(\beta \sigma^2 m \left|\sum_{\mu\leq s}\bm{\xi}_{\mu}\right|\right)}.
\end{equation}
Here $m$ denotes the Mattis magnetization of the symmetric $s$-mixture.
Second, we also rewrite the saddle point equation for $m$:
\begin{equation}
\label{eq:sp_defm_a0_again}
    m=\overline{\g{d}\left(\beta \sigma^2 m \left|\sum_{\mu=1}^s\;\bm{\xi}_{\mu}\right|\right)\frac{\bm{\xi}_{\mu}\cdot\sum_{\nu=1}^s\;\bm{\xi}_{\nu}}{\left|\sum_{\nu=1}^s\;\bm{\xi}_{\nu}\right|}}.
\end{equation}

The Hessian of \eqref{eq:eq_free_en_dens_a0}, when evaluated on a symmetric mixture, reads \footnote{To correctly evaluate the Hessian, one should rewrite the free energy in eq. \eqref{eq:eq_free_en_dens_a0} as a function of the full configuration of Mattis magnetizations $\bm{m}_{\mu}=(m_1,\dots, m_P)$, do the derivatives and only after that restrict to symmetric mixtures of order $s$.}
\begin{equation}
\label{eq:hessian_mixtures_alphazero_general}
    \frac{1}{\sigma^2}\frac{\partial^2 f}{\partial m_{\mu}\partial m_{\nu}}\Biggl|_{\{m_{\mu}\equiv m\}}\,=\,\delta_{\mu\nu}-\beta\sigma^2\overline{\bm{\xi}_{\mu}^T\left[g_d'(\beta \sigma^2 m |\bm{z}| )\mm{P}_{\parallel}\left(\frac{\bm{z}}{|\bm{z}|}\right)
    + \frac{g_d(\beta \sigma^2 m |\bm{z}| )}{\beta \sigma^2 m |\bm{z}|}\mm{P}_{\perp}\left(\frac{\bm{z}}{|\bm{z}|}\right)
    \right]\bm{\xi}_{\nu}}
\end{equation}
where $\bm{z}\equiv \sum_{\mu=1}^s \bm{\xi}_{\mu}$.
We recall that $\K{d}(\cdot)$ and $\g{d}(\cdot)$ are defined in eq. \eqref{eq:Kd_gd}, while $\mathbb{P}_{\parallel}$ and $\mathbb{P}_{\perp}$ are respectively longitudinal and orthogonal projectors. In all eqs. \eqref{eq:eq_free_en_dens_a0_again}, \eqref{eq:sp_defm_a0_again} and \eqref{eq:hessian_mixtures_alphazero_general}, the average $\overline{\cdot}$ is understood to be over the patterns of the mixture $\{\bm{\xi}_{\mu}\}_{\mu=1}^s$.

The Hessian is a $P\times P$ matrix with only three distinct non-zero entries, respectively for the three set of indices $\mu\neq\nu\leq s$, $\mu=\nu\leq s$ and $P>\mu=\nu>s$. In the following, we explain how to carry on the disorder average over the patterns of the mixture.

\subsection{Average over mixtures}

We decomposed the mixture vector into two contributions, $\bm{z}=\sum_{\mu\le s}\bm{\xi}_{\mu}=\sum_{\mu\le s-1}\bm{\xi}_{\mu}+\bm{\xi}_s\equiv \bm{w}+\bm{\xi}_s$. One has $|\bm{z}|^2=|\bm{w}|^2+1+2|\bm{w}|u$ with $u=\frac{\bm{w}}{|\bm{w}|}\cdot \bm{\xi}_s$. The average over the disorder is thus expressed as an average over the random variables $\bm{w}$ and $u$. They represent respectively the mixture vector of an order $s-1$ symmetric mixture and the scalar product of two random unit vectors.
The pdfs of $u$ and $w=|\bm{w}|$ are the following:
\begin{equation}
    P_u(u)\,=\,\frac{(1-u^2)^{\frac{d-3}{2}}}{\int_0^{\pi}(\sin\vartheta)^{d-2}d\vartheta}\qquad\qquad -1\leq u \leq 1
\end{equation}
\begin{equation}
    P_w^{(s)}(w)=\mathcal{S}_{d-1}(1)w^{d-1}\rho_{s-1}(w)
\end{equation}
where $\rho_{s}(z)\equiv P_{\bm{z}}(\bm{z})$, the pdf of the mixture vector $\bm{z}$, satisfies the recursive equation
\begin{equation}
\label{eq:iterative_formula_pdf_of_mixtures}
\begin{gathered}
\rho_1(z)\,=\,\frac{\delta(z-1)}{\mathcal{S}_{d-1}(1)} \\
\; \\
\rho_s(z)\,=\,\frac{\int_{0}^{\pi}\;d\vartheta\sin^{d-2}(\vartheta)\;\rho_{s-1}(\sqrt{z^2+1-2 z \cos\vartheta})}{\int_{0}^{\pi}\;d\vartheta\sin^{d-2}(\vartheta)}\qquad s>1
\end{gathered}
\end{equation}
Note that $\rho_s(z)$ is the distribution of the position vector of a discrete-time random walk in $d$ dimensions at time $s$, where at each time step the walker makes a unit-length move in a random direction.

For simplicity, we consider only the cases $s=2, 3$: for these values, the distributions $\rho_2(z)$ and $\rho_3(z)$ read
\begin{equation}
    \rho_2(z)\,=\,\frac{1}{\mathcal{S}_{d-1}(1)}\frac{\left(4-z^2\right)^{\frac{d-3}{2}}\vartheta(z)\vartheta(2-z)}{2^{d-3}z\int_{0}^{\pi}\;d\vartheta\sin^{d-2}(\vartheta)}
\end{equation}
\begin{equation}
    \rho_3(z)\,=\,\frac{\int_0^{\Theta(z)}\;\frac{d\vartheta\;\sin^{d-2}(\vartheta)\;\left[4-(z^2+1-2z\cos\vartheta)\right]^{\frac{d-3}{2}}}{2^{d-3}\sqrt{z^2+1-2z\cos\vartheta}}}{\mathcal{S}_{d-1}(1)\left[\int_{0}^{\pi}\;d\vartheta\sin^{d-2}(\vartheta)\right]^2},\qquad \Theta(z)\,=\,\pi\vartheta(1-z)+\arcos\left(\frac{z}{2}-\frac{3}{2 z}\right)\vartheta(z-1)\vartheta(3-z).
\end{equation}

We could not find explicit expressions for $s>3$ and generic $d$, but only for odd values of the spin dimension $d$. For instance, for $d=3$ one has
\begin{equation}
    \rho_3(z;\;d=3)\,=\,\frac{1}{8\pi}\theta(1-z)+\frac{3-z}{16\pi z}\theta(z-1)\theta(3-z)
\end{equation}
\begin{equation}
    \rho_{4}(z;\;d=3)\,=\,\frac{1}{16\pi}\left(1-\frac{z}{2}\right)\theta(2-z)+\frac{1}{16\pi z}\theta(z-2)\theta(4-z)
\end{equation}

\subsection{Eigenvalues of the Hessian}

With the notation introduced in the last paragraph, the saddle point eq. for Mattis magnetization $m$, eq. \eqref{eq:sp_defm_a0}, and the elements of the hessian in eq. \eqref{eq:hessian_mixtures_alphazero_general} can be rewritten as
\begin{equation}
\label{eq:m_a0_sym_mix}
    m\,=\,\int_0^{s-1}\;dw P_w^{(s)}(w)\int_{-1}^1\;du P_u(u)\;\g{d}\left(\beta \sigma^2 m\sqrt{1+w^2+2 w u} \right)\frac{1+u w}{\sqrt{1+w^2+2w u}}
\end{equation}
\begin{itemize}
    \item For $\mu=\nu\leq s$
    \begin{equation}
    \label{eq:hessian_a0_sym_mix_a}
    \begin{gathered}
        \frac{1}{\rho_1^2 \rho_2^2}\frac{\partial^2 f}{\partial m_{\mu}\partial m_{\mu}}\,=\,1-\beta\sigma^2 I_1\equiv a \\
        \; \\
        I_1 = \int_0^{s-1}\;dw P_w^{(s)}(w)\int_{-1}^1\;du P_u(u)\;
        \Bigl[\frac{(1+u w)^2}{1+w^2+2 w u}\g{d}'\left(\beta \sigma^2 m\sqrt{1+w^2+2 w u}\right) \\
        \;\\
        +\left(1-\frac{(1+u w)^2}{1+w^2+2 w u}\right)\frac{\g{d}\left(\beta \sigma^2 m\sqrt{1+w^2+2 w u}\right)}{\beta \sigma^2 m\sqrt{1+w^2+2 w u}}\Bigr]
    \end{gathered}
    \end{equation}
    \item For $\mu=\nu>s$
    \begin{equation}
    \label{eq:hessian_a0_sym_mix_c}
    \begin{gathered}
        \frac{1}{\sigma^2}\frac{\partial^2 f}{\partial m_{\mu}\partial m_{\mu}}\,=\,1-\beta\sigma^2 I_2\equiv c \\
        \; \\
        I_2=\frac{1}{d}\int_0^{s}\;dw P_w^{(s+1)}(w)\left[\g{d}'(\beta \sigma^2 m w)+(d-1)\frac{\g{d}(\beta \sigma^2 m w)}{\beta \sigma^2 m w}\right]
    \end{gathered}
    \end{equation}
    \item For $\mu\neq \nu\leq s$ by computing $\sum_{\mu\nu}^{1,s}\frac{\partial^2 f}{\partial m_{\mu}\partial m_{\nu}}$ using \eqref{eq:hessian_mixtures_alphazero_general} and knowing that $\frac{\partial^2 f}{\partial m_{\mu}\partial m_{\nu}}$ are expected to be all equal, one can show that
    \begin{equation}
    \label{eq:hessian_a0_sym_mix_b}
    \begin{gathered}
    \frac{1}{\sigma^2}\frac{\partial^2 f}{\partial m_{\mu}\partial m_{\nu}}\,=\,\frac{\beta\sigma^2}{s-1} \left(I_1-\frac{I_3}{s}\right)\equiv b \\
    \; \\
    I_3=\int_0^{s}\;dw P_w^{(s+1)}(w) \g{d}'(\beta \sigma^2 m w)w^2
    \end{gathered}
    \end{equation}
    \item For $\mu\neq\nu>s$, the Hessian entries are zero.
\end{itemize}




We now proceed to evaluate the Hessian spectrum. The Hessian \eqref{eq:hessian_mixtures_alphazero_general} has a simple block structure (compare with eqs \eqref{eq:hessian_a0_sym_mix_a}, \eqref{eq:hessian_a0_sym_mix_b}, \eqref{eq:hessian_a0_sym_mix_c})

\[
\begin{pNiceArray}{cc|cc}
  && \\
  a & \mathbf{b} \,\,\,\,\,\, & \Block{2-2}<\Large>{\mathbf{0}} \\
  \;\;\;\;\;\;\;\;\;\;\;\;\ddots & &  \\
  \mathbf{b} & a\,\,\,\,\,\, \\
  && \\
  \hline
  & & \\
  \Block{2-2}<\Large>{\mathbf{0}} && c & {\Large \mathbf{0}}\,\,\,\,\,\, \\
  & & \;\;\;\;\;\;\;\;\;\;\;\;\ddots \\
  && {\Large \mathbf{0}} & c\,\,\,\,\,\, \\
  & &
\end{pNiceArray}
\]
Due to this simple structure, the Hessian has three distinct eigenvalues
\begin{itemize}
    \item The nondegenerate eigenvalue \begin{equation}     \label{eq:lambda1}         \lambda_1\,=\, a+(s-1)b \,=\,1-\frac{\beta\sigma^2}{s}I_3 \end{equation}
    This eigenvalue is related to longitudinal fluctuations of the symmetric mixture, i.e. fluctuations that preserve the degree and symmetry of the mixture.
    \item The eigenvalue
    \begin{equation}
    \label{eq:lambda2}
        \lambda_2\,=\,c\,=\,1-\beta\sigma^2 I_2
    \end{equation}
    with degeneracy $M-s$. This eigenvalue is related to fluctuations that increase the degree of the mixture, without breaking its symmetry.
    \item The eigenvalue
    \begin{equation}
    \label{eq:lambda3}
        \lambda_3\,=\,a-b\,=\,1-\frac{s}{s-1}\beta\sigma^2 I_1+\frac{\beta\sigma^2}{s(s-1)}I_3
    \end{equation}
     with degeneracy $s-1$. This eigenvalue represents fluctuations that break the symmetry of the mixtures.
\end{itemize}
\begin{figure}
    \centering
    \includegraphics[width=0.4\linewidth]{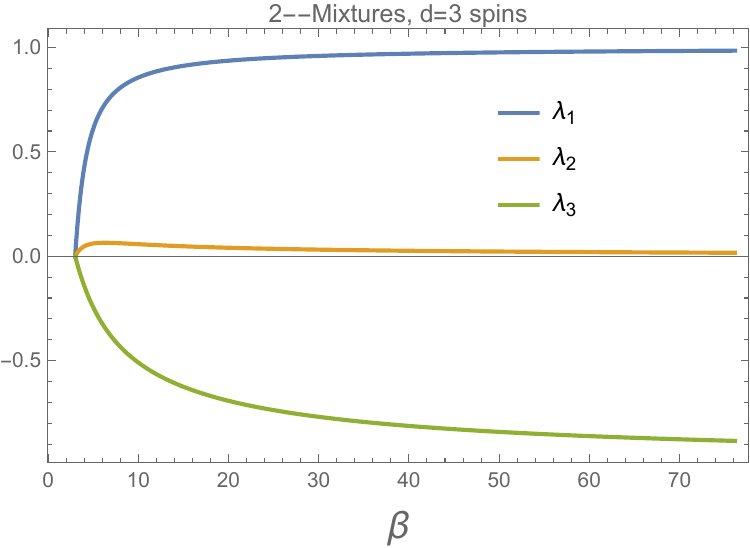}
    \includegraphics[width=0.4\linewidth]{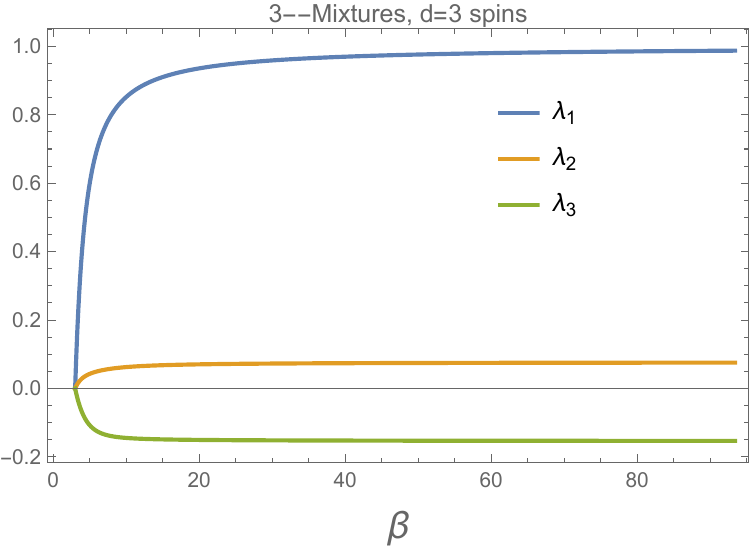}
    \caption{The three distinct eigenvalues of the $\alpha=0$ free energy Hessian in \eqref{eq:hessian_mixtures_alphazero_general} evaluated on symmetric mixtures, as functions of inverse temperature $\beta$. The left panel is for $s=2$ mixtures, the right one for $s=3$ ones. Mixtures of patterns are never stable states: the instability is due to the eigenvalue $\lambda_3$, related to fluctuations that break the symmetry of the mixture.}
    \label{fig:instability_sym_mixtures_a0}
\end{figure}
In figure \ref{fig:instability_sym_mixtures_a0} we show for the $d=3$ VHM that symmetric mixtures with $s=2, 3$ are unstable for any $T\geq 0$: the instability is due to the eigenvalue $\lambda_3$, therefore, to fluctuations that break the symmetry of the mixture. For the SHM ($d=1$) one has instead that odd\footnote{Symmetric mixtures of even orders are always unstable in the SHM, because of the possibility of having null mixtures with finite probability.} symmetric mixtures stabilize, the eigenvalue $\lambda_3$ becoming positive for sufficiently low temperatures \cite{amit1985spin}. The cause of this different behavior is the different nature of the degrees of freedom: continuous for VHM and discrete for SHM. Mathematically, in the SHM model ($d=1$) and for odd-order mixtures the integrals $I_1, I_3$ appearing in \eqref{eq:lambda3} vanish exponentially fast in the zero temperature limit, resulting in $\lambda_3\rightarrow 1$. Conversely, for $d>1$ the integral $I_1$ decays only as a power-law, $I_1\simeq  \frac{\widetilde{I}_1}{\beta\sigma^2 m}$. 

\subsection{Large $s$ and large $d$ limits}

We now consider the instability of mixtures in the limits $s\rightarrow\infty,\,d=\Oo(1)$ and $d\rightarrow\infty,\,s=\Oo(1)$ respectively.
For both these limits, we find that solutions of symmetric mixtures become marginally stable. Let us consider the two cases in the following:
\begin{itemize}
    \item Large $s$, fixed $d$:
    Here we find that the the distribution of $|\bm{z}|$ is
    \begin{equation}
        P_z^{(s)}(z)=\frac{\mathcal{S}_{d-1}(1)z^{d-1}e^{-\frac{d z^2}{2 s}}}{(2\pi \frac{s}{d})^{d/2}}
    \end{equation}
    where finite $s$ corrections are present close to the upper edge $z_{max}=s$.
    Let us focus on the $T=0$ case: we have
    \begin{equation}
        m\simeq \frac{1}{\sqrt{s}}(1-1/d)\sqrt{\frac{d}{2}}\frac{\Gamma(\frac{d-1}{2})}{\Gamma(\frac{d}{2})}\equiv \frac{k_d}{\sqrt{s}}
    \end{equation}
    \begin{eqnarray}
        &\beta I_1=1+O(1/s) \\
        &\beta I_2=1-O(1/\sqrt{s}) \\
        &\beta I_3=\frac{\beta I_1}{s-1}=\frac{1}{s}+O(1/s^2)
    \end{eqnarray}
    The eigenvalues \eqref{eq:lambda1},\eqref{eq:lambda2}, \eqref{eq:lambda3} are
    \begin{equation}
    \begin{gathered}
        \lambda_1=1-\frac{c_1}{s} \\
        \; \\
        \lambda_2=\frac{c_2}{\sqrt{s}} \\
        \; \\
        \lambda_3=-\frac{c_3}{s}
    \end{gathered}
    \end{equation}
    for some constants $c_1, c_2, c_3>0$.
    For any $d$, in the large $s$ limit, mixtures become marginally stable.
    \item Large $d$, fixed $s$:
    in this limit, the distribution of $|\bm{z}|$ is trivial, $P_z^{(s)}(z)\simeq \delta(z-\sqrt{s})$. The distribution of the scalar products is $P_u(u)\simeq \delta(u)$. Thus, we find for $d\rightarrow\infty$ (compare with eq. \eqref{eq:m_a0_sym_mix})
    \begin{equation}
        m=\frac{1}{\sqrt{s}}\sqrt{1-\frac{1}{\beta}}
    \end{equation}
    \begin{equation}
    \begin{gathered}
        \beta I_1=\frac{1}{s(2\beta-1)}+1-\frac{1}{s} \\
        \; \\
        \beta I_2=1 \\
        \; \\
        \beta I_3=\begin{cases}
            0\qquad s=1 \\
            \; \\
            \frac{\beta I_1}{s-1}-\frac{1}{(s-1)(2\beta-1)}\qquad s>1
        \end{cases}
    \end{gathered}
    \end{equation}
    Thus we find for the three eigenvalues \eqref{eq:lambda1},\eqref{eq:lambda2}, \eqref{eq:lambda3}
    \begin{equation}
    \begin{gathered}
        \lambda_1=1-\frac{1}{2\beta-1}-\frac{c_1}{d} \\
        \; \\
        \lambda_2=\frac{c_2}{d} \\
        \; \\
        \lambda_3=\begin{cases}
            \lambda_1\qquad s=1 \\
            \; \\
            -\frac{c_3}{d}\qquad s>1
        \end{cases}
    \end{gathered}
    \end{equation}
    for some constants $c_1, c_2, c_3>0$.
    In the large $d$ limit, symmetric mixtures of $s$ examples become marginally stable.
\end{itemize}

\section{Scaling limits}
\label{sec:scaling_limits_computations}

We consider the RS solution \eqref{eq:free_energy_density_RS}, \eqref{eq:energy_density_RS} and the related saddle point equations \eqref{eq:sp_eqs_T}, \eqref{eq:sp_eqs_T0} in the limit $d\rightarrow\infty$. We consider this limit as simultaneous with $N\rightarrow\infty$ and $P\rightarrow\infty$, identifying two scaling regimes:
\begin{enumerate}[label=(\alph*)]
    \item $N, P, d\rightarrow\infty$ with $\alpha'=\frac{P}{N d}=\Oo(1)$.
    \item $N, P, d\rightarrow\infty$ with $\alpha''=\frac{P d}{N}=\Oo(1)$.
\end{enumerate}

The first regime corresponds to storing $P'=\Oo(N d)\gg \Oo(N)$ examples, while in the second one instead has $1\ll P''=\Oo(N/d)\ll \Oo(N)$. Note that in this latter case one can store an infinite number of examples, provided that $N/d\rightarrow \infty$.

When $d\rightarrow\infty$, the correct scaling for the spins norm such that the free energy in eq. \eqref{eq:free_energy_density_RS} is well defined is $\sigma=\Oo(\sqrt{d})$: in particular, $\sigma=\sqrt{d}$ fixes the $\alpha=0$ critical temperature to unit for any $d$, so we choose this value for spin norms in our large $d$ limit computations. 

\paragraph{$\alpha'$ scaling}: expressing the free energy in \eqref{eq:free_energy_density_RS} it in terms of $\alpha'$ and fixing $\sigma=\sqrt{d}$, it reads
\begin{equation}
\label{eq:fe_larged_intermediate_passage}
\begin{gathered}
    \frac{f}{d}\,=\,\frac{1}{2}m^2 -\frac{1}{\beta d}\int_0^{\infty}dh\; P_h\left(h\right)|_{\alpha=\alpha'd}\log   \left[d^{\frac{d-1}{2}}\K{d}\left(\beta \sigma h\right)\right]   \\
    \\
    +\frac{\alpha'}{2}\left\{
    1
    +\frac{1}{\beta}\log\left[1-\left(1-q\right)\right]\right.
    + \left. \frac{\beta (1-q)q}{\left[1-\beta\left(1-q\right)\right]^2}
    -\frac{q}{1-\beta\left(1-q\right)}
    \right\}
\end{gathered}
\end{equation}
The distribution of cavity field in this last equation is
\begin{equation}
    P_h(h)|_{\alpha=\alpha'd}=\frac{h^{d-1}e^{-\frac{h^2}{2 \alpha' r'}}}{(2\pi \alpha' r')^{d/2}}e^{-\frac{m^2d}{2\alpha' r'}}\K{d}\left(\frac{m \sqrt{d}}{\alpha' r'}h\right)
\end{equation}
where we set $r'=r d$.
To perform the limit $d\rightarrow\infty$, we shall evaluate the integral in \eqref{eq:fe_larged_intermediate_passage}. The function $\K{d}(\cdot)$ defined in \eqref{eq:Kd_gd} has the following asymptotic behavior for large $d$ \cite{abramowitz1965handbook}
\begin{equation}
\label{eq:Kd_asymp_larged}
    d^{\frac{d-1}{2}}\K{d}(d\,x)=\int_{\sph{d}(\sqrt{d)}}d\bm{r}e^{\bm{r\cdot\bm{x}d}}\sim \left(\frac{1}{\sqrt{1+4 x^2}+1}\right)^{\frac{d-2}{2}}\frac{2^{d-1}e^{\frac{d}{2}}\pi^{\frac{d-1}{2}}e^{d\frac{2 x^2}{1+\sqrt{1+4 x^2}}}}{(1+4 x^2)^{1/4}}
\end{equation}
After rescaling $h\rightarrow\sqrt{d}\,h$, by using the expression in eq. \eqref{eq:Kd_asymp_larged} we can evaluate the integral in \eqref{eq:fe_larged_intermediate_passage} as a saddle point integral for large $d$, obtaining
\begin{equation}
\begin{gathered}
    \lim_{d\rightarrow\infty}\frac{1}{d}\int_0^{\infty}dh\; P_h\left(h\right)|_{\alpha=\alpha'd}\log   \left[d^{\frac{d-1}{2}}\K{d}\left(\beta \sigma h\right)\right] \\
    \, \\
    =\frac{1}{2}
        \left[\sqrt{1+4 \beta^2 \left(\alpha' r'+m^2\right)}-\log\left(\sqrt{1+4\beta^2  \left(\alpha' r'+m^2\right)}+1\right)+\log 4\pi\right]
\end{gathered}
\end{equation}
Finally, we obtain the free energy density
\begin{equation}
\label{eq:large_d_free_energy_alpha1}
\begin{gathered}
    \lim_{d\rightarrow \infty}\frac{f}{d}\Bigl|_{\alpha=\alpha'd}\,=\,\frac{1}{2}m^2-\frac{1}{2\beta}
        \left[\sqrt{1+4 \beta^2 \left(\alpha' r'+m^2\right)}-\log\left(\sqrt{1+4\beta^2  \left(\alpha' r'+m^2\right)}+1\right)+\log 4\pi\right] \\
    \\
    +\frac{\alpha'}{2}\left\{1+\frac{1}{\beta}\ln[1-\beta (1-q)]+\frac{q\beta (1-q)}{[1-\beta (1-q)]^2}-\frac{q}{1-\beta (1-q)}\right\}
\end{gathered}
\end{equation}
The saddle point equations in eqs \eqref{eq:sp_eqs_T} converge to the following equations
\begin{equation}
\label{eq:sp_eq_r'_infty_d_alpha1}
    r' = \frac{q}{1-\beta (1-q)}
\end{equation}
\begin{equation}
\label{eq:sp_eq_m_infty_d_alpha1}
    m=\frac{2 \beta m}{1+\sqrt{1+4 \beta^2\left(\alpha' r'+m^2\right)}}
\end{equation}
\begin{equation}
\label{eq:sp_eq_q_infty_d_alpha1}
    q = \frac{4 \beta^2 (\alpha' r'+m^2)}{[1+\sqrt{1+4 \beta^2 (\alpha' r'+m^2)}]^2}
\end{equation}
where we proceeded as before for the integral using the asymptotic expansion of $\g{d}(\cdot)$:
\begin{equation}
    \label{eq:gd_asymp_larged}
    \g{d}(d\, x)= \frac{2 x}{1+\sqrt{1+4 x^2}}+\Oo\left(\frac{1}{d}\right).
\end{equation}

The only physical solution of eqs \eqref{eq:sp_eq_r'_infty_d_alpha1}, \eqref{eq:sp_eq_m_infty_d_alpha1}, \eqref{eq:sp_eq_q_infty_d_alpha1} for $\alpha'>0$ is the no retrieval solution, with $m=0$ and $q=1-\beta_c(\alpha')/\beta$ for $\beta>\beta_c(\alpha')=1/(1+\sqrt{\alpha'})$, $q=0$ otherwise. This solution, when casted into eq. \eqref{eq:large_d_free_energy_alpha1}, returns the equilibrium free energy of the spherical Hopfield model with pairwise interactions \cite{bolle2003spherical}.

\paragraph{$\alpha''$ scaling}: by proceeding analogously to the previous case, one gets the limiting free energy
\begin{equation}
\label{eq:large_d_free_energy_alpha2}
\begin{gathered}
    \lim_{d\rightarrow \infty}\frac{f}{d}\Bigl|_{\alpha=\frac{\alpha''}{d}}\,=\,\frac{1}{2}(m^2+\alpha'' r'')-\frac{1}{2\beta}
        \left[\sqrt{1+4 \beta^2 \left(\alpha'' r''+m^2\right)}-\log\left(\sqrt{1+4\beta^2 \left(\alpha'' r''+m^2\right)}+1\right)+\log 4\pi\right].
\end{gathered}
\end{equation}
where we set $r''=r/d$.
The saddle point equations read
\begin{equation}
\label{eq:sp_eq_r_infty_d_alpha2}
    r'' = \lim_{d\rightarrow\infty}\frac{1-1/\beta}{[1-\beta(1-\langle\g{d}^2(\beta \sqrt{d} h)\rangle_{h, \alpha=\alpha''/d})]^2d^2}
\end{equation}
\begin{equation}
\label{eq:sp_eq_m_infty_d_alpha2}
    m=\frac{2 \beta m}{1+\sqrt{1+4 \beta^2\left(\alpha'' r''+m^2\right)}}
\end{equation}
\begin{equation}
\label{eq:sp_eq_q_infty_d_alpha2}
    q=1-\frac{1}{\beta} = \frac{4 \beta^2 (\alpha'' r''+m^2)}{[1+\sqrt{1+4 \beta^2 (\alpha'' r''+m^2)}]^2}
\end{equation}
Equations \eqref{eq:sp_eq_m_infty_d_alpha2}, \eqref{eq:sp_eq_q_infty_d_alpha2} admit the retrieval solution
\begin{equation}
    \label{eq:solution_sp_eqs_alpha2}
    m(\alpha'', \beta)=\sqrt{1-\frac{1}{\beta}-\alpha''r''(\alpha'', \beta)}.
\end{equation}
We did not derive the expression of $r''$ as a function of $\alpha''$ and $\beta$, as it involved next-to-leading-order corrections to the expression in \eqref{eq:Kd_asymp_larged}, \eqref{eq:gd_asymp_larged}, resulting computationally quite demanding. However, numerically, by extrapolating the expression in \eqref{eq:sp_eq_r_infty_d_alpha2} to $d\rightarrow\infty$ we found that $r''$ at $T=0$ is real in the range $0<\alpha''<\alpha_c''(0)=4/27\simeq 0.148181$.

\section{Details on the retrieval dynamics}
\label{sec:details_retrieval_dynamics}

The retrieval dynamics follows eq. \eqref{eq:retrieval_versor_rule}, denoted by us as \emph{versor rule}. We write it again in this appendix for the convenience of readers
\begin{equation}
    \bm{S}_i(t+1)=\frac{\sigma\sum_{j:j\neq i}\mm{J}_{ij}\bm{S}_j(t)}{|\sum_{j:j\neq i}\mm{J}_{ij}\bm{S}_j(t)|}=\frac{\sigma\bm{\eta_i}(t)}{|\bm{\eta_i}(t)|}.
\end{equation}
For simplicity, let us assume $\sigma=1$ for the rest of this Appendix.
We stop the retrieval algorithm when the spins on average have aligned above a certain threshold: the stopping time is defined as
\begin{equation}
\label{eq:stopping_time}
    t_{stop}=\min\left\{t\in \mathbb{N}/\{0\}: \frac{1}{N}\sum_{i=1}^N \bm{S}_i(t)\cdot \bm{S}_i(t-1)>1-\epsilon,\;\; \bm{S}_i(t)=\frac{\sum_{j:j\neq i}\mm{J}_{ij}\bm{S}_j(t)}{|\sum_{j:j\neq i}\mm{J}_{ij}\bm{S}_j(t)|}\right\}.
\end{equation}
In all our simulations, we always used $\epsilon=10^{-12}$.
In Algorithm \ref{alg:retrievalAlgo}, we report the pseudocode for the retrieval algorithm. Here we consider the retrieval of patterns, but eq. \eqref{eq:retrieval_versor_rule} is a minimization algorithm for the energy function \eqref{eq:ModelHam} and thus can be used for arbitrary initial states.

\begin{algorithm}[H] 
\caption{Retrieval dynamics algorithm}\label{alg:retrievalAlgo}
\begin{algorithmic}[1]
    \State Initialize $\conf{S}$ to a random pattern $\conf{\xi}^{(\mu)}$ sampled uniformly from $\{\conf{\xi}^{(1)},\dots,\conf{\xi}^{(P)}\}$
    \State Initialize $q=0$ and $t=0$
    \While{$q<1-\epsilon$}
    \State Save old state $\conf{S}_{old}\leftarrow \conf{S}(t)$
    \For{$i\in\operatorname{RandPermutation(\{1,\dots, N\})}$}
    \State $\bm{S}_i(t+1)\leftarrow \frac{\sum_{j:j\neq i}\mathbb{J}_{ij}\bm{S}_j(t)}{|\sum_{j:j\neq i}\mathbb{J}_{ij}\bm{S}_j(t)|}$
    \EndFor
    \State Compute $q\leftarrow \frac{1}{N}\conf{S}(t)\cdot \conf{S}_{old}$
    \State Increment time $t\leftarrow t+1$
    \EndWhile
\end{algorithmic}
\end{algorithm}

\subsection{Statistical analysis of stopping time and quality of memory states}

We performed numerical experiments to study the relation between the quality of the memory state (high Mattis magnetization) and the stopping time defined in eq. \eqref{eq:stopping_time}. These simulationsa are apart from those performed in section \ref{sec:linear_stability} to study the spectrum of memory states. We generated several samples of the $d=3$ VHM and run the retrieval dynamics as described in Algorithm \ref{alg:retrievalAlgo}, but using as a convergence criterion the stronger condition $\min_i \bm{S}_i\cdot \bm{\xi}_i^{(\mu)}>1-\epsilon$. Even though in the different batch of simulations we run to study the spectrum in \ref{sec:linear_stability} we used the average overlap criterion, with this stronger condition we can characterize spins alignment rates locally and thus compare the average convergence time (which corresponds to the stopping time defined in eq. \eqref{eq:stopping_time}) and the maximal convergence time.

\begin{figure}
    \centering
    \includegraphics[width=0.9\linewidth]{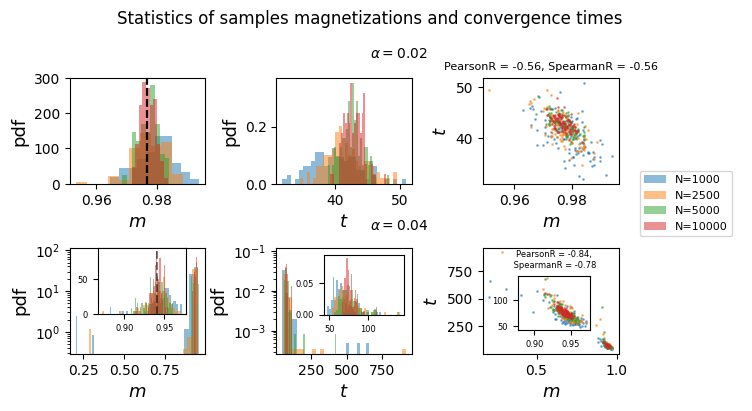}
    \caption{Statistics of Mattis magnetizations and stopping times, for sizes $N=1000, 2500, 5000, 10000$. The top row is for capacity $\alpha=0.02$, the bottom row for $\alpha=0.04$. Left figures are the pdf of Mattis magnetizations, center figures of the stopping times and right figures are scatter plots highlighting the correlation between the two. The correlation coefficients reported in the figure are for the largest size $N=10000$, while the dashed vertical line on the left figures marks the position of the asymptotic value, predicted by equations \eqref{eq:sp_eqs_T0}.}
    \label{fig:Mattis_magnetizations_VS_stopping_times}
\end{figure}

We focused on capacity values $\alpha=0.02$ and $\alpha=0.04$, representing memory states as global ($\alpha=0.02<\alpha_m\simeq 0.251$) and local minima ($\alpha=0.04>\alpha_m$).
Since these simulations are for retrieval at zero temperature, Mattis magnetizations are simply calculated as
\begin{equation}
    m_{\mu}=\frac{1}{N}\conf{S}\cdot \conf{\xi}^{(\mu)}.
\end{equation}
In fig. \ref{fig:Mattis_magnetizations_VS_stopping_times}, we report our results.
For the largest sizes, in both cases the distribution of Mattis magnetizations is concentrated around the asymptotic value, given by the solution of equations \eqref{eq:sp_eqs_T0}. For $\alpha=0.04$, we observe that at the smallest sizes the retrieval dynamics occasionally converges to spurious states, with magnetization $m\simeq 0.25\ll m_{theor}\simeq 0.947$. 
In studying the properties of memory states in section \ref{sec:linear_stability}, we only selected states with high enough magnetization: for the specific case $\alpha=0.04$, for instance, we found that the threshold $m\simeq 0.8$ was a good compromise.

When retrieval dynamics does not converge to memory states, stopping times increase significantly. Moreover, stopping times are anti-correlated to Mattis magnetizations and thus to the quality of the memory state, as predictable.

\subsection{Statistical analysis of local effects: differential alignments and noisy spins}

\begin{figure}
    \centering
    \includegraphics[width=0.9\linewidth]{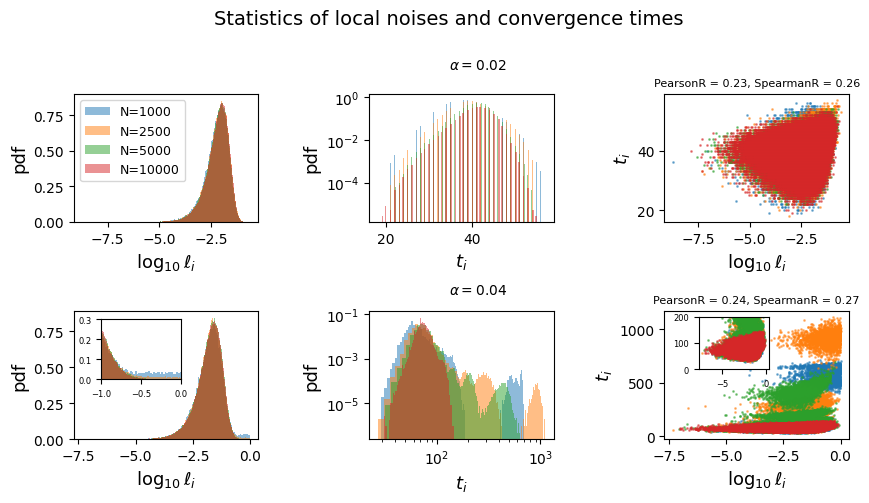}
    \caption{Statistical analysis of local noises and local convergence times. Top row refers to $\alpha=0.02$, while bottom row to $\alpha=0.04$. Left plots are the distributions of log-local noises, eq. \eqref{eq:local_noise_def}, central plots the pdfs of local convergence times \eqref{eq:local_stopping_time}, 
    right plots are scatter plots of the two quantities. The colour legend is common for all the plots. The correlation coefficients in the scatter plots refer to the largest size analyzed $N=10000$.
    }
    \label{fig:local_noises_versus_local_stopping_tinmes}
\end{figure}

We considered the different rates at which spins converged, defining local stopping times
\begin{equation}
\label{eq:local_stopping_time}
    t_i^{stop}=\min\left\{t\in \mathbb{N}/\{0\}:  \bm{S}_i(t)\cdot \bm{S}_i(t-1)>1-\epsilon,\;\; \bm{S}_i(t)=\frac{\sum_{j:j\neq i}\mm{J}_{ij}\bm{S}_j(t)}{|\sum_{j:j\neq i}\mm{J}_{ij}\bm{S}_j(t)|}\right\}
\end{equation}
and measured their correlation with the local noises, eq. \eqref{eq:local_noise_def}, whose definition we report for convenience here
\begin{eqnarray}
    \ell_i = \frac{1-\bm{S}_i\cdot \bm{\xi}_i}{2}
\end{eqnarray}
We report our results in figure \ref{fig:local_noises_versus_local_stopping_tinmes}. First, we comment on the pdfs of the log-local noises. For $\alpha=0.02$, finite-size effects are absent; for $\alpha=0.04$, we observe finite size effects on the upper tail of the distributions. These effects are caused by the dynamics occasionally converging to spurious states, as discussed in the previous paragraph (see figure \ref{fig:Mattis_magnetizations_VS_stopping_times}).
The different features observed for the values $\alpha=0.02, 0.04$ are reflected in the behavior of the distributions of local convergence times: in the former case, they are single supported, showing that retrieval dynamics consistently converges to memory states; in the latter, for $N<10000$ we observe multi-supported distributions, an indication that retrieval dynamics not always converges to the closest memory states. Notably, it appears that there are states that are as close to the starting pattern as typical memory states but are more noisy and require longer runs to convergence (see the scatter plot in figure \ref{fig:local_noises_versus_local_stopping_tinmes} in the lower right corner). Regarding correlation, we observe a positive correlation between local convergence times and local noises, yet weaker than that observed in figure \ref{fig:Mattis_magnetizations_VS_stopping_times} between Mattis magnetizations and convergence times. 

\subsection{Computational cost of slow spins}

\begin{figure}
    \centering
    \includegraphics[width=0.9\linewidth]{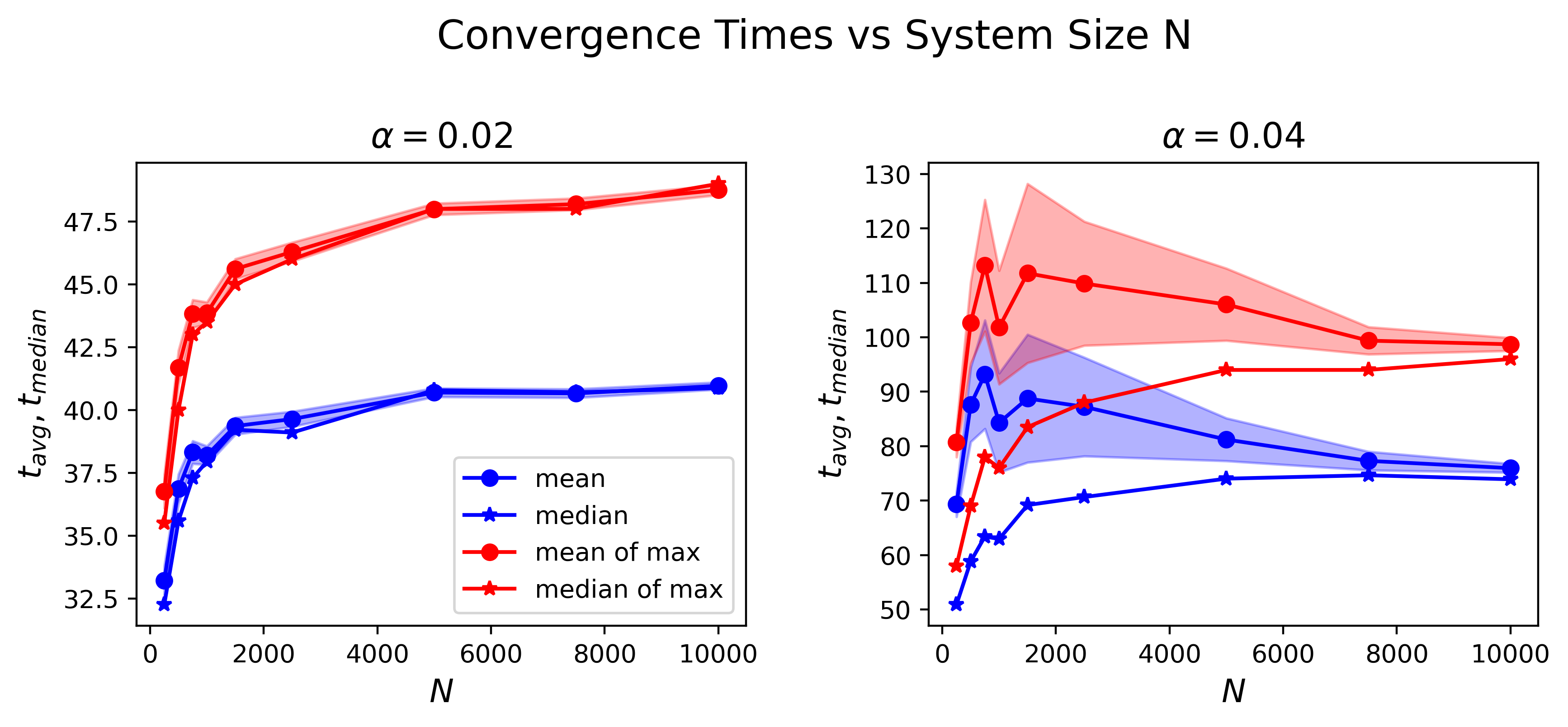}
    \caption{Means and medians of the samples convergence times and means and medians of the maximal convergence times of each sample.}
    \label{fig:convergence_times_versus_N}
\end{figure}

Finally, we characterize the computational cost of slow spins. In the previous analysis, we showed that slow spins correspond to noisy spins. In this paragraph, we want to check the scaling of convergence times with system size. We compare the average and median of the sample convergence times (eq. \eqref{eq:stopping_time}) with the average and median of the maximal convergence times (the time the slowest spin needs to reach the desired alignment accuracy) of the different samples. 

In figure \ref{fig:convergence_times_versus_N}, we show our results.
We first remark that the difference between the two metrics (average and median) is apparent for $\alpha=0.04$, in agreement with with the fact that in this case the dynamics occasionally converges to spurious states that require longer times to convergence.
In both cases, we find that the gap between typical and maximal times seems to saturate, or at most to grow very slowly: with the current accuracy and range of sizes, we are not yet able to single out one of the two options. Anyway, for any practical purposes, the gap essentially stabilizes at large enough sizes. This means that the additional computational cost of slow spins comes only from the cost related to the additional iterations. Since our retrieval dynamics in eq. \eqref{eq:retrieval_versor_rule} can be parallelized, slow spins carry an additional $\mathcal{O}(n N^2/b)$ computational cost compared to the average, where $n$ is the number of additional iterations and $b$ the number of batches of parallelization. Therefore, the optimal choice to avoid unnecessary slowdowns is to stop the retrieval dynamics according to the average overlap, as we did in the simulations we run for the spectral analysis in \ref{sec:linear_stability}. For sufficiently small thresholds $\epsilon$, the loss of accuracy that one has on consecutive alignments of slow spins is negligible.

\section{Further results for the hessian spectrum (I)}

This appendix contains several derivations for the Hessian spectrum that were not included in the main body of the text.

\subsection{The distribution of local fields}
\label{sec:local_fields_pdf_study}

The local fields
\begin{equation}
    \eta_i=|\sum_{j:j\neq i}\mm{J}_{ij}\bm{S}_j^*|
\end{equation}
are distributed according to
\begin{equation}
\label{eq:pEta}
    P_\eta(\eta)=\vartheta\left(\eta-p\right)P_{h}\left(\eta-p\right),\qquad p=\frac{\alpha\sigma}{d}\frac{\chi}{\frac{d}{\sigma^2}-\chi} 
\end{equation}
where $p$ is the Onsager reaction term, with $\chi$ solving the $T=0$ saddle points equations in \eqref{eq:sp_eqs_T0}, and $P_h(h)$ is the pdf of cavity fields defined by eq. \eqref{eq:cavity_fields_pdf}.

We measured the empirical pdfs of local fields over numerical samples
\begin{equation}
\label{eq:pEta_emp}
    P_{\eta}^{(emp)}(\eta)=\frac{1}{N}\sum_{i=1}^N\overline{\delta\left(\eta_i-\left|\sum_{j:j\neq i}\mm{J}_{ij}\bm{S}_j^*\right|\right)}
\end{equation}
and compared it with the prediction yielded by \eqref{eq:pEta}, for both Mattis states and spurious spin glass states.
In figure we show our results, for the $d=3$ VHM at capacity $\alpha=0.04$. The left panel shows the pdf for Mattis states, the right one that for no-retrieval states. In the former case, we observe a perfect agreement between the theoretical prediction \eqref{eq:pEta} and the empirical one \eqref{eq:pEta_emp}; in the latter case, it can be seen that our prediction is not exact. This is due to the RS ansatz used for our solution in \eqref{eq:energy_density_RS}, \eqref{eq:sp_eqs_T0},\eqref{eq:cavity_fields_pdf}, 
that in the case of spurious spin glass states is not correct because of the RSB transition.

\begin{figure*}
    \centering
    \includegraphics[width=\linewidth]{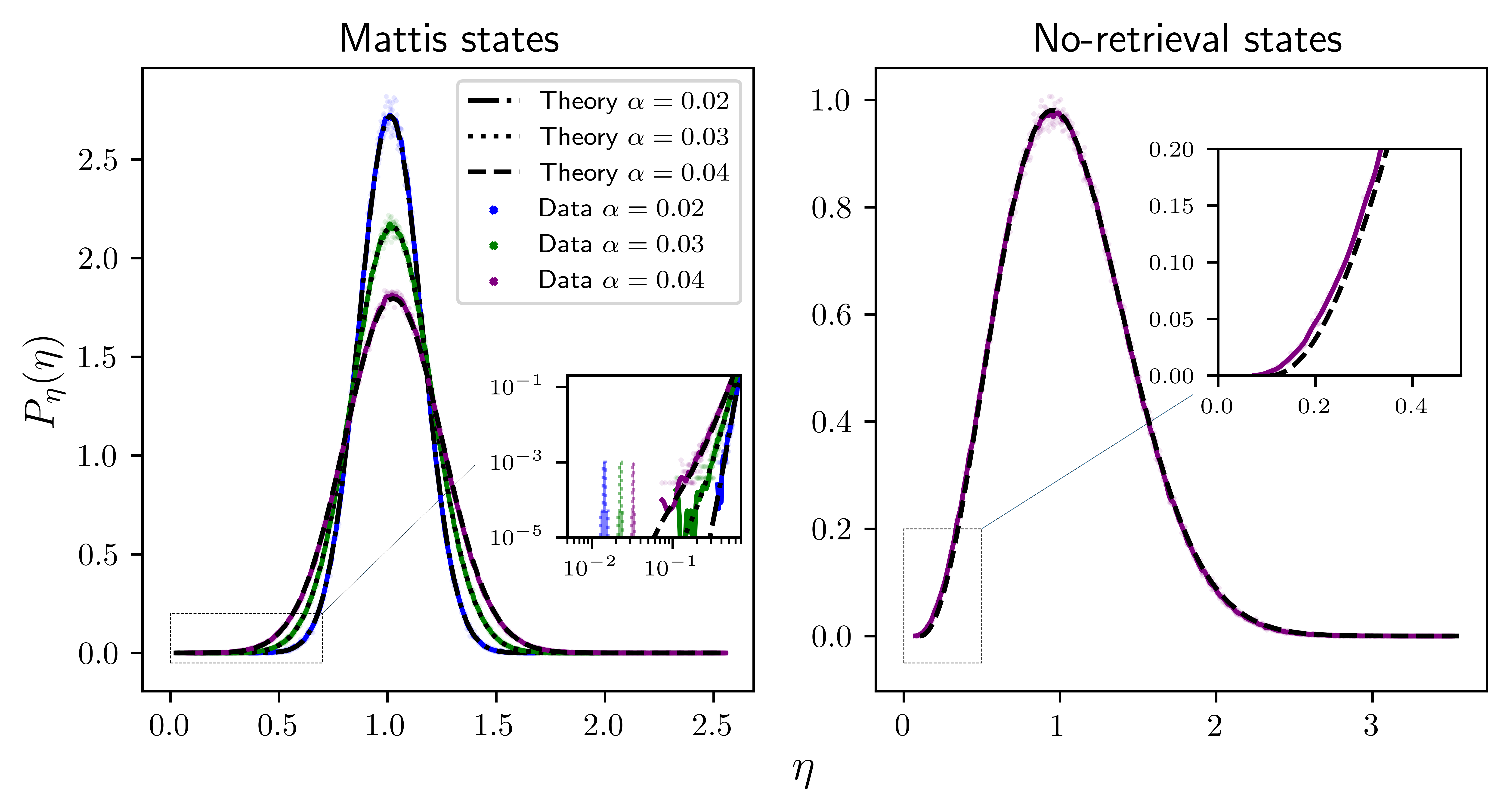}
    \caption{The distribution of local fields of the $d=3$ VHM. We compare our theoretical prediction \eqref{eq:local_fields_norms_pdf} with empirical pdf from numerical simulations of $N=4000$ spins. \textbf{Left}: the distribution for Mattis states. We show theoretical and empirical pdfs for capacities $\alpha=0.02, 0.03, 0.04$. There is a good agreement between theory and measurements. As shown in the inset in log-log scale, Onsager reactions (whose value is indicated by colored arrows) are quite small: one has $p\simeq 0.014, 0.023, 0.032$ respectively for $\alpha=0.02, 0.03, 0.04$. \textbf{Right}: the pdf for no-retrieval states only for capacity $\alpha=0.04$ (the curves of different capacities would be too close). As expected, there is no good agreement between the RS theory and experiment: the inset highlights that RSB corrections seems to move low local fields to lower values. Here the RS Onsager reaction is $p\simeq 0.11$.}
    \label{fig:distribution_local_fields}
\end{figure*}

\subsection{Derivation of the Hessian}
\label{sec:derivation_Hessian}
The Hessian in \eqref{eq:Hessian} can be derived straightforwardly from the Hamiltonian \eqref{eq:ModelHam}.
Consider a configuration of spins $\conf{S}_*$ that is a stationary point of $\mathcal{H}$: any configuration sufficiently close to it can be parametrized as follows
\begin{equation}
\label{eq:parametrize_with_pions}
    \bm{S}_i=\sigma\bm{\pi}_i+\sqrt{1-\pi_i^2}\bm{S}_i^*    
\end{equation}
where $|\bm{S}_i^*|=\sigma$ and $\bm{\pi}_i\cdot \bm{S}_i=0$. For any spin, we introduced an orthogonal "pion" vector: it is clear that \eqref{eq:parametrize_with_pions} describes for any site $i$ a random rotation of the original spin $\bm{S}_i^*$ within its half of hypersphere: indeed, $0\leq\frac{1}{\sigma^2}\bm{S}_i\cdot\bm{S}_i^*=\sqrt{1-\pi_i^2}\leq 1$. 
By plugging \eqref{eq:parametrize_with_pions}
 into \eqref{eq:ModelHam} we get
 \begin{equation*}
 \begin{gathered}
     \mathcal{H}(\conf{S})=-\frac{1}{2}\sum_{i\neq j}\sqrt{(1-\pi_i^2)(1-\pi_j^2)}\;\bm{S}_i^*\cdot\mm{J}_{ij}\bm{S}_j^*-\frac{\sigma^2}{2}\sum_{i\neq j}\bm{\pi}_i^\intercal \mm{J}_{ij}\bm{\pi}_j-\sigma\sum_{i\neq j}\sqrt{1-\pi_j^2}\;\bm{\pi}_i^\intercal \mm{J}_{ij}\bm{S}_j^* \\
     \; \\
     = \mathcal{H}(\conf{S}_*)-\sigma \sum_{i=1}^N \bm{\pi}_i\cdot \sum_{j:j\neq i} \mm{J}_{ij}\bm{S}_j^*+\frac{\sigma^2}{2}\sum_{i j}\bm{\pi}_i^\intercal \left[- \mm{J}_{ij}+\delta_{ij}\mm{I}_d\frac{\bm{S}_i^*}{\sigma^2}\cdot \sum_{j:j\neq i} \mm{J}_{ij}\bm{S}_j^* \right] \bm{\pi}_j+\Oo(\conf{\pi}^3) \\
     \; \\
     \simeq \mathcal{H}(\conf{S}_*)+\frac{\sigma^2}{2}\sum_{ij}\bm{\pi}_i^\intercal \left(-\mm{J}_{ij}+\delta_{ij}\frac{\eta_i}{\sigma}\mm{I}_d\right)\bm{\pi}_j
 \end{gathered}
 \end{equation*}
 where $\mm{I}_d$ is the $d$-dimensional identity matrix  and between the third and the fourth side of this last equation we applied the condition of stationarity \eqref{eq:stationarity_modelHam}: local field vectors $\bm{\eta}_i=\sum_{j:j\neq i}\mm{J}_{ij}\bm{S}_j^*$ are parallel to spins on a stationary point. The Hessian obtained translates into \eqref{eq:Hessian} when extended to the whole space of configurations:
 \begin{equation*}
     \mm{M}_{ij}(\conf{S}_*)=\mm{P}_{\perp}(\bm{S}_i^*)\left(-\mm{J}_{ij}+\frac{\eta_i}{\sigma}\delta_{ij}\mm{I}_d\right)\mm{P}_{\perp}(\bm{S}_j^*)
 \end{equation*}
where $\mm{P}_{\perp}(\bm{x})=\mm{I}_d-\frac{\bm{x}}{x}\frac{\bm{x}^\intercal}{x}$ is an orthogonal projector.
 
\subsection{The spectrum of the interaction matrix}
\label{sec:spectrum_interaction_matrix}
We consider the off-diagonal part of the Hessian, having elements
\begin{equation}
    W_{ij}^{\gamma\delta}\,=\,-(\mm{P}_{i}^{\perp}\mm{J}_{ij}\mm{P}_{j}^{\perp})_{\gamma\delta}(1-\delta_{ij})
\end{equation}
We compute its resolvent function
\begin{equation}
    g(z)\,=\,
    \frac{1}{N\;d}\operatorname{Tr}(\underline{\mm{W}}-z\underline{\mm{I}}_{N\;d})^{-1}
\end{equation}
through simplification of the exact Schur formula
\begin{equation}
\label{eq:Schur}
    (\mm{G}_{11}^{-1})_{\beta\gamma}\,=\,-z\delta_{\beta\gamma}-\sum_{j,k>1}\sum_{\delta\epsilon}^{1,d}W_{1j}^{\beta\delta}\widetilde{G}_{jk}^{\delta\epsilon}W_{k1}^{\epsilon\gamma}
\end{equation}
We define $\widetilde{\underline{\mm{G}}}$ as the cavity resolvent (the resolvent in the system with a cavity on site $1$).
We found that the spectrum of the interaction matrix does not depend on the type of minimum (pattern-correlated or spin glass). In the following, the calculation: introducing $v_{i,\gamma}^{\mu}=(\mm{P}_{i}^{\perp}\bm{\xi}_i^{\mu})_{\gamma}$, we expand eq.\eqref{eq:Schur}
\begin{equation}
\label{eq:Schur_1}
(\mm{G}_{11}^{-1})_{\beta\gamma}\,=\,-z\delta_{\beta\gamma}-\sum_{j,k>1}\sum_{\delta\epsilon}^{1,d}\frac{1}{N^2}\sum_{\mu,\nu}^{1, M}v_{1,\beta}^{\mu}v_{j,\delta}^{\mu}\widetilde{G}_{jk}^{\delta\epsilon}v_{k,\epsilon}^{\nu}v_{1,\gamma}^{\nu}
\end{equation}
The second term in the r.h.s. can be easily manipulated (note that in the following $A_{\mu\nu}$ is uncorrelated to $v_{1,\beta}^{\mu}, v_{1,\gamma}^{\nu}$ by construction)
\begin{eqnarray}
\label{eq:passaggi}
    &\frac{1}{N}\sum_{\mu,\nu}^{1, M}v_{1,\beta}^{\mu}\underbrace{\left(\frac{1}{N}\sum_{j,k>1}\sum_{\delta\epsilon}^{1,d}v_{j,\delta}^{\mu}\tilde{G}_{jk}^{\delta\epsilon}v_{k,\epsilon}^{\nu}\right)}_{A_{\mu\nu}}v_{1,\gamma}^{\nu}
    \,=\,
    \frac{1}{N}\sum_{\mu,\nu}v_{1,\beta}^{\mu}\;A_{\mu\nu}\;v_{1,\gamma}^{\nu}=
    \frac{1}{N}\sum_{\mu}
    v_{1,\beta}^{\mu}v_{1,\gamma}^{\mu}A_{\mu\mu}+O(1/\sqrt{N}) \\
    & \nonumber \\
    & \simeq \frac{(\mm{P}_{1}^\perp)_{\beta\gamma}}{N\;d}\sum_{j,k>1}\sum_{\delta\epsilon}\left(\frac{1}{N}\sum_{\mu}v_{j,\delta}^{\mu}v_{k,\epsilon}^{\mu}\right)\widetilde{G}_{jk}^{\delta\epsilon}\,=\,
    \frac{(\mm{P}_{1}^\perp)_{\beta\gamma}}{N\;d}\sum_{j,k>1}\sum_{\delta\epsilon}(-W_{jk}^{\delta\epsilon}+K_{j}^{\delta\epsilon}\delta_{jk})\widetilde{G}_{jk}^{\delta\epsilon} \nonumber \\
    & \nonumber \\
    & = \frac{(\mm{P}_{1}^\perp)_{\beta\gamma}}{N\;d}(-\operatorname{Tr}\underline{\widetilde{\mm{W}}}\;\underline{\widetilde{\mm{G}}}+\sum_{j>1}\operatorname{Tr} \mm{K}_j\widetilde{\mm{G}_j})\simeq 
    -(\mm{P}_{1}^\perp)_{\beta\gamma}-(\mm{P}_{1}^\perp)_{\beta\gamma}zg(z)+\frac{(\mm{P}_{1}^\perp)_{\beta\gamma}}{N d}\sum_{j>1}\operatorname{Tr}\mm{K}_j\widetilde{\mm{G}}_{jj}\nonumber
\end{eqnarray}
The tensor $K_j^{\gamma\delta}$ is
\begin{equation}
    K_{j}^{\gamma\delta}=\frac{1}{N}\sum_{\mu=1}^M v_{j,\gamma}^{\mu}v_{j,\delta}^{\mu}\simeq ... \simeq \frac{\alpha}{d}(\mm{P}_{j}^\perp)_{\gamma\delta}
\end{equation}
The last term in the last r.h.s. of \eqref{eq:passaggi} is
\begin{equation}
    \frac{1}{N d}\sum_{j>1}\operatorname{Tr} \mm{K}_j\widetilde{\mm{G}}_{jj}\,=\frac{\alpha}{N d^2}\sum_{j>1}\operatorname{Tr} \mm{P}_{j}^\perp\widetilde{\mm{G}}_{jj}\simeq\,\alpha\frac{(d-1)}{d^2}g_{\perp}(z)
\end{equation}
where we used the idem-potency of the Projector and the properties of trace to write $(1/Nd)\sum\operatorname{Tr}\mm{P}_{j}^\perp\widetilde{\mm{G}}_{jj}=(1/Nd)\sum\operatorname{Tr}(\mm{P}_{j}^\perp\widetilde{\mm{G}}_{jj}\mm{P}_{j}^\perp)=(1/Nd)\sum\operatorname{Tr}\widetilde{\mm{G}}_{jj}^{\perp})\simeq (1-1/d)g_{\perp}(z)$.
Collecting everything and using $g(z)=(1-1/d)g_{\perp}(z)+(1/d)g_{\parallel}(z)$, we end up with the expression
\begin{equation}
    (\mm{G}_{11}^{-1})_{\beta\gamma}\,=\,-z\delta_{\beta\gamma}
    -\left[-1-\frac{z}{d}g_{\parallel}(z)-z\left(1-\frac{1}{d}\right)g_{\perp}(z)+\frac{\alpha}{d}\left(1-\frac{1}{d}\right)g_{\perp}(z)\right](\mm{P}_{j}^\perp)_{\beta\gamma}+O(N^{-1/2})\end{equation}
Or equivalently, introducing $C(z)=1+\frac{z}{d}g_{\parallel}(z)+\left(z-\frac{\alpha}{d}\right)\left(1-\frac{1}{d}\right)g_{\perp}(z)$
\begin{equation}
    \mm{G}_{11}\,=\,-\left(z\mm{I}_d-C(z)\mm{P}_{1}^\perp\right)^{-1}
\end{equation}
The resolvent function is thus
\begin{equation}
    g(z)\,=\,-\frac{1}{N d}\sum_{i=1}^N\sum_{\beta=1}^d\;\left(z\mm{I}_d-C(z)\mm{P}_{i}^\perp\right)_{\beta\beta}^{-1}
\end{equation}
We can easily compute the r.h.s. of this last expression, knowing that the orthogonal projector has $1$ null eigenvalue and $d-1$ unit eigenvalues:
\begin{equation}
\label{eq:res_perp_parall}
    g(z)\,=\,-\frac{d-1}{d}\frac{1}{z-C(z)}-\frac{1}{d\;z}=(1-1/d)g_{\perp}(z)+(1/d)g_{\parallel}(z)
\end{equation}
The last equation shows that there are $N$ null eigenvalues out of $N\times d$ (there is a first order pole in $z=0$ with weight $1/d$). Let us consider from now on only $g_{\perp}(z)$, the resolvent function in the sector orthogonal to $\conf{S}$. From eq. \eqref{eq:res_perp_parall}, we obtain
\begin{equation}
\label{eq:equation_resolvent_interaction_matrix}
    g_{\perp}(z)\,=\,-\frac{1}{z-\left(1-\frac{1}{d}\right)-\left(z-\frac{\alpha}{d}\right)(1-1/d)g_{\perp}(z)}
\end{equation}
which yields a second order algebraic equation in $g(z)$ whose correct\footnote{It is the solution having the expected analytical properties for the resolvent function defined. In particular, it should behave as $g(z)\sim -1/z$ for large $|z|$.} solution is the one with the '+'
\begin{equation}
\label{eq:resolvent_interaction_matrix}
    g_{\perp}(z)\,=\,\frac{z-\left(1-\frac{1}{d}\right)}{2(z-\alpha/d)\left(1-\frac{1}{d}\right)}
    +\frac{\sqrt{\left[z-\left(1-\frac{1}{d}\right)\right]^2+4(z-\alpha/d)(1-1/d)}}{2|z-\alpha/d|\left(1-\frac{1}{d}\right)}
\end{equation}
The spectrum is given from the first order poles of the resolvent function and from its cuts on the real line. It is not hard to show that the resolvent has a pole on the point $\lambda_p=+\alpha/d$ when $\alpha<d-1$, with weight $1-\alpha/(d-1)$. For the continuous part of the spectrum, we shall compute the imaginary part of $g_{\perp}(z)$. Let $z=\lambda+i0_+$: one has that the argument of the square root in \eqref{eq:resolvent_interaction_matrix} becomes negative when $\lambda_{-}<\lambda<\lambda_{+}$, with
\begin{equation}
    \lambda_{\pm}\,=\,-[1-1/d\mp2\sqrt{(1-1/d)(\alpha/d)}]
\end{equation}
The final expression for the spectral density is a shifted Marchenko-Pastur law
\begin{equation}
\label{eq:shifted_nodiag_Wishart}
    \rho_{\perp}(\lambda)\,=\,(1-\alpha/(d-1))\vartheta(d-1-\alpha)\delta(\lambda-\lambda_p)+\frac{\sqrt{(\lambda_+-\lambda)(\lambda-\lambda_{-})}}{2\pi|\lambda-\lambda_p|(1-1/d)}\vartheta(\lambda-\lambda_{-})\vartheta(\lambda_+-\lambda)
\end{equation}
Interestingly, we recover the spectrum of a shifted (with no diagonal) Wishart matrix in the large $d$ limit if and only if one fixes $\alpha'=\alpha/d$.

\subsection{Derivation of the spectral equation}
\label{sec:derivation_spectral_equation}

We derive the spectral equation \eqref{eq:self_consistent_equation_resolvent}
using random matrix techniques. In the large $N$ limit, we treat the Hessian \eqref{eq:Hessian} as an instance of a random matrix ensemble, in this case the deformed Wishart ensemble, where a sort of Wishart matrix (the off-diagonal part or interaction matrix in \eqref{eq:Hessian}
) is summed to a disordered diagonal matrix (the local fields in \eqref{eq:Hessian}
).

We assume that the two random matrices that compose Hessian \eqref{eq:Hessian} are free: two random matrices are free when their respective eigenbasis are uncorrelated \cite{potters2020first}. Free random matrices have the so-called R-transform of the resolvent function $\mathcal{R}(\mathcal{G})=z(\mathcal{G})+\frac{1}{\mathcal{G}}$ additive, where $z(\mathcal{G})$ is the inverse of the resolvent function. 
Given a matrix $\mm{A}$ and a perturbation $\mm{B}=\operatorname{diag}\conf{v}$, one can show that the resolvent of the sum $\mm{A}+\mm{B}$, thanks to $\mathcal{R}_{A+B}=\mathcal{R}_A+\mathcal{R}_B$, satisfies a self-consistent equation \cite{potters2020first}
\begin{equation}
\label{eq:Zee}
    \mathcal{G}(z)=\int \frac{dv\;P_v(v)}{v-z+R_A(\mathcal{G}(z))}
\end{equation}
Therefore, if one knows the distribution of the diagonal elements (for our Hessian, $P_\eta$), it suffices to know the expression of the R-transform of matrix $\mm{A}$ (the off-diagonal part of \eqref{eq:Hessian} in our case).

The self-consistent equation for the Green function $\Green(z)$ of the off-diagonal part of the Hessian is eq. \eqref{eq:equation_resolvent_interaction_matrix}: from it, we see that the inverse of $\Green(z)$ is
\begin{equation}
    z(\Green)\,=\,-\frac{1}{[1-(1-1/d)\Green]\Green}+\frac{(1-1/d)(1-\alpha \Green/d)}{1-(1-1/d)\Green}    
\end{equation}
and consequently for the R-transform
\begin{equation}
    R(\Green)\,=\,z(\Green)+\frac{1}{\Green}\,=\,-\frac{(\alpha/d)(1-1/d)\Green}{1-(1-1/d)\Green}
\end{equation}
It then follows, plugging this last equation into \eqref{eq:Zee}, that the resolvent function is given by
\begin{equation}
    \mathcal{G}(z)=\int \frac{d\eta P_{\eta}(\eta)}{\frac{\eta}{\sigma}-\lambda-\frac{(\alpha/d)(1-1/d)\mathcal{G}(z)}{1-(1-1/d)\mathcal{G}(z)}}
\end{equation}
The resolvent in this is last equation is defined as $\Green(z)=\frac{\sigma^2}{N d}\operatorname{Tr}(\underline{\mm{M}}-z\;\mm{I}_{N d})^{-1}$:  by changing definition into $\mathcal{G}(z)=\frac{d-1}{N d}\operatorname{Tr}(\underline{\mm{M}}-z\;\mm{I}_{N d})^{-1}$, one gets
\begin{equation*}
\mathcal{G}(z)=\frac{d-1}{\sigma^2}\int \frac{d\eta P_{\eta}(\eta)}{\frac{\eta}{\sigma}-\lambda-\frac{(\alpha/d)\mathcal{G}(z)}{\frac{d}{\sigma^2}-\mathcal{G}(z)}}
\end{equation*}
which is eq. \eqref{eq:self_consistent_equation_resolvent}.

\subsection{Numerical solution of spectral equation}
\label{sec:spectral_equation_numerical}

In this Appendix we explain how to solve numerically the spectral equation \eqref{eq:self_consistent_equation_resolvent}. Written in its real and imaginary part, eq. \eqref{eq:self_consistent_equation_resolvent} translates into
\begin{equation}
\label{eq:self_consistent_equation_resolvent_RE}
    \Re \Green(\lambda)\,=\,\frac{(d-1)}{\sigma}\int\frac{d\eta\;P_{\eta}(\eta)\left[\eta-\lambda\sigma-\frac{\alpha\sigma}{d}\frac{(\Re \Green(\lambda)\;\frac{d}{\sigma^2}-|\Green(\lambda|^2)}{|\frac{d}{\sigma^2}-\Green(\lambda)|^2}\right]}{\left|\eta-\lambda \sigma-\frac{\alpha\sigma}{d}\frac{\Green(\lambda)}{\frac{d}{\sigma^2}-\Green(\lambda)}\right|_{\mathbb{C}}^2}.
\end{equation}
\begin{equation}
\label{eq:self_consistent_equation_resolvent_IM}
    1\,=\,\frac{\alpha (d-1)}{\sigma^2|\frac{d}{\sigma^2}-\Green(\lambda)|^2}\int\frac{d\eta\;P_{\eta}(\eta)}{\left|\eta-\lambda \sigma-\frac{\alpha\sigma}{d}\frac{\Green(\lambda)}{\frac{d}{\sigma^2}-\Green(\lambda)}\right|_{\mathbb{C}}^2}
\end{equation}
On principle, one has two solve eqs \eqref{eq:self_consistent_equation_resolvent_RE} and \eqref{eq:self_consistent_equation_resolvent_IM} simultaneously in order to determine $\Green(\lambda)$. We used a simplification that reduces the number of equations to be solved to one: we introduce the auxiliary resolvent $\widetilde{\Green}(\lambda)=\frac{\Green(\lambda)}{\frac{d}{\sigma^2}-\Green(\lambda)}$, which satisfy the self consistent equation
\begin{equation}
\label{eq:self_consistent_auxiliary_resolvent}
    \widetilde{\Green}(\lambda)\,=\,\frac{\frac{(d-1)}{\sigma}\int \frac{dh\;P_h(h)}{h+x(\lambda)}}{\frac{d}{\sigma^2}-\frac{(d-1)}{\sigma}\int \frac{dh\;P_h(h)}{h+x(\lambda)}}
\end{equation}
where we used \eqref{eq:local_fields_norms_pdf} to express integrals in terms of cavity fields pdf $P_h(h)$ and set $x(\lambda)=\frac{\alpha\sigma}{d}\widetilde{\Green}(0)-\lambda\sigma-\frac{\alpha\sigma}{d}\widetilde{\Green}(\lambda)$.
The equations for the real and imaginary parts of \eqref{eq:self_consistent_auxiliary_resolvent} are
\begin{equation}
\label{eq:self_consistent_auxiliary_resolvent_RE}
    \Re \widetilde{\Green}(\lambda)\,=\,\widetilde{\Green}(0)-\frac{d}{\alpha\sigma}\left(\lambda\sigma+\Re x\right)\,=\,\frac{\Re\mathcal{I}(x)\frac{d}{\sigma^2}-|\mathcal{I}(x)|_{\mathbb{C}}^2}{|\frac{d}{\sigma^2}-\mathcal{I}(x)|_{\mathbb{C}}^2}
\end{equation}
\begin{equation}
\label{eq:self_consistent_auxiliary_resolvent_IM}
    1\,=\,\frac{\alpha (d-1) \mathcal{J}(x)}{\sigma^2|\frac{d}{\sigma^2}-\mathcal{I}(x)|_{\mathbb{C}}^2}
\end{equation}
\begin{equation}
\label{eq:I_def}
    \mathcal{I}(x)=\int \frac{dh P_h(h)}{h+x}
\end{equation}
\begin{equation}
\label{eq:J_def}
    \mathcal{J}(x)=\int \frac{dh P_h(h)}{|h+x|_{\mathbb{C}}^2}
\end{equation}
For any value of $\Re x$ (respectively $\Im x$), one can solve \eqref{eq:self_consistent_auxiliary_resolvent_IM} to determine the correspondent $\Im x$ (respectively $\Re x$) and then compute $\lambda$ through \eqref{eq:self_consistent_auxiliary_resolvent_RE}: in this way one is able to obtain the auxiliary resolvent $\widetilde{\Green}(\lambda)$ for any given $\lambda>0$. Finally, the original resolvent can be obtained through
\begin{equation}
    \Re \Green(\lambda)\,=\,\frac{(\Re\widetilde{\Green}(\lambda)+|\widetilde{\Green}(\lambda)|_{\mathbb{C}}^2)d}{(1+\Re\widetilde{\Green}(\lambda))^2+(\Im\widetilde{\Green}(\lambda))^2}
\end{equation}
\begin{equation}
    \Im \Green(\lambda)\,=\,\frac{d}{\sigma^2}\frac{\Im\widetilde{\Green}(\lambda)}{(1+\Re\widetilde{\Green}(\lambda))^2+(\Im\widetilde{\Green}(\lambda))^2}
\end{equation}
We solved numerically eq \eqref{eq:self_consistent_auxiliary_resolvent_IM} using Mathematica\texttrademark, using as input values of $\Re x$ ranging in $-5<\Re x<0$. The theoretical curves shown in figures \ref{fig:fig_spec_dens_Mattis_d=3_theor_vs_emp} were obtained using the method described in this Appendix to solve \eqref{eq:self_consistent_equation_resolvent}.

\section{Further results for the hessian spectrum (II)}

\subsection{Crossover from non-trivial to trivial bulk}
\label{sec:crossover_nontrivial_trivial_bulk}

The spectrum of the Hessian is composed of a trivial and nontrivial bulk (see eq. \eqref{eq:spectral_density_effective}). The nontrivial bulk spectral density $\rho_{nt}$ is generated by the convolution of the Marchenko-Pastur term in \eqref{eq:shifted_nodiag_Wishart} with $P_\eta(\eta)$ in \eqref{eq:local_fields_norms_pdf}, while the trivial contribution is given by the free convolution of the Dirac delta term in \eqref{eq:shifted_nodiag_Wishart} with \eqref{eq:local_fields_norms_pdf}.

Let us show how to obtain the trivial bulk from the spectral equation \eqref{eq:self_consistent_equation_resolvent}.
In order to consider the effect of the pole $\lambda_p=\alpha/d$ in \eqref{eq:shifted_nodiag_Wishart} to the solution of \eqref{eq:self_consistent_equation_resolvent}, let us consider the behavior of the equation for $\Re g\rightarrow\infty$: by expanding, one finds
\begin{equation}
    \label{eq:self_consistent_resolvent_approx}
    g(\lambda)\,\simeq\,(d-1)\int\frac{d\eta\;P_{\eta}(\eta)}{\eta-(\lambda-\alpha/d)-i\epsilon}
\end{equation}
where we individuated the small parameter $\epsilon=\alpha\frac{\Im g}{(\Re g)^2}\ll 1$. In this limit, then one finds ($\frac{\epsilon}{x^2+\epsilon^2}\simeq \pi\delta(x)$ and $\rho = \frac{1}{(d-1)\pi}\Im g$)
\begin{equation}
\label{eq:approx_img_smaller_reg}
    \rho(\lambda) \approx P_{\eta}(\lambda-\alpha/d)
\end{equation}
This regime is expected to hold with very good approximation for sufficiently large $\lambda$, being more accurate the larger the $\lambda$. 

There are two ways to define the crossover value: the first is based on the weight born by the singular term in \eqref{eq:shifted_nodiag_Wishart}: by comparison, it is easy to see that the equation for $\widehat{\lambda}$ reads
\begin{equation}
\label{eq:lambda_st_1}
    1-\frac{\alpha}{d-1}=\int_{\lambda_*}^{\infty}\;d\lambda\;P_\eta(\lambda-\alpha/d)
\end{equation}
Eq. \eqref{eq:lambda_st_1} assigns a weight equal to that of the pole to the total probability of being in the region where the approximation \eqref{eq:self_consistent_resolvent_approx} holds. Alternatively, one can estimate the crossover value from the breaking of the regime where \eqref{eq:lambda_st_1} holds, i.e. $\epsilon=1$ or 
\begin{equation}
    1\,=\,\alpha\frac{\Im g(\hat{\lambda}_*)}{(\Re g(\hat{\lambda}_*))^2}.
\end{equation}

\subsection{Asymptotic expansions of the spectrum close to the lower edge}
\label{sec:asymptotic_expansions_spectrum_lower_edge}

\subsubsection{Expansion of $\rho(\lambda)$ for $\lambda\rightarrow 0_+$ and $d>2$}

Let us derive the behavior of the spectral density and the eigenvector moments in the vicinity of the lower edge $\lambda=0$. We consider the spectrum in the case of both Mattis states and spin-glass states. In the following paragraphs, we will use the notation $\langle\cdot\rangle\equiv \int dh P_h(h)(\cdot)$ for averages over cavity fields. Before doing the expansion, we show that the lower spectral edge must be $\lambda=0$. We consider only the case of Mattis states, as for spin-glass states of vector spin glass systems it is well known that the spectrum is gapless.

We can easily show for Mattis states that $\lambda_{min}=0$ in the thermodynamic limit, using a \emph{ab absurdum} short proof: suppose that $\rho(\lambda)=\pi\Im\Green(\lambda)=0$ for small enough $\lambda>0$,
which is equivalent to asking that the spectrum be gapped. By expanding eq. \eqref{eq:self_consistent_equation_resolvent} for $\lambda\rightarrow 0$, we get
\begin{equation}
    \label{eq:self_consistent_resolvent_small_lambda}
    \Re\Green(\lambda)\simeq \frac{(d-1)}{\sigma}\int_{0}^{\infty}\frac{d h \;P_h(h)}{h-\frac{\lambda\sigma}{\Lambda_0}}
\end{equation}
where $\Lambda_0$ is the $T=0$ Replicon defined in eq. \eqref{eq:Replicon_T0}.
Since $\Lambda_0>0$ for a Mattis state, the integral in \eqref{eq:self_consistent_resolvent_small_lambda} exists if and only if $\lambda=0$, contradicting the initial condition $\Im\Green(\lambda)=0$ for $\lambda>0$. It follows that one must necessarily have $\rho(\lambda)=\pi\Im\Green(\lambda)>0$ for any $\lambda>0$.

We proceed with the expansion.
\begin{itemize}
    \item Mattis states: here the Replicon, as defined by eq. \eqref{eq:Replicon_T0}, is strictly positive. Let us begin: by plugging eq. \eqref{eq:self_consistent_equation_resolvent_IM} into \eqref{eq:self_consistent_equation_resolvent_RE}, we get

    \begin{equation}
    \label{eq:self_consistent_equation_resolvent_RE_bis}
        \Re \Green(\lambda) \,=\, \frac{(d-1)}{\sigma}\left\langle \frac{h}{|h+x|_{\mathbb{C}}^2}\right\rangle+\frac{\sigma|\frac{d}{\sigma^2}-\Green(\lambda)|_{\mathbb{C}}^2\;\Re x(\lambda)}{\alpha}
    \end{equation}
    
    where
    
    \begin{equation}
    \label{eq:x_def}
        x(\lambda)\,=\,\frac{\alpha\sigma}{d}\frac{\chi}{\frac{d}{\sigma^2}-\chi}-\lambda\sigma-\frac{\alpha\sigma}{d}\frac{\Green(\lambda)}{\frac{d}{\sigma^2}-\Green(\lambda)}.
    \end{equation}
    
    Now, knowing that $\Re \Green(\lambda)\,=\,\chi+\chi_{SG}\lambda+O(\lambda^2)$, and $\chi\,=\,\frac{(d-1)}{\sigma}\langle 1/h \rangle$, after a few straightforward simplifications eq. \eqref{eq:self_consistent_equation_resolvent_RE_bis} rewrites
    \begin{equation}
    \label{eq:self_consistent_resolvent_small_lambda_0}
        -\left(\chi_{SG}\lambda+\frac{\sigma(\frac{d}{\sigma^2}-\chi)^2}{\alpha}\Re x(\lambda)\right)\,=\,\frac{(d-1)}{\sigma}|x(\lambda)|_{\mathbb{C}}^2\left\langle\frac{1}{h|h+x|_{\mathbb{C}}^2}\right\rangle
    \end{equation}
    Since the r.h.s. of this last equation is positive, the l.h.s. is positive if and only if $\Re x<0$: this is indeed the case. Indeed, $x(0)=0$ and one can show that
    \begin{equation}
        \chi_{SG}=\frac{d\Green}{d\lambda}(0)\,=\,-\frac{(d-1)}{\sigma}\left\langle \frac{1}{h^2}\right\rangle \frac{d\Re x}{d\lambda}(0)
    \end{equation}
    so that
    \begin{equation}
      \frac{d\Re x}{d\lambda}(0)\,=\,-\frac{\sigma \chi_{SG}}{(d-1)\left\langle \frac{1}{h^2}\right\rangle}\,=\,-\frac{\sigma}{\Lambda}.  
    \end{equation}
    We have then that the l.h.s. of eq. \eqref{eq:self_consistent_resolvent_small_lambda_0} is
    \begin{equation}
        -\chi_{SG}\lambda-\frac{\sigma(\frac{d}{\sigma^2}-\chi)^2}{\alpha}\Re x\simeq -\frac{\sigma^2(\frac{d}{\sigma^2}-\chi)^2}{\alpha\Lambda}(1-\Lambda)\lambda+\frac{\sigma^2(\frac{d}{\sigma^2}-\chi)^2}{\alpha\Lambda}\lambda\,=\,\frac{\sigma^2(\frac{d}{\sigma^2}-\chi)^2}{\alpha}\lambda.
    \end{equation}
    With this, eq. \eqref{eq:self_consistent_resolvent_small_lambda_0} rewrites
    \begin{equation}
        1\,=\,\frac{\alpha (d-1) |x|_{\mathbb{C}}^2}{\sigma^3(\frac{d}{\sigma^2}-\chi)^2\lambda}\left\langle\frac{1}{h|h+x|_{\mathbb{C}}^2}\right\rangle
    \end{equation}
    To evaluate the integral in the r.h.s. of this last expression, we assume that $|\Im x|\ll |\Re x|$: by doing this, we can evaluate the integral easily using $\frac{\epsilon}{x^2+\epsilon^2}\simeq \pi \delta(x)$
    \begin{equation}
    \label{eq:simplification}
        \int\frac{dh P_h(h)}{h|h+x|^2}\,=\,\frac{\pi}{|\Im x|}\int\frac{|\Im x| P_h(h) dh}{\pi h[(h+\Re x)^2+(\Im x)^2]}\simeq \frac{\pi}{|\Im x|}\frac{P_h(|\Re x|)}{|\Re x|}
    \end{equation}
    By combining the last two equations and expanding until leading order, we finally get ($|\Im x|\simeq \frac{\alpha}{\sigma^2(\frac{d}{\sigma^2}-\chi)^2}\Im \Green$ and $\rho=\frac{\Im \Green}{\pi (d-1)}$)
    \begin{equation}
    \label{eq:rho_close_to_lower_edge}
    \begin{gathered}
        \rho(\lambda)\simeq \frac{\sigma}{\Lambda}P_h\left(\frac{\lambda \sigma}{\Lambda}\right)=a_d\lambda^{d-1} \\
        \; \\
        a_d = \frac{\mathcal{S}_{d-1}(\sigma)\sigma\;e^{-\frac{m^2(d-1)\langle 1/h^2\rangle \sigma^2}{2(1-\Lambda)}}(1-\Lambda)^{d/2}}{[2 \pi (d-1)\langle 1/h^2\rangle]^{d/2}\Lambda^d}
    \end{gathered}
    \end{equation}
where we used the identity $\alpha r_0 = \frac{1-\Lambda}{(d-1)\langle 1/h^2\rangle}$ coming from eq. \eqref{eq:Replicon_T0}.
The prefactor $a_d$ is exponentially suppressed as $\Lambda\rightarrow 1$, which corresponds to the sub-linear capacity limit $\alpha\rightarrow 0$: this corresponds to the trivialization of the spectrum in that limit, $\rho(\lambda)\rightarrow \delta(\lambda-1)$.

Let us check a posteriori the validity of the assumption $|\Im x|\ll |\Re x|$: combining eq. \eqref{eq:simplification} and $\Re x\simeq -\lambda/\Lambda$, we get
\begin{equation}
\label{eq:Img_less_Reg}
\begin{gathered}
    \frac{|\Im x|}{|\Re x|}\simeq b_d\lambda^{d-2}\ll 1\Longrightarrow \lambda_{cr}=\frac{1}{b_d^{\frac{1}{d-2}}} \\
    \; \\
    b_d = \frac{\pi\alpha (d-1) a_d \Lambda}{\sigma^4 \left(\frac{d}{\sigma^2}-\chi\right)^2}
\end{gathered}
\end{equation}
The prefactor $b_d$ is proportional to $a_d$, therefore the crossover eigenvalue $\lambda_{cr}$ grows exponentially fast as $\alpha\rightarrow 0$, consistently with the trivialization of the spectrum in this limit $\rho(\lambda)\rightarrow \delta(\lambda-1)$ (with a delta peak spectrum, the imaginary part of $x$ is identically null for any $\lambda$).

We finally complete this section with the estimation of the pseudogap eigenvalue. To do that, we rearrange the terms in eq. \eqref{eq:x_def}, obtaining for small $|x|_{\mathbb{C}}$
\begin{equation}
\label{eq:x_def_rewritten}
    -\lambda-\Lambda_0 x=\left(\frac{d-1}{\sigma}\right)\alpha r_0 \left[\left\langle\frac{1}{h^3}\right\rangle+\left(\frac{d-1}{\sigma}\right)\sigma\sqrt{r_0}\,\left\langle \frac{1}{h^2}\right\rangle^2\right]x^2
\end{equation}
By neglecting the imaginary part of $x$, we can solve eq. \eqref{eq:x_def_rewritten} as a second order algebraic equation for its real part: to estimate the pseudogap, we compute the value of $\lambda$ that cancels the discriminant of said equation, obtaining
\begin{equation}
    \label{eq:pseudogap_width}
    \lambda_{pg}=\frac{\Lambda_0^2}{4 W}
\end{equation}
where $W$ is the coefficient of the $x^2$ term in eq. \eqref{eq:x_def_rewritten}. In the $d=3$ case, the term $\langle 1/h^3\rangle$ is divergent and the coefficient is replaced by a singular term $W\propto |\log \Lambda_0|$.

\item Spin glass states: although we do not know the Full Replica Symmetry Breaking phase (fRSB) expression of the pdf of local fields on a local minimum of the energy, we can altogether characterize the scaling of $\rho(\lambda)$ at small eigenvalues. In order to obtain a physical solution for the spectrum, we need to impose $\Lambda_0=0$ for any $\alpha>0$: in fact, the unphysical solution with a spectral gap in the right panel of figure \ref{fig:fig_spec_dens_Mattis_d=3_theor_vs_emp} is a consequence of having $\Lambda_0<0$ from eq. \eqref{eq:Replicon_T0} when using the RS $P_h(h)$. In the following lines, the average over the cavity fields is assumed to be over the fRSB pdf. 

Let us consider eq. \eqref{eq:self_consistent_equation_resolvent_RE_bis}: by assuming $|\Re x|\ll |\Im x|$ and $|\Re x|=\Oo(\Im x)^2$, the integral can be expanded and we get at leading order
\begin{equation}
    \Re\Green(\lambda)=\chi-\frac{2\Re x}{\alpha r_0 \sigma}-\frac{d-1}{\sigma}\left\langle \frac{1}{h^3}\right\rangle\,(\Im x)^2+\frac{\sigma}{\alpha}\left|\frac{d}{\sigma^2}-\Green(\lambda)\right|^2\Re x
\end{equation}
In deriving this last expression we imposed the nullity of the Replicon in \eqref{eq:Replicon_T0} and thus set $1=\alpha r (d-1)\langle 1/h^2\rangle$. Here, $\langle \cdot \rangle $ denotes an average over the unknown $P_h^{RSB}(h)$.
Note that the relation between the distribution of local fields and that of cavity fields in \eqref{eq:local_fields_norms_pdf} still holds, though with a value of the Onsager reaction dictated by the unknown fRSB solution and a different expression of the fRSB distribution of cavity fields.

If we further assume that $\Re \Green(\lambda)=\chi+o(\Im x)^2$ and thus $\Re x=-\lambda+o(\lambda)$, we get at leading order
\begin{equation}
    \frac{d-1}{\sigma}\left\langle \frac{1}{h^3}\right\rangle (\Im x)^2 = \frac{\lambda}{\alpha r_0\sigma}\Longrightarrow \rho(\lambda)=\frac{1}{\pi}\sqrt{\frac{\lambda}{(\alpha r_0)^3 (d-1)\langle 1/h^3\rangle}}\equiv \widehat{a}_d \,\sqrt{\lambda}
\end{equation}
where we used $|\Im x|=\alpha r_0 |\Im \Green|$ and $\rho=\frac{1}{\pi}|\Im \Green|$. The prefactor $\widehat{a}_d$ is well defined if $P_h^{RSB}(h)=o(h^2)$ for small $h$. On contrary, it should be replaced by a $\lambda$-dependent singular term. Assuming that $P_h^{RSB}(h)$ retains the low cavity field of the RS solution, $P_h^{RSB}(h)\approx h^{d-1}$, we get the following scaling for the spectral density at low eigenvalues
\begin{equation}
    \rho(\lambda)\approx \begin{cases}
        \sqrt{\lambda}\qquad d>3 \\
        \, \\
        \sqrt{\frac{\lambda}{|\log \lambda|}}\qquad d=3
    \end{cases}
\end{equation}

\end{itemize}

\subsubsection{Expansion of $\rho(\lambda)$ for $\lambda\rightarrow 0_+$ for $d=2$}

In the $d=2$ case the integral defining the Replicon in eq. \eqref{eq:Replicon_T0} is logarithmically divergent to $-\infty$, if close to $h=0$ the scaling $P_h(h)\propto h$ is assumed. Given that the retrieval phase has fRSB as well, we expect both for Mattis and spin glass states to have $\Lambda=0$ and thus $P_h^{(RSB)}(h)\ll h$ close to $h=0$.

Here we cannot find the scaling of the spectrum without knowing the fRSB distribution of fields. We conjecture that in the case of Mattis states the behavior of the exponent is continuous, so that one has by comparing with eq. \eqref{eq:rho_close_to_lower_edge} 
\begin{equation}
    \rho(\lambda) \approx \lambda\qquad d=2
\end{equation}
For spin glass minima, instead we conjecture a behavior akin to that of the $d=3$ case, with a square root behavior with logarithmic corrections.

\subsubsection{Expansion of $\kappa(\lambda)$ for $\lambda\rightarrow 0_+$}
\label{sec:expansion_rescaled_IPR_close_to_lower_edge} In this section, we expand eq. \eqref{eq:rescaled_IPR_theory} for the rescaled IPR $\kappa(\lambda)=NI(\lambda)$ close to the lower edge.

Preliminary to that, we derive eq. \eqref{eq:rescaled_IPR_theory} starting from eq. \eqref{eq:eigenvector_weights_local_fields}, which we report here for the convenience of the readers
\begin{equation}
\begin{split}
    \label{eq:eigenvector_weights_local_fields}
    N\langle|\bm{\psi}_i^{(k)}|^2\rangle=\frac{\alpha/\sigma^2}{\left|\frac{d}{\sigma^2}-\Green(\lambda_k)\right|_{\mm{C}}^2}\frac{d-1}{\left|\eta_i-\lambda_k\sigma-\frac{\alpha\sigma}{d}\frac{\Green(\lambda_k)}{\frac{d}{\sigma^2}-\Green(\lambda_k)}\right|_{\mm{C}}^2}    
\end{split}
\end{equation}
where $|\cdot|_{\mathbb{C}}$ it the complex norm.
Eq. \eqref{eq:eigenvector_weights_local_fields} should be understood as an eq. for the average value of eigenvector weights at fixed local fields $\eta_k$. Indeed, eigenvector components in deformed ensembles $\psi_{i,a}^j$ for a fixed configuration of $\eta_k$ are gaussian variables with variance given by \eqref{eq:eigenvector_weights_local_fields} with a factor $1/d$ (see \cite{truong2016eigenvectors} for the case of deformed Wigner ensembles).
The rescaled IPR 
$\kappa(\lambda)=NI(\lambda)$ can be computed combining eqs \eqref{eq:IPR} and \eqref{eq:eigenvector_weights_local_fields} and exploiting the gaussian nature of the $\psi_{i,a}^j$
\begin{equation}
\label{eq:rescaled_IPR_theory_bis}
\begin{gathered}
    \kappa(\lambda)=N\sum_{i=1}^N\langle |\bm{\psi}_i(\lambda)|^4 \rangle=3N\sum_{i=1}^N \langle |\bm{\psi}_i(\lambda)|^2 \rangle^2 \\
    \, \\
    \overset{N\rightarrow\infty}{=} 3\left(1-\frac{1}{d}\right)\left[\frac{(d-1)\alpha}{\sigma^2|\frac{d}{\sigma^2}-\Green(\lambda)|_{\mathbb{C}}^2}\right]^2
    \int\frac{d\eta P_\eta(\eta)}{\left|\eta-\lambda\sigma-\frac{\alpha\sigma}{d}\frac{\Green(\lambda)}{\frac{d}{\sigma^2}-\Green(\lambda)}\right|_{\mm{C}}^{4}}
\end{gathered}
\end{equation}

We now proceed to expand \eqref{eq:rescaled_IPR_theory_bis} in the limit $\lambda\rightarrow 0_+$. We consider the case of Mattis states of $d>2$ VHM. By proceeding similarly to the previous section, we find
\begin{equation}
\label{eq:rescaled_IPR_loweredge_scaling_patological}
    \kappa(\lambda)\approx \frac{3\left(d-1\right)^3}{d}(\alpha r_0)^2\frac{\pi P_h(-\Re x)}{2(\Im x)^3}\propto \lambda^{-2(d-1)}
\end{equation}
where we defined $x$ in \eqref{eq:x_def}. The scaling in \eqref{eq:rescaled_IPR_loweredge_scaling_patological} leads to $I(\lambda)\propto N\lambda^{2(d-1)}\propto N^{1-\frac{2}{d}}$ for the IPR, where we used the scaling $\lambda\sim N^{-1/d}$, valid for lower edge eigenvalues. Since by definition the IPR is bounded by unit (compare with the definition in \eqref{eq:IPR}), formula \eqref{eq:rescaled_IPR_theory_bis} cannot be true for the softest modes.
To highlight the origin of this paradox, consider equation \eqref{eq:eigenvector_weights_local_fields}. Applying it to the normalization of eigenvectors $\sum_k |\bm{\psi}_k^j|^2=1$, it returns
\begin{equation}
\label{eq:normalization_condition_in_terms_of_spec}
    1 = \frac{\alpha (d-1)}{\sigma^2\left|\frac{d}{\sigma^2}-\Green(\lambda)\right|_{\mm{C}}^2}\int \frac{d\eta P_\eta(\eta)}{\left|\eta-\lambda\sigma-\frac{\alpha\sigma}{d}\frac{\Green(\lambda)}{\frac{d}{\sigma^2}-\Green(\lambda)}\right|_{\mm{C}}^2}
\end{equation}
For any $\lambda>0$, this identity is valid: it is just the imaginary part of the spectral equation \eqref{eq:self_consistent_equation_resolvent}. However, at $\lambda=0$ the r.h.s. of \eqref{eq:normalization_condition_in_terms_of_spec} is equal to $1-\Lambda_0$, where $\Lambda_0<1$ is the $T=0$ Replicon introduced by eq. \eqref{eq:Replicon_T0}. Since for $d>2$ one has $\Lambda_0>0$ for Mattis states, it follows that \eqref{eq:normalization_condition_in_terms_of_spec} is wrong in the limit $\lambda\rightarrow 0_+$. The solution of such an impasse 
is that soft modes of Mattis states undergo condensation: by assuming the existence of a condensate weight $|\bm{\psi}_{i_*}(0)|^2=\Oo(1)$ for some node $i_*$, we get by using eq. \eqref{eq:eigenvector_weights_local_fields} with the condensate weight
\begin{equation}
    \label{eq:normalisation_at_lambda0}
    1=|\bm{\psi}_{i_*}(0)|^2+\frac{\alpha (d-1)}{\sigma^4\left[\frac{d}{\sigma^2}-\Green(0)\right]^2}\int \frac{d\eta P_\eta(\eta)}{\left[\eta-\frac{\alpha}{d}\frac{\Green(0)}{\frac{d}{\sigma^2}-\Green(0)}\right]^2}.
\end{equation}
The second addendum in the rhs of this last equation is just $1-\Lambda_0$ (see eqs \eqref{eq:local_fields_norms_pdf} and \eqref{eq:Replicon_T0}); therefore, we find that the condensate is equal to the Replicon $|\bm{\psi}_{i_*}(0)|^2=\Lambda_0$.

The presence of condensation in soft modes of continuous disordered systems is mathematically akin to a Bose-Einstein condensation, and is sometimes referred in literature as Random Matrix Bose-Einstein Condensation (RMBEC) \cite{Ikeda2023Bose}.
The RMBEC condition of our model can be obtained from eq.\eqref{eq:eigenvector_weights_local_fields} by imposing $|\bm{\psi}_{i_*}^{(k)}|^2=\Oo(1)$:
\begin{equation}
\label{eq:condensation_condition}
    \frac{\alpha}{\sigma^2\left|\frac{d}{\sigma^2}-\Green(\lambda_k)\right|_{\mm{C}}^2}\frac{d-1}{\left|\eta_{i_*}-\lambda_k\sigma-\frac{\alpha\sigma}{d}\frac{\Green(\lambda_k)}{\frac{d}{\sigma^2}-\Green(\lambda_k)}\right|_{\mm{C}}^2}=\Oo(N)
\end{equation}
that is valid in the regime $\lambda_k\sim N^{-1/d}$. It can be shown that eq. \eqref{eq:condensation_condition} implies an order statistics between low-rank eigenvalues and low-rank cavity fields of the kind $\lambda\sim \Lambda_0h$ \cite{lee2016extremal, franz2022delocalization}.

We can estimate the scale of eigenvalues in $N$ at which formula \eqref{eq:rescaled_IPR_theory_bis} does not need to be rectified with RMBEC. Let us consider lower edge modes with subextensive rank $k=\widetilde{k}N^{a}$ for $0<a<1$: then $\lambda_k=\widetilde{\lambda}N^{-\frac{1-a}{d}}$ and from eq. \eqref{eq:rescaled_IPR_loweredge_scaling_patological} it follows that RMBEC disappears at ranks with an exponent
\begin{equation}
    \lambda^{-2(d-1)}\leq N\Rightarrow a_*= \frac{d-2}{2(d-1)}
\end{equation}
For $a> a_*$, one has $I(\lambda)=\kappa(\lambda)/N\leq1$ and therefore for modes with rank $k\gg N^{a_*}$ eq. \eqref{eq:rescaled_IPR_theory_bis} is correct. 

\subsection{Limit $d\rightarrow\infty$ for the spectrum}
\label{sec:spectrum_infinite_d_limit}
Let us consider the large $d$ limit of the spectral equation \eqref{eq:self_consistent_equation_resolvent} in the two regimes $P\propto N d$ and $P\propto N/d$ described in Appendix \ref{sec:scaling_limits_computations}.
Let us begin with the first, where the capacity is $\alpha'=\frac{\alpha}{d}$. In this case, one has $P_h(h)\rightarrow \delta(h-\sqrt{m^2+\alpha' r'})$, and the spectral equation becomes (again, we use $\sigma=\sqrt{d}$ in the derivation)
\begin{equation}
    \label{eq:spectral_eq_SupLin}
    \Green(\lambda)=\frac{1}{\sqrt{m^2+\alpha' r_0'}+\frac{\alpha' \chi}{1-\chi}-\lambda-\frac{\alpha' \Green(\lambda)}{1-\Green(\lambda)}}
\end{equation}
Considering the no-retrieval physical solution $m=0$, $\chi=\frac{1}{1+\sqrt{\alpha'}}$, knowing that $r_0=\frac{1}{(1-\chi)^2}$, the solution of \eqref{eq:spectral_eq_SupLin} reads
\begin{equation}
\label{eq:spectral_eq_SupLin_sol}
    \Green(\lambda)=\frac{2(1+\sqrt{\alpha'})-\lambda}{2(1+2\sqrt{\alpha'}+\alpha'-\lambda)}+\frac{\sqrt{\lambda\,(\lambda-4\sqrt{\alpha'})}}{2(1+2\sqrt{\alpha'}+\alpha'-\lambda)}
\end{equation}
The spectrum is given by the continuous bulk (the imaginary part of \eqref{eq:spectral_eq_SupLin_sol}) and the pole in $\lambda=1+2\sqrt{\alpha'}+\alpha'$, which has weight $1-\alpha'$ and exists for $0\leq\alpha'\leq 1$. The final expression of the spectral density is thus
\begin{equation}
\label{eq:spectrum_SupLin}
    \rho(\lambda)=\frac{\sqrt{\lambda\,(4\sqrt{\alpha'}-\lambda)}}{2(1+2\sqrt{\alpha'}+\alpha'-\lambda)}+(1-\alpha')\vartheta(1-\alpha')\delta(\lambda-1-2\sqrt{\alpha'}-\alpha')
\end{equation}
This is the spectrum that one obtains for a $p=2$ spherical Hopfield model: this system is marginally stable having $\Lambda_0=0$, as it should be since the lower tail in eq. \eqref{eq:spectrum_SupLin} is $\propto \sqrt{\lambda}$. As a consequence, the corresponding eigenvectors are fully delocalized with $\kappa(\lambda)=\Oo(1)$ for all $0\leq \lambda\leq4\sqrt{\alpha'}$ or $\lambda=1+2\sqrt{\alpha'}+\alpha'$.

In the second case the capacity reads $\alpha''=\alpha d$. The spectrum becomes trivial for $d\rightarrow\infty$: again, $P_h(h)\rightarrow\delta(h-\sqrt{m^2+\alpha''r''})$, but the spectral equation is simply
\begin{equation}
    \Green(\lambda)=\frac{1}{\sqrt{m''^2+\alpha'' r_0''}-\lambda}
\end{equation}
In the SubLin regime, one has at equilibrium $\sqrt{m^2+\alpha''r_0''}=1$ and therefore $\rho(\lambda)=\delta(\lambda-1)$, as in the $\alpha=0$ case for finite $d$.

\section{Details on STN method to predict first-step denoising}
\label{sec:dyn_appendix}

\begin{figure}[h!]
    \centering
    \includegraphics[width=0.45\linewidth]{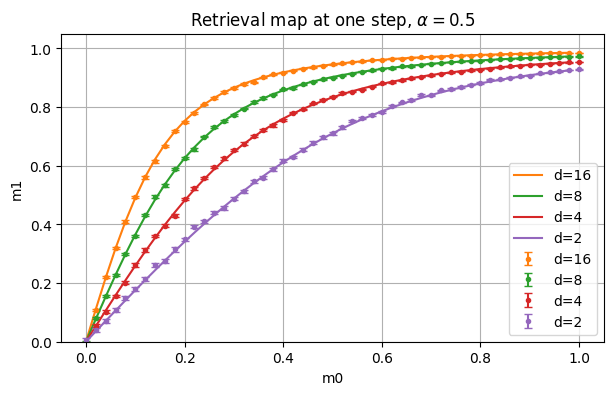}
    \includegraphics[width=0.45\linewidth]{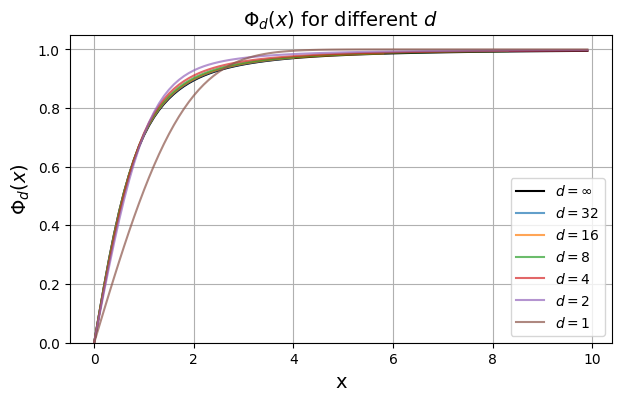}
    \caption{\textbf{Left} Overlap $m_1$ after one step, starting from an initial overlap $m_0$: comparison between numerical results [points] and analytical prediction from Signal-To-Noise (STN) [solid lines]. \textbf{Right} Function $\Phi_d(x)$ direct dependence on $d$ is weak already from modest values of $d$: the main dependence is through the argument $x=m_0/\sqrt{\alpha'}$.}
    \label{fig:transient_basins}
\end{figure}

\paragraph{Patterns}

Let us consider a fixed point $\conf{S}$ of eq. \eqref{eq:retrieval_versor_rule} with magnetization $m$ with the pattern $\conf{\xi}^1$: the local field vector on this point reads
\begin{equation}
\label{eq:signal_to_noise_stationary}
\begin{gathered}
    \frac{1}{\sigma}\sum_{j:j\neq i}\mm{J}_{ij}\bm{S}_j=\frac{1}{N\sigma}(\conf{\xi}^1\cdot\conf{S})\;\bm{\xi}_i^1+\frac{1}{N\sigma}\sum_{\mu=2}^P(\conf{\xi}^\mu\cdot\conf{S})\bm{\xi}_i^{\mu}+\Oo\left(\frac{1}{\sqrt{N}}\right) \simeq m\bm{\xi}_i^1+\bm{n}_i
\end{gathered}
\end{equation}
The vector $\bm{n}_i$ is an isotropic gaussian vector with zero mean and variance $\frac{\alpha r_0}{\sigma^2}$ (compare with def. \eqref{eq:noise_order_parameter}), where $r_0$ is the order parameter defined in eqs. \eqref{eq:sp_eqs_T0}.
Thanks to the stationarity condition \eqref{eq:stationarity_modelHam}, the spins are $\bm{S}_i=\frac{(m\bm{\xi}_i^1+\bm{r}_i)\sigma}{|m\bm{\xi}_i^1+\bm{r}_i|}$, hence the Mattis magnetization $m=\frac{1}{N\sigma}\conf{S}\cdot\conf{\xi}^1$ for large $N$ satisfies
\begin{equation}
\label{eq:sp_from_noise_to_signal_analysis}
\begin{gathered}
     m\,=\,\frac{1}{N\sigma}\sum_{i=1}^N\bm{\xi}_i^1\cdot\bm{S}_i=\int \frac{d\bm{n}\; e^{-\frac{n^2 \sigma^2}{2\alpha r_0}}}{(2\pi \alpha r/\sigma^2)^{d/2}}\int\frac{d\bm{\xi}\delta(|\bm{\xi}|-1)}{\mathcal{S}_{d-1}(1)}\frac{m\bm{\xi}+\bm{n}}{|m\bm{\xi}+\bm{n}|}\cdot\bm{\xi} \\
     \;\\
    =\int_0^{\infty}dn\; P_h\left(n\right)\g{d}\left(\frac{m \sigma n}{\alpha r_0}\right)=\Phi_d\left(\frac{m}{\sqrt{\alpha r_0}}\right)
\end{gathered}
\end{equation}
and we derived again the third of eqs. \eqref{eq:sp_eqs_T0}. We introduced a function $\Phi_d(\cdot)$ to represent in a compact fashion the integral in the last r.h.s. of \eqref{eq:sp_from_noise_to_signal_analysis}. For the standard Hopfield model, when $d=1$, one can show that $\Phi_1(\cdot)=\operatorname{Erf}\left(\frac{\cdot}{\sqrt{2}}\right)$.

We remark that the noise vectors $\bm{n}_i$ have the same statistics as the cavity fields of the RS solution, but they are not exactly cavity vectors since in cavity theory the cavity fields $\bm{h}_i$ vectors must be parallel to the spins $\bm{S}_i$. 

\vspace{10pt}
\paragraph{Mixture states} We proceed in a similar way as in the previous paragraph, but now consider as initial condition $  \frac{1}{\sigma}\bm{S}_i(0)=m_0\frac{\bm{z}_i}{z_i}+\sqrt{1-m_0^2}\;\bm{\pi}_i$, where $\bm{\pi}_i$ are unit vectors satisfying $\bm{\pi}_i\cdot \bm{z}_i=0$ and $\bm{z}_i=\sum_{\mu=1}^s c_\mu \bm{\xi}_i^{\mu}$ with $s=\Oo(1)$. We found

\begin{equation}
     \label{eq:m1_mixtures}
     \begin{split}
         \frac{1}{\sigma}&\sum_{j:j\neq i}\mm{J}_{ij}\bm{S}_j(0)=\widetilde{m}_0 \bm{z}_i+\bm{n}_i \\
         \; \\
         m_1&=\frac{1}{N}\sum_{i=1}^N \frac{\widetilde{m}_0\bm{z}_i+\bm{n}_i}{|\widetilde{m}_0\bm{z}_i+\bm{n}_i|}\cdot \frac{\bm{z}_i}{z_i}=\\
         &=\mathcal{S}_{d-1}(1)\int_0^s dz z^{d-1}\rho_s(z)\Phi_d\left(\frac{\widetilde{m}_0 z}{\sqrt{\alpha/d}}\right)
     \end{split}
 \end{equation}
 where the noise vector is again an isotropic gaussian vector with zero mean and variance $\alpha/d^2$, $\rho_s(z)$ is the pdf of the vector $\bm{z}=\sum_{\mu=1}^s c_{\mu}\bm{\xi}_{\mu}$ and $\widetilde{m}_0=\langle 1/z \rangle m_0$. We discuss how to compute the distributions $\rho_s(z)$ in Appendix \ref{sec:Instability_mixtures}.

\end{document}